# A fluorescence microscopy platform for time-resolved studies of spin-correlated radical pairs in biological systems

**Authors:**
Noboru Ikeya[1] and Jonathan R. Woodward[1]
1. Graduate School of Arts and Sciences, The University of Tokyo, Tokyo, Japan

**Corresponding Author:**
Jonathan R. Woodward
e-mail: jrwoodward@g.ecc.u-tokyo.ac.jp

**Abstract**

The importance of spin-correlated radical pairs in biology is increasingly recognized, with roles in biological effects of weak magnetic fields and emerging quantum spin-based biomedical applications. Fluorescence microscopy offers sufficient sensitivity to study magnetic field effects on radical pair reactions in living cells, but conventional techniques cannot directly resolve their dynamics because most biologically relevant radical pairs are non-emissive. To overcome this challenge, we introduce two novel microscopy techniques: single color pump–probe (PP) and pump-field-probe (PFP) fluorescence. Here, we describe their working principles, provide their mathematical formulation, and validate both techniques through theoretical analysis and experiments on well-established flavin-based magnetic field sensitive reactions. These approaches offer a sensitive and broadly applicable platform for quantifying and visualizing the quantum spin dynamics of radical pair chemical reactions in biological systems.

## Introduction

Spin-correlated radical pairs (SCRPs) are short-lived chemical reaction intermediates that render reactions that proceed through their formation and reaction sensitive to magnetic fields. Their effects were first observed in magnetic resonance spectra around 6 decades ago [1-3] and extensive studies throughout the ensuing period comprise the field known as spin chemistry [4]. In recent years they have drawn increasing attention due to their significance in solid state photoactive devices [5-7] and due to the radical pair mechanism (RPM) hypothesis that implicates them as the magnetosensitive component in the geomagnetic sensing abilities of many animals and in particular migratory birds [8-11]. Their potential importance in biology more generally is a current topic receiving much attention due to the enormous number of published but unexplained biological responses to magnetic fields [12] and indeed we have demonstrated magnetic field dependent RP-based photochemistry taking place in living cells observed via their natural autofluorescence [13]. Therefore, developing new tools to allow the direct, time-resolved observation of RPs in biological systems is paramount to making progress in understanding their detailed roles and significance to biological function.

To date, two optical strategies dominate time-resolved magnetic field effect (MFE) detection - transient absorption (TA) [14-19] and fluorescence [13, 20-23]. TA (flash photolysis) detects RPs directly and resolves their spectra and kinetics, making it the bulk-scale method of choice for mechanism elucidation [24]. However, at cellular and subcellular scales, ultrashort optical pathlengths, scattering backgrounds, and the difficulty of integrating cavity-enhanced schemes through objectives limit sensitivity. Fluorescence offers intrinsically low background and high sensitivity, enabling direct observation of MFEs from endogenous chromophores in cells. However, in biological systems (non-emissive RPs), existing fluorescence-based MFE readouts monitor precursors that become magnetically sensitive through both spin-state mixing and spin-selective RP recombination, rendering fluorescence indirect. This method requires an equilibrium state, no time-resolved information of short-lived RPs can be accessed—information crucial for defining their identity and dynamics. Also, this equilibrium constraint makes the observed MFE strongly dependent on excitation intensity, complicating reproducibility of MFEs in cells [25, 26].

To overcome this challenge, we here introduce two novel fluorescence microscopy techniques: single-color pump–probe (PP) and pump-field–probe (PFP) fluorescence. The PP technique monitors the dynamics of the total dark-state population, providing a platform for fluorescence-based, time-resolved detection of RPs. The PFP technique combines the PP method with rapidly switched magnetic-field techniques [27-31] to directly monitor RP dynamics.

The goal of this study was to fully develop and characterize these new tools. These techniques provide the same detailed time-resolved information on RPs currently only accessible by transient optical absorption detection techniques, while exploiting the greater sensitivity of fluorescence detection and thus providing a best-of-both worlds solution.

**Principle**

Here we describe the principle of the new techniques using a general RP-based magnetic field sensitive photochemical reaction scheme (**Fig. 1a**) as an example. Here the RP is born in the triplet state, but the scheme and analysis are also fully compatible with a singlet born RP. A more rigorous mathematical description is given in the Supporting Information (SI).

**Fig. 1a** shows a general, slightly simplified flavin RP photochemical reaction scheme [24]. Flash-photolysis fluorescence detects no MFE if the sample is refreshed between flashes, because the fluorescence arises from molecules that proceed RP formation. Therefore, existing MFE-based fluorescence microscopy uses continuous or pseudo-continuous excitation to establish an equilibrium between precursors and transient states [13, 20-23]. An external field modulates singlet–triplet (ST) mixing in the RP, and singlet RPs regenerate the precursor ground state spin-selectively, making the fluorescence yield magnetic field sensitive. However, this approach makes the observed MFE highly dependent on excitation intensity as a result of the equilibrium position and not the inherent magnetic field sensitivity of the RP reaction. Moreover, no time-resolved information of short-lived RPs can be accessed.

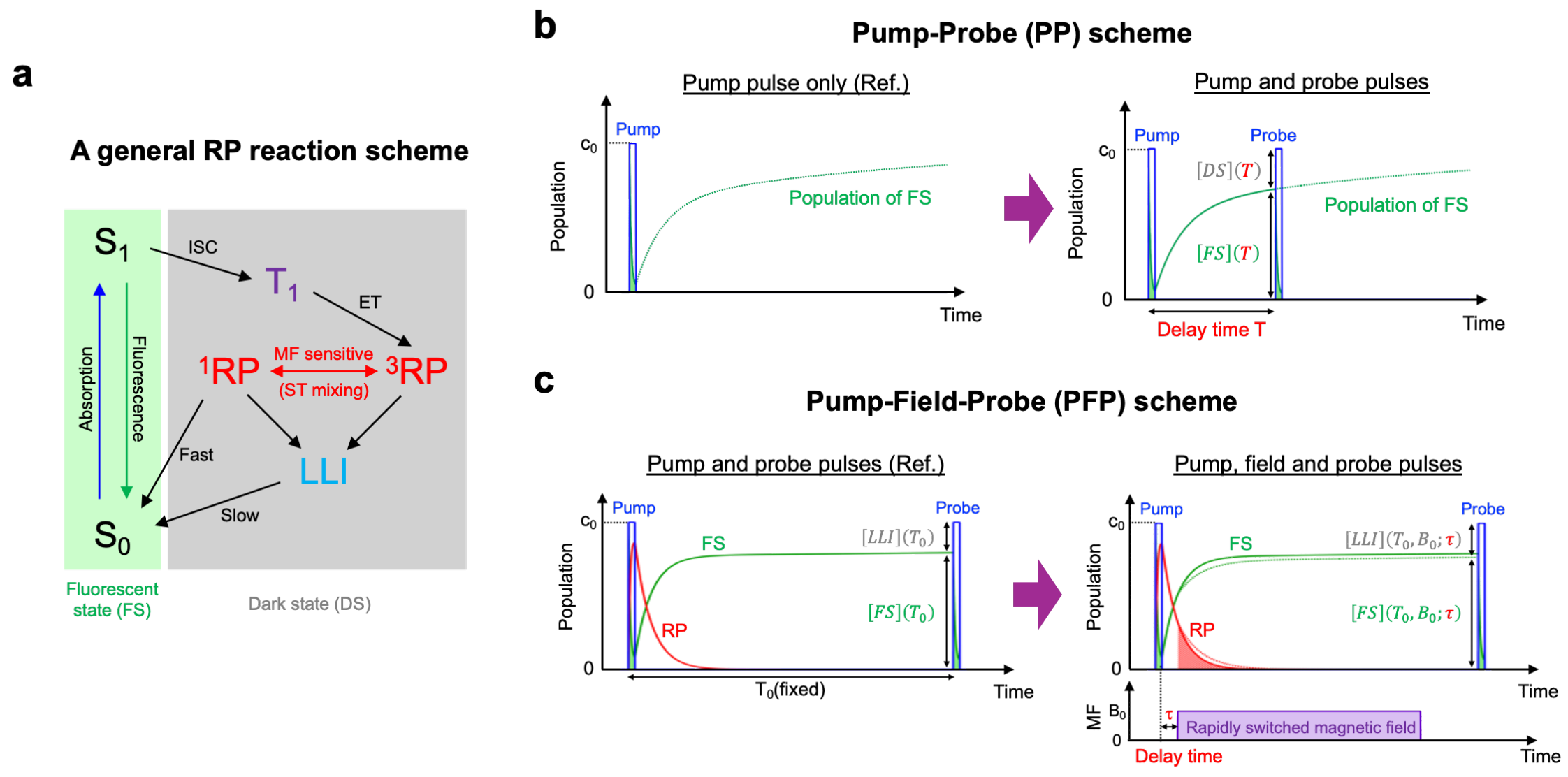

**Figure 1. Schematic representation of the principles of pump-probe (PP) and pump-field-probe (PFP) fluorescence techniques.** (a) A general triplet-born RP based reaction scheme. Upon photoexcitation, the molecule undergoes a transition from the singlet ground state ($S_0$) to the singlet excited state ($S_1$) and returns to $S_0$ producing fluorescence. In competition with fluorescence (and internal conversion), the $S_1$ can undergo intersystem crossing (ISC) to form an excited triplet state ($T_1$). This can accept an electron from an electron donor to generate a spin-correlated triplet RP ($^3$RP) via electron transfer (ET). This RP undergoes magnetic field sensitive coherent singlet-triplet spin state mixing (ST mixing), and the resulting singlet RP ($^1$RP) can undergo spin-selective, rapid back electron transfer to regenerate $S_0$. Alternatively, radicals can escape from the pair generating a magnetically insensitive long lived intermediate (LLI) state such as free radicals or reaction products that are returned to the ground state on much slower time scales or not at all. An external magnetic field can alter both the rate and extent of ST mixing of the RP, thereby changing the concentration of the $S_0$ – $^1$RP precursor. Under short-pulse excitation, $S_0$ and $S_1$ collectively are classified as the fluorescent state (FS), whereas $T_1$, RP, and LLI are classified as the dark state (DS). (b) PP fluorescence detection scheme. In PP, A pump pulse drives molecules into the non-emissive dark state. After a delay time T, a second (probe) pulse excites the sample and the resulting fluorescence is recorded. The reduction of the probe fluorescence relative to the pump-only fluorescence reflects the dark-state population at time T. Using the pump-only fluorescence signal as a reference, scanning the probe delay measurement monitors the time evolution of the dark-state population. (c) PFP fluorescence detection scheme. In PFP, a pump pulse generates RPs and a probe pulse excites the sample at a fixed delay time $T_0$ that is longer than the $T_1$/RP existence time but shorter than the LLI existence time, so the probe fluorescence reports only the LLI concentration. If a rapidly switched magnetic field (RSMF) is applied within the RP existence time, the number of RPs exposed to the field modulates the subsequent LLI yield, whereas applying the field after all RPs have decayed produces no change. Using the probe fluorescence signal in an absence of the RSMF as a reference, scanning the RSMF delay measurement directly monitors the dynamics of the RP population.

To overcome these limitations, we use a combination of two single color laser pulses. The key requirement is that the width of the laser pulses is longer than the fluorescence lifetime but shorter than the total duration of the formation and lifetime of the dark state species (typically nanoseconds to microseconds). This enables a fully binary classification of states.

Under pulsed excitation on the timescale of nanoseconds, the molecule undergoes fluorescence through repetitive transitions between the ground state ($S_0$) and the excited singlet state ($S_1$) until it undergoes transition to another excited state (i.e. by intersystem crossing or direct photochemical reaction). Once the molecule makes this transition, it does not return to the ground state during the excitation pulse, because the pulse duration is short relative to the lifetime of the other excited states. Therefore, molecules that undergo this transition do not emit fluorescence from that point on. This leads us to classify $S_0$ and $S_1$ collectively as the *fluorescent state (FS)* and $T_1$, RP, and LLI as the *dark state (DS)*:

**Fluorescent state (FS)** = {$S_0$, $S_1$}, **Dark state (DS)** = {$T_1$, RP, LLI}

Consequently, the total fluorescent and dark state population can be defined as:

$$[FS](t) = [S_0](t) + [S_1](t) \quad \text{(eq. 1)}$$

$$[DS](t) = [T_1](t) + [RP](t) + [LLI](t) \quad \text{(eq. 2)}$$

where $[X](t)$ denotes the population state $X \in \{S_0, S_1, T_1, RP, LLI\}$. In principle, the total population is conserved. At $t = 0$, the beginning of the excitation, all molecules can be regarded to be in the singlet ground state:

$$[FS](t) + [DS](t) = c_0, \qquad [S_0](0) = c_0 \quad \text{(eq. 3,4)}$$

From these definitions, in the case of two single color laser pulses, both of width *w*, the fluorescence intensity from the pump pulse, $F_{pu}$, and from the probe pulse at a pump-probe delay time *T*, $F_{pr}(T)$, can be calculated as follows (SI for further details):

$$F_{pu} = \int_0^w k_F\,[S_1](t)dt \propto [FS](0) = c_0 \quad \text{(eq.5)}$$

$$F_{pr}(T) = \int_T^{T+w} k_F\,[S_1](t)dt \;\propto [FS](T) = c_0 - [DS](T) \quad \text{(eq 6)}$$

These relations form the basis of two new detection methodologies: PP and PFP fluorescence detection (**Fig. 1b and 1c**).

<u>*Pump probe (PP) fluorescence detection*</u>

Fluorescence is recorded sequentially under pump-probe and pump-only excitation (**Fig.1b**). The reduction of the probe fluorescence relative to the pump-only fluorescence reflects the dark state population at delay *T*. On this basis, to isolate the probe-only fluorescence signal, the difference signal in the PP is defined as:

$$\Delta F_{PP}(T) = F_{pu+pr}(T) - F_{pu} \quad \text{(eq. 7)}$$

where $F_{pu+pr}(T) = F_{pu} + F_{pr}(T)$. For equal pump and probe pulse widths and intensities, the proportionality constants of (eq.5) and (eq.6) are identical (see SI). Consequently, the normalized fluorescence difference signal is expressed as follows:

$$\Delta\overline{F_{PP}}(T) = \frac{\Delta F_{PP}(T)}{F_{pu}} = 1 - \overline{[DS]}(T) \quad \text{(eq. 8)}$$

Thus, the population dynamics of the total dark state can be monitored as a function of the fluorescence difference signal with respect to the pump probe delay time *T*.

Time resolved MFEs of the total dark state species can be obtained from the difference of the normalized fluorescence difference signals, $\Delta\Delta\overline{F_{PP}}$, with and without an applied magnetic field ($B_0$):

$$\Delta\Delta\overline{F_{PP}}(T,B_0) = \Delta\overline{F_{PP}}(T,B_0) - \Delta\overline{F_{PP}}(T,0) = -\left(\overline{[DS]}(T,B_0) - \overline{[DS]}(T,0)\right) \quad \text{(eq. 9)}$$

Alternatively, the difference of the fluorescence difference signals, $\Delta\Delta F_{PP}$, can be obtained from the difference between the pump probe fluorescence signals with and without an applied magnetic field:

$$\Delta\Delta F_{PP}(T,B_0) = F_{pu+pr}(T,B_0) - F_{pu+pr}(T,0) \propto -\left([DS](T,B_0) - [DS](T,0)\right) \quad \text{(eq. 10)}$$

which avoids error propagation of double differencing (eq. 7 and 9) and increases the precision.

In this model (**Fig.1a**), the population of the excited triplet state is not changed with magnetic field application, therefore $\Delta\Delta F_{PP}$ monitors the time dependence of the MFEs of the combined RP and LLI states.

$$\Delta\Delta F_{PP}(T,B_0) \propto -\left(\Delta RP(T,B_0) + \Delta LLI(T,B_0)\right) \quad \text{(eq. 11)}$$

where $\Delta X(t,\mathrm{B}_0) = X(t,\mathrm{B}_0) - X(t,0)$.

Pump-field-probe (PFP) fluorescence detection

Here (**Fig.1c**), fluorescence under pump-probe excitation is recorded sequentially with and without a rapidly switched magnetic field (RSMF) at a fixed pump-probe delay time $T_0$, while varying the RSMF switching delay τ. To isolate the probe-only fluorescence signal, the difference signal in the PFP is defined as:

$$\Delta F_{PFP}(\tau,B_0) = F_{pu+pr}(T_0,B_0;\tau) - F_{pu+pr}(T_0,0) \quad \text{(eq. 12)}$$

The pump-probe delay time $T_0$ is set much longer than the $T_1$ and RP lifetimes but shorter than the LLI lifetime. Consequently, at delay $T_0$, the $T_1$ and RP populations are negligible, and the dark state equals the LLI:

$$\Delta F_{PFP}(\tau,B_0) \propto -\left([LLI](T_0,B_0;\tau) - [LLI](T_0,0)\right) \quad \text{(eq. 13)}$$

Thus, $\Delta F_{PFP}$ monitors the change of the LLI (i.e. the reaction yield) from the RP with the application of a RSMF. If the RSMF is applied within the RP existence time, the number of RPs exposed to the field modulates the subsequent LLI yield, whereas applying the field after all RPs have decayed produces no change. Therefore, scanning the RSMF delay directly monitors the time evolution of the RP population.

**Instrument**

A new single-color laser pump-probe excitation system and a rapidly switched magnetic field system are introduced into the custom-built fluorescence microscope presented in a previous study [13]. Fig. 2 shows the schematic of the instrument. To achieve arbitrary delay times between pump and probe pulses, two independent, identical 450 nm nanosecond pulse lasers are used for single-color pump-probe excitation. These pulses are combined into a multimode fiber using a knife-edge prism mirror and collimated for optimal spatial overlap on a sample. In PP measurements, the static magnetic field is generated by a projected field electromagnet (GMW5204, GMW Associates). In PFP measurements, the rapidly switching magnetic field (RSMF) is generated by a capacitor bank-based custom rapid risetime pulser circuit and a homemade solenoid coil (5 turns, 4 mm diameter). This setup allows sub-30 ns rise-time switching and provides a flat magnetic field output on the microsecond timescale [29,30]. The fluorescence is captured using an sCMOS camera through a 100×/NA 1.49 objective lens with a dichroic mirror, a reflection mirror, a long-pass filter, and a tube lens. The timing for the two laser pulses, the static magnetic field, the RSMF, and the camera are controlled by a custom circuit based on the Raspberry Pi Pico microcontroller with data acquisition programs written in Micropython (and PIO assembly language) and LabVIEW code. Building on this setup, all measurements used a laser repetition rate $f_{rep}$ low enough to suppress residual LLIs and their MFEs. Fluorescence images were recorded with exposure Δt = 200 ms, (i.e., 0.2*$f_{rep}$ excitations per frame). The integrated fluorescence signal was defined as the ROI-averaged intensity to reduce pixel-to-pixel fluctuations fluctuations. Details of the data analysis are described in the SI.

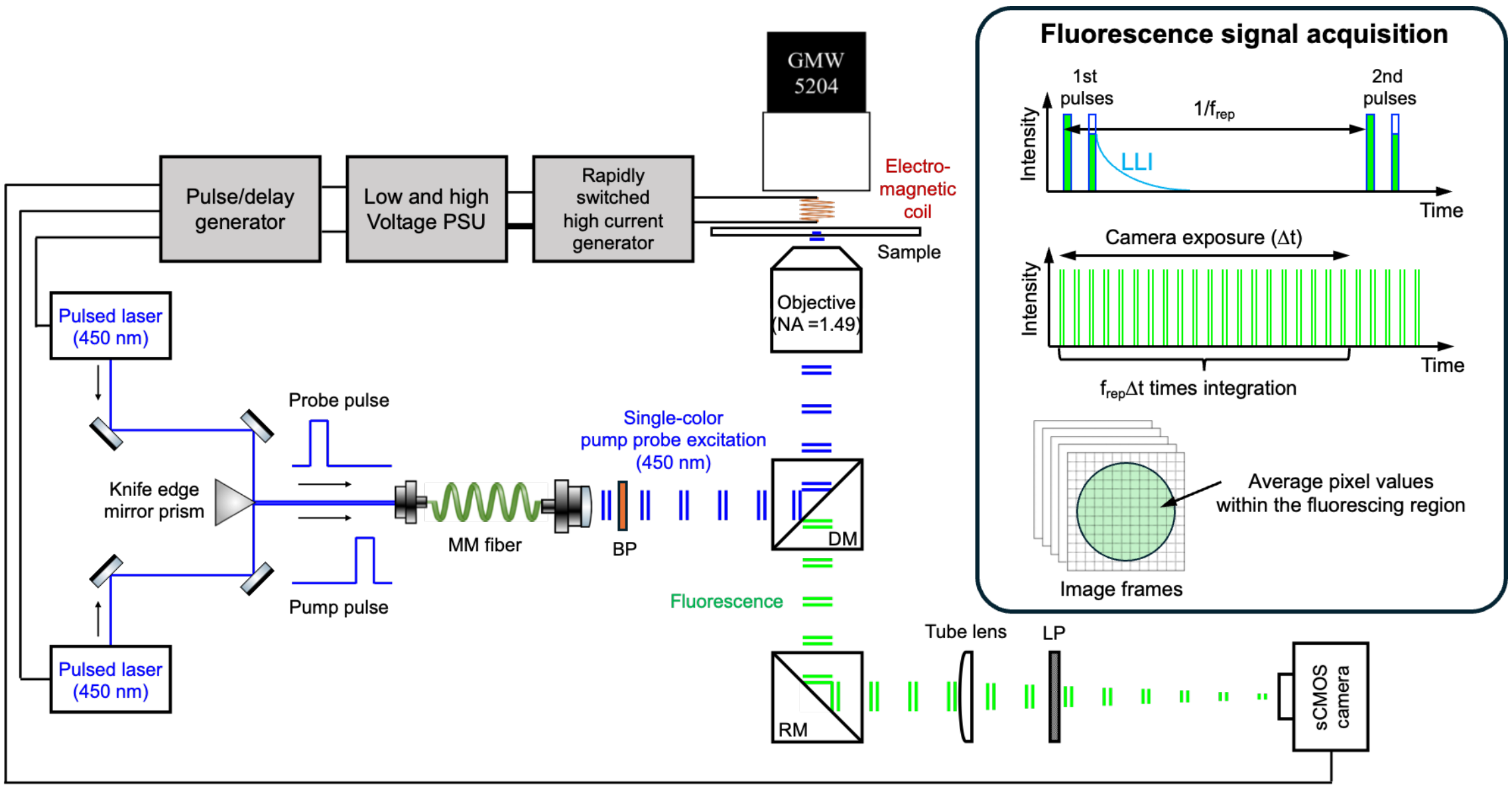

**Figure 2. Schematic of microscope setup.** For fluorescence signal acquisition, samples are excited at a repetition rate of $f_{rep}$ set sufficiently low to allow the reaction to complete, or at least to minimize signal distortion from residual long-lived intermediates (LLIs). Subsequently, fluorescence signals are continuously integrated by camera-based detection over an exposure time (Δt). The integrated fluorescence intensity is defined as the average pixel intensity within the fluorescing region of interest (ROI), thereby reducing pixel-to-pixel fluctuations caused by spatial non-uniformities in excitation intensity and providing a more accurate value of fluorescence intensity.

## Results

To validate the techniques and their analyses, we studied well-characterized flavin photochemical reactions with increasingly complex intermediates. A key aim is using these techniques to measure how an applied magnetic field affects the kinetics of photoinduced reactions in living cells. All measurements were thus performed using flavins at typical endogenous concentrations (~µM, [32]) within measurement volumes of typical cells (1.2~4.29 pL, [33]), within typical adherent cell thickness (<9.0 µm, [34]).

*FMN system*

To show the PP fluorescence technique can track the total dark state population, we used flavin mononucleotide (FMN) in isotropic solution. This simple photoreaction system forms no RPs and shows no MFE. Photoexcitation of the singlet ground state ($^1$FMN) yields the singlet excited state ($^1$FMN$^*$) , which fluoresces, converts internally, or undergoes intersystem crossing (ISC) to the excited triplet state ($^3$FMN$^*$) (**Fig. 3a**). $^3$FMN$^*$ then either decays back to $^1$FMN (microseconds) or photobleaches, forming stable, non-fluorescent photoproducts (e.g., lumiflavin and lumichrome) [35].

**Fig. 3b** plots $1 - \Delta\overline{\mathrm{F_{PP}}}$ versus pump-probe delay, T at different laser repetition rates. The decay of $1 - \Delta\overline{\mathrm{F_{PP}}}$ represents the dynamics of the dark state species and here corresponds to the return of $^3$FMN$^*$ to $^1$FMN. At T = 0, the value equals the fraction of dark state population created by the pump. The non-zero signal saturation, which increases with laser repetition rate, results from the accumulation of long-lived photo-excited FMN intermediates or products, as diffusion is insufficient for complete ground-state recovery between excitations. (see SI for details).

In this measurement, the low concentration of FMN (10 μM) minimizes quenching effects due to intermolecular interactions. Consequently, the observed decay can be described by first-order reaction kinetics. The decay is attributed to $^3$FMN$^*$ and fitted with a single exponential function as the long-lived species have much longer lifetimes:

$$\overline{[T_1]}(t) = \overline{[T_1]}_0 \mathbf{e}^{-(k_{T_1}+k_B)t} \approx \overline{[T_1]}_0 \mathbf{e}^{-k_{T_1}t} \qquad \text{(eq. 14)}$$

The fitted rate coefficients for the triplet state lifetimes of 2.64 ± 0.05 μs, 2.74 ± 0.05 μs, and 2.64 ± 0.05 μs at laser repetition rates of 30 Hz, 50 Hz, and 100 Hz, respectively and are consistent with the reported value in non-degassed solution (2.93 ± 0.02 μs, ref.16).

The saturation value of $\Delta\overline{F_{PP}}$ at long delays enables estimation of the photobleaching quantum yield and if photobleaching occurs only from $^3$FMN$^*$, satisfies:

$$1 - \Delta\overline{F_{PP}}(T_{long}) \approx \frac{k_B}{k_{T_1} + k_B}\overline{[T_1]}_0 = \phi_B \qquad \text{(eq. 15)}$$

Thus, the photobleaching quantum yield ($\phi_B$) can be directly estimated by measuring the $\Delta\overline{F_{PP}}$ signal at long delays and low laser repetition rates. **Fig. 3c** shows $1 - \Delta\overline{F_{PP}}$ for $T_{long}$ = 30 μs at different laser repetition rates. Below 20 Hz, $1 - \Delta\overline{F_{PP}}(T_{long})$ is minimized with a photobleaching quantum yield of 0.028 in agreement with the 40 ms dwell time of molecules in the detection volume at 25Hz (see SI).

Thus, PP can directly monitor the dynamics of $^3$FMN$^*$ and estimate the photobleaching quantum yield per excitation. This has broad application to many fluorescent molecules, even when RPs are not involved.

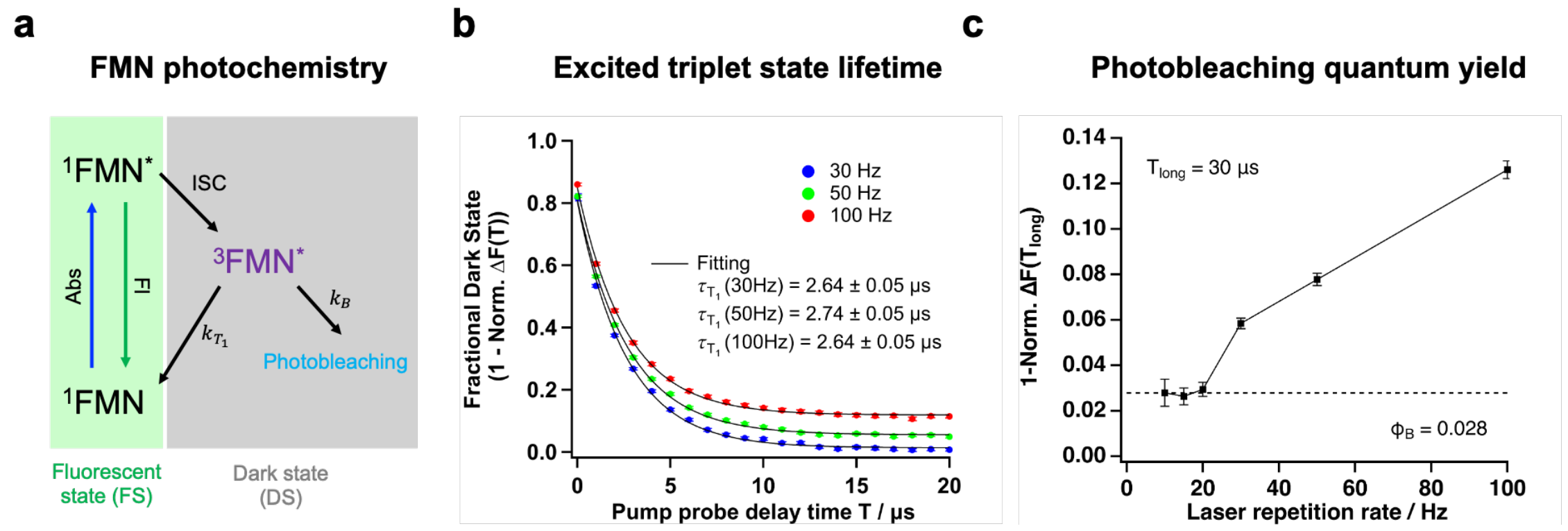

**Figure 3. The pump-probe (PP) measurements of 10 μM FMN in PBS buffer (pH 7.4, Sample thickness = 4.95 μm).** (a) Photoreaction scheme of FMN. Upon photoexcitation, the singlet ground state FMN ($^1$FMN) initially generate the singlet excited state ($^1$FMN$^*$) which undergoes fluorescence (and interconversion) to the ground state or intersystem crossing to the excited triplet state ($^3FMN^*$). In the reaction cycle, the excited triplet state decays back to the singlet ground state on a microsecond timescale but also undergoes photobleaching to generate stable photoproducts, which do not fluoresce. (b) The excited triplet state lifetime estimation. $1 - \Delta\overline{F_{PP}}$ is plotted as a function of pump-probe delay at different laser repetition rates (30 Hz, 50 Hz and 100 Hz). Curve fitting is performed with a mono exponential function (eq.14). (c) Photobleaching quantum yield measurement. The $1 - \Delta\overline{F_{PP}}$ at a long pump probe delay ($T_{long}$ = 30 µs) is plotted with different laser repetition rates. At excitation repetition rates sufficiently low compared to diffusion, the signal from the accumulation of photobleached species is negligible, and the measured value reflects the photobleaching quantum yield.

*FAD system*

We employ the well-characterized FAD photochemistry in acidic solution (which forms an intramolecular RP and exhibits MFEs [15,37-40]) to demonstrate the PP technique for time-resolved RP studies. Our preliminary data suggest that previously reported MFEs for FAD at physiological pH [40] are likely not due to an intramolecular RP mechanism, as these effects only appear at high FAD concentrations and high excitation repetition rates. Therefore, we focus on the established kinetics of FAD in acidic solution to rigorously validate the PP technique.

**Fig. 4a** displays the low-pH (>3.6) FAD photochemistry proposed by Murakami et al. [15]. Following photoexcitation, a protonated triplet state (PTS, $^3FH^+-AH_2^+$) is formed by intersystem crossing/protonation [15]. The PTS generates a triplet radical pair (T-RP, $^3\{FH^\bullet - AH^{\bullet+}\}$) via coupled electron transfer/deprotonation, which remains in pseudo-equilibrium with the PTS. The open FAD conformation at low pH allows ST-mixing. RPs then undergo rapid spin-selective back electron transfer, restoring the fluorescent FAD ground state.

**Fig. 4b** plots $1 - \Delta\overline{F_{PP}}$ as a function of pump-probe delay time at laser repetition rates of 50 Hz and 100 Hz with and without a magnetic field. In both cases, the decay of $1 - \Delta\overline{F_{PP}}$ is slower in the presence of the magnetic field, consistent with a triplet-born RP. As for FMN, $\Delta\overline{F_{PP}}$ saturates at increasing values at higher repetition rates indicating minor accumulation of LLI, (e.g. from photobleaching). However, the $\Delta\overline{F_{PP}}$ signal in the presence and absence of a magnetic field, $\Delta\Delta\overline{F_{PP}}$, shows the same time dependence at 50 Hz and 100 Hz, indicating that LLIs do not affect RP dynamics (Fig. **S7-1 a**).

$\Delta\Delta F_{PP}$ is the difference between pump-probe fluorescence signals with and without a magnetic field. **Fig. 4c** compares $\Delta\Delta F_{PP}$(Eq. 10) with $\Delta\Delta\overline{F_{PP}}$ (Eq. 9) at 50Hz and confirms the methods agree, while also demonstrating the superior precision of the Eq. 10 approach. Furthermore, $\Delta\Delta F_{PP}$is independent of the laser repetition rate (Fig. **S7-1 b**), confirming that RP dynamics are not affected by these rates.

To validate the robustness of this approach, we compared PP with transient absorption (TA) detection. The magnetic field dependent dark states are the protonated triplet state, $^3FH^+$, and the neutral radical state, $FH^\bullet$. Therefore, the $\Delta\Delta F_{PP}$ signal is expressed as follows:

$$\Delta\Delta \mathrm{F_{PP}}(\mathrm{T}, \mathrm{B_0}) \propto -\left(\Delta[^3\mathrm{FH}^+](\mathrm{T}, \mathrm{B_0}) + \Delta[\mathrm{FH}^\bullet](\mathrm{T}, \mathrm{B_0})\right) \qquad \text{(eq. 16)}$$

where $\Delta X(t, \mathrm{B_0}) = X(t, \mathrm{B_0}) - X(t, 0)$. The optical absorption detection signal, ΔΔA, which represents the MFEs, is given by [15].

$$\Delta\Delta A(T, B_0, \lambda) \propto \varepsilon_T(\lambda)\Delta[^3\mathrm{FH}^+](T, B_0) + \varepsilon_R(\lambda)\Delta[\mathrm{FH}^\bullet](T, B_0) \qquad \text{(eq. 17)}$$

where $\varepsilon_{T/R}(\lambda)$ is the absorption coefficient of the triplet/radical state at a given wavelength (λ). The $\Delta\Delta F_{PP}$ and ΔΔA signals differ only in sign and the scaling by molar absorption coefficients. For equal triplet and radical states absorption coefficients, $\Delta\Delta F_{PP}$ and ΔΔA are negatively proportional.

$$\varepsilon_T(\lambda_0) = \varepsilon_R(\lambda_0) \Rightarrow \Delta\Delta F(T, B_0) \propto -\Delta\Delta A(T, B, \lambda_0) \qquad \text{(eq. 18)}$$

In flavins, this condition approximately true in the wavelength range 500 - 550 nm [15].

A comparison of $\Delta\Delta F_{PP}$ (**Fig. 4b, 4c**) with ΔΔA from TA [15] and our TOAD microscope [40] initially showed different time dependencies. This was traced to changes in pH near the sample cover glass surface (see SI), not the detection method. When ΔΔA was remeasured with the same 3-micron thickness in TOAD, the result matched $\Delta\Delta F_{PP}$ (**Fig. 4d**). This confirms that ΔΔA and $\Delta\Delta F_{PP}$ yield identical time-resolved data (exceptions discussed below). Critically, the fluorescence method demonstrated much superior signal-to-noise ratio despite a 20 times reduction in concentration. With longer averaging, concentrations as low as a few hundred nanomolar can be detected.

Careful comparison reveals small but important differences between TA (ΔA and ΔΔA) and fluorescence ($\Delta F_{PP}$ and $\Delta\Delta F_{PP}$) measurements on the same sample. Both ΔA and $\Delta F_{PP}$ do not return to zero, indicating a long-lived intermediate (LLI1). However, since ΔΔA does return to zero, is not magnetically field-sensitive and likely forms before formation. Conversely, $\Delta\Delta F_{PP}$ does not return to zero, suggesting a second species (LLI2) is formed after RP formation. The return of ΔΔA to zero further implies LLI2 does not absorb the probe wavelengths (532 nm and 598 nm). This difference highlights that fluorescence measures the total non-fluorescent dark state species, while TA only detects absorbing species. The generalized features of PP detection and TA detection are shown in **Fig.4e**. LLI1 is tentatively identified as lumiflavin, a long-lived, absorbing species (532 nm, 598 nm) generated before RP formation [41]; LLI2 may be $FADH_2$, produced after RP formation via electron transfer and non-absorbing at these wavelengths [42]. A detailed analysis is future work.

While PP fluorescence directly quantifies the total dark state species (avoiding reliance on molar absorption coefficients for unknowns), TA provides necessary spectral information to resolve multiple intermediates. These distinct advantages make the two techniques complementary tools for studying complex photochemistry.

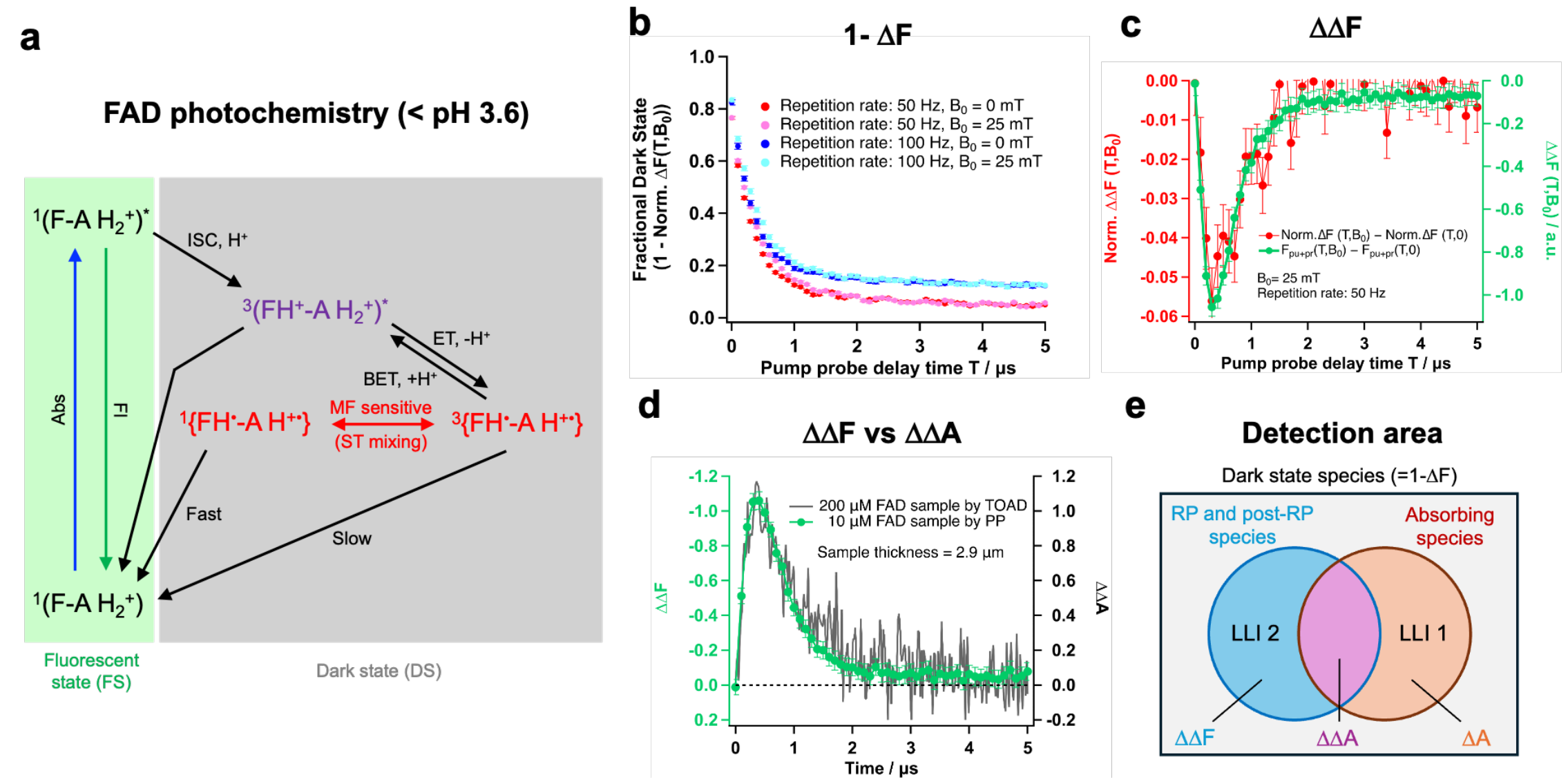

**Figure 4. The pump-probe (PP) measurements of 10 μM FAD in an acidic solution (pH 2.3).**
(a) The proposed photoreaction scheme of FAD at low pH (<pH 3.6). At low pH (<3.6), the adenine group of ground state FAD is protonated ($^1F{-}AH_2^+$). Upon photoexcitation, the excited singlet state FAD ($^1F^*{-}AH_2^+$) generates the protonated triplet state ($^3FH^+{-}AH_2^+$) through intersystem crossing and proton transfer from the adenine moiety to the flavin part in its excited triplet state. This protonated triplet state undergoes intramolecular electron transfer from the adenine moiety to the flavin part, forming the triplet RP ($^3\{FH^\bullet{-}AH^{+\bullet}\}$). This triplet RP only regenerates the protonated excited triplet state through electron transfer coupled with protonation. Under these low pH conditions, the FAD molecule adopts an open, flexible conformation that allows sufficient spatial separation of the flavin and adenine radical moieties, enabling the RP to undergo ST-mixing. Singlet RPs can undergo spin-selective back electron transfer to regenerate the ground state of FAD. (b) Plots of $1-\Delta\overline{F_{PP}}$ as a function of pump-probe delay at laser repetition rates of 50 Hz and 100 Hz, both in the presence (25 mT) and absence of an external magnetic field. (c) Comparison of $\Delta\Delta F_{PP}$ measured by two different schemes (eq.10 and eq.11). Laser repetition rate = 50 Hz. The $\Delta\Delta\overline{F_{PP}}$ is obtained from the data shown in **Fig.4b**. (e) Comparison of the difference in fluorescence difference signals, $\Delta\Delta F_{PP}$, and with the difference in transient absorption difference signal, $\Delta\Delta A$, measured by transient optical absorption detection (TOAD) microscope [18,19]. The FAD concentrations were 10 μM (PP) and 200 μM (TOAD). Sample thickness = 2.9 μm. $B_0$ = 20 mT. Due to the thin sample, a 594 nm probe was used to maximize the detectable PTS/RP absorbance signal, though $\varepsilon_T(\lambda) \approx 1.5\varepsilon_R(\lambda)$. (e) General feature of PP and TA detection area. ΔF monitors all dark-state molecular species. ΔA monitors only those dark-state species that absorb light. ΔΔF monitors the MFEs on the RPs and on post-RP species. ΔΔA monitors the MFEs on the absorbing RPs and post-RP species.

*FMN/tryptophan system*

The simple intermolecular RP system of FMN and tryptophan [43] is employed to demonstrate PFP's ability to selectively monitor RP existence time. The acidic reaction scheme is shown in **Fig. 5a**. Upon short-pulse excitation, $^3FMN^*$ can be quenched by intermolecular electron transfer to tryptophan, generating a RP composed of FMN•⁻/FMNH• and TrpH•⁺/Trp• in a pH-dependent manner. As $pK_a$(FMN$^{\bullet-}$/FMNH•) ≈ 8–8.5 [44] and $pK_a$ (TrpH•⁺/Trp•) ≈ 4.3–4.5 [45], at pH 2.3 the RP predominantly comprises the neutral flavin semiquinone and the protonated tryptophan radical, $\{FMNH^\bullet{-}TrpH^{+\bullet}\}$. In water, the RP is short-lived due to rapid diffusion, but FMN carries an overall negative charge (deprotonated phosphate), so electrostatic attraction between the oppositely charged partners can prolong the solvent-cage lifetime. By contrast, at pH 6.4, TrpH•⁺ rapidly deprotonates to Trp•, giving a RP (FMNH$^\bullet$– Trp$^\bullet$) with no Coulomb attraction and a shorter lifetime. Consistent with this, lowering pH from 6.4 to 2.3 increases the MFE for FMN/tryptophan (**Fig. S8-1**). Accordingly, we validated the PFP framework in acidic solution. In this system, the $\Delta F_{PFP}$ signal corresponds to the change in concentration of FMNH$^{\cdot}$ with and without the application of the RSMF:

$$\Delta F_{PFP}(\tau, B_0) = -([\text{FMNH}^{\cdot}](T_0, B_0; \tau) - [\text{FMNH}^{\cdot}](T_0, 0)) \qquad \text{(eq. 19)}$$

**Fig 5b and 5c** present normalized $\Delta F_{PFP}$ from Off-On and On-Off shift measurements across different concentrations. In the Off-On mode (**Fig. 5b**), applying the RSMF after delay $\tau_{01}$ results in a decreasing $\Delta F_{PFP}$ as $\tau_{01}$ increases, reflecting fewer RPs for field on. This decay rate accelerates with increasing tryptophan concentration due to a faster RP formation rate. Conversely, the On-Off mode (**Fig. 5c**), where the RSMF is removed after $\tau_{01}$, shows a rising $\Delta F_{PFP}$ as $\tau_{01}$ increases, reflecting more RPs exposed to the field. This rise rate also increases with tryptophan concentration. Note that the signal at zero delay is caused by the RSMF's slow fall time (~30 ns, see SI), which leaves a small residual field present just after the pump pulse.

With 10 μM FMN, escaped-radical concentrations are very low and f-pair reactions are negligible on our observation timescale. In PFP, an RSMF is applied only during a chosen delay after excitation and the probe fluorescence signal is detected at a fixed delay (3 μs). Therefore, the PFP signals can be represented by a four-state model and fit using a biexponential function by introducing a magnetic dependent recombination rate parameter, $k_{rec}(B)$. (see SI):

$$\Delta F_{PFP}^{(OFF-ON)}(\tau_{01}, B_0) = -\left[C_{T_1}^{(OFF-ON)} \mathrm{e}^{-k_{T_1}\tau_{01}} + C_{RP}^{(OFF-ON)} \mathrm{e}^{-k_{RP}(0)\tau_{01}}\right] \qquad \text{(eq. 20)}$$

$$\Delta F_{PFP}^{(ON-OFF)}(\tau_{10}, B_0) = -\left[C_{T_1}^{(ON-OFF)}\left(1 - \mathrm{e}^{-k_{T_1}\tau_{10}}\right) + C_{RP}^{(ON-OFF)}\left(1 - \mathrm{e}^{-k_{RP}(B_0)\tau_{10}}\right)\right] \qquad \text{(eq. 21)}$$

The electron transfer (quenching) rate and radical pair lifetime can be estimated from these functions, confirming that PFP correctly monitors the existence time. This is validated here using Off-On RSMF shift measurements, with similar principles applicable to On-Off shift measurements.

**Fig 5d** shows decay rates from single-exponential fits of the Off–On $\Delta F_{PFP}$ signals versus tryptophan concentration. When tryptophan is low ($k_{RP} \gg k_{T_1}$), the RP lifetime's impact on its existence time is negligible. Consequently, $\Delta F_{PFP}$ can be accurately fitted using a single-exponential function:

$$\Delta F_{PFP}^{(OFF-ON)}(\tau_{01}, B_0) \approx -C_{T_1}^{(OFF-ON)} \mathrm{e}^{-k_{T_1}\tau_{01}}, \qquad \text{(eq. 22)}$$

$$k_{T_1} = k_q[\text{Trp}] + k_{DT} \qquad \text{(eq. 23)}$$

The decay rates increase linearly from 0 to 3 mM tryptophan. A linear fit yields the second-order quenching rate constant, $k_q = 2.65 \pm 0.11 \times 10^9$, consistent with literature values (2.0-3.0 × $10^9$ $M^{-1}s^{-1}$, ref.46). Above 5 mM, the rates plateau, indicating the RP formation rate ($k_{T_1}$) is becoming comparable to or exceeding the RP decay rate ($k_{RP}$). At 10 mM, a single-exponential fit predominantly reflects the RP decay rate, yielding an RP lifetime of 58.7 ± 9.9 ns. A biexponential fit (Eq. 23) using the determined $k_q$ confirms this, giving a similar lifetime of 48.4 ±13.7 ns (**Fig. S8-2**). This result is reasonable, given the estimate from kinetic simulation at pH 6.4 (~33 ns, [43]) and the longer lifetime suggested by the larger $\Delta\Delta F_{PP}$ signal at pH 2.3 (**Fig. S8-1**).

The PFP fluorescence technique directly monitors non-emissive radical pair (RP) existence times down to ~50 ns, maintaining the sensitivity and spatial resolution of fluorescence microscopy. This allows highly sensitive exploration of electron-transfer rates and RP lifetimes in biological systems. Direct measurement of RP lifetimes is crucial for investigating biological geomagnetic MFEs, as sensitivity for RP-based reactions is theorized to require RP coherence times of ~700 ns [11].

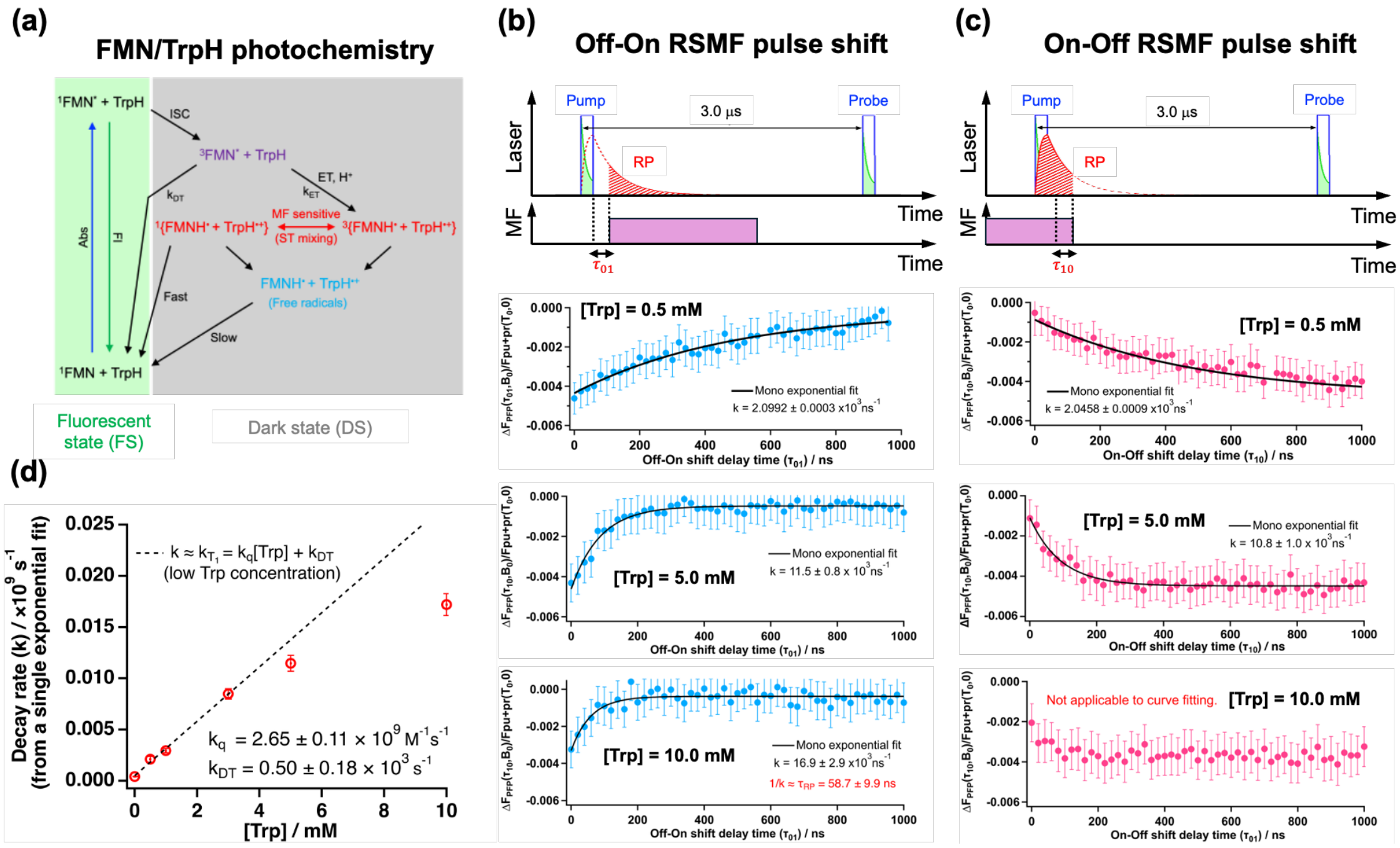

**Figure 5. The pump-field-probe measurements of 10 µM FMN with different tryptophan (TrpH) concentrations in an acidic solution (pH = 2.3, Sample thickness = 5.0 µm).** (a) Photoreaction scheme of FMN and tryptophan at low pH. In this case, upon short-pulse excitation, $^3FMN^*$ is quenched by intermolecular electron transfer to tryptophan, generating the radical pair ($FMNH^{\bullet}$−$TrpH^{\bullet+}$). The radicals that escape the pair serve as long-lived intermediates (LLIs). (b) Off-On RSMF shift measurement at different tryptophan concentrations (0.5 mM, 5.0 mM, and 10.0 mM). In the Off-On shift, the RSMF is applied after a delay, $\tau_{01}$, following the pump pulse excitation. Curve fitting is performed using a single exponential function. (c) On-Off RSMF shift measurements at increasing tryptophan concentrations (0.5 mM, 5.0 mM, and 10.0 mM). In the the On-Off shift, the RSMF is applied during the delay, $\tau_{10}$, after the pump pulse excitation. Curve fitting is performed using a single exponential function. (d) Tryptophan concentration vs decay rate obtained by single exponential fitting of the $\Delta F_{PFP}$ signal from the Off-On RSMF shift measurement.

## *FAD/tryptophan system*

To elucidate RP reactions in complex cellular environments where multiple components can be involved, methods to disentangle these contributions are indispensable. Comparing complementary PP and PFP measurements achieves this – demonstrated here by studying FAD's reaction with tryptophan in acidic solution (**Fig. 6a**) [21]. Photoexcitation of FAD at low pH (<3.6) rapidly yields an intramolecular RP ({$FH^{\bullet}$−$AH^{\bullet+}$}) in pseudo-equilibrium with the PTS. In the presence of tryptophan, both the RP and PTS are potentially quenched, forming a intermolecular radical pair (FADH• and $TrpH^{\bullet+}$) [21]. Rapid radical separation makes this RP short-lived (electrostatic interaction is absent as the majority of FADH• radicals are uncharged at this pH). The magnetic field effect (MFE) originates in the initial FAD RP and is transferred to the escaping, long-lived radicals. We can identify two RP types: quenched and unquenched. Since PFP fluorescence monitors only quenched RPs, while PP fluorescence detects MFEs on all dark state species, comparing the two signals (**Fig. 6b**) enables distinguishing effects on the unquenched RPs.

**Fig.6c** shows the delay-time dependence of the normalized PP and PFP fluorescence signals (independent of absolute fluorescence magnitude) for 10 μM FAD at tryptophan concentrations ranging from 0 to 5.0 mM:

$$PP{:}\frac{F_{pu+pr}(T,B_0)-F_{pu+pr}(T,0)}{F_{pu+pr}(T,0)},PFP{:}\frac{F_{pu+pr}(T_0,B_0;\tau_{10})-F_{pu+pr}(T_0,0)}{F_{pu+pr}(T_0,0)} \qquad \text{(eq. 24)}$$

Both signals converge at 3 μs (the probe pulse time), as here they both detect MFEs on the long-lived species. Their time dependence, however, differs significantly. PP measures the cumulative MFE across all dark-state species (PTS, intramolecular and intermolecular RPs, and long-lived radicals). In contrast, PFP selectively detects only the MFE generated in the two RPs transferred to the quenched, long-lived radicals. This system was studied using PP and PFP to observe changes in both the MFE and its magnetic field dependence (MARY) over time.

To gain deeper insight into the FAD RP and protonated triplet state (PTS) quenching behavior, we simulated the data using a first-order kinetic model coupled to Schulten–Wolynes semiclassical spin dynamics [47]. Simulations were performed globally across all tryptophan concentrations using the RadicalPy framework [48], modified to support two RPs and both PP and PFP measurements (full details in SI). The resulting single parameter set reproduced all the experimental data (kinetics and MARY) accurately without scaling. The best fit was obtained assuming tryptophan quenched only the RP state. Models assuming exclusive PTS quenching or equal quenching of PTS and RP failed to capture the global behavior.

In the absence of tryptophan, the MFE is governed by the competition between spin-selective back electron transfer (BET, $k_{BET}$) and protonation ($k_{-1}[H^+]$), which regenerates the PTS. Our data suggest a smaller $k_{BET}$ than previously reported [15], as larger values overestimated the absolute MFE magnitude. With tryptophan added, the intramolecular RP can undergo (1) singlet BET to ground state FAD, (2) triplet-selective protonation/electron transfer to PTS, or (3) non-spin-selective quenching to the intermolecular RP and then escaping radicals. Since the PTS can regenerate the RP, triplet-born RPs are continuously produced. The major dark state species shifts over time from PTS/RP to free radicals, so the fractional contribution to $B_{1/2}$ shifts from unquenched to quenched RPs.

Increasing tryptophan concentration decreases the PP signal (shorter RP lifetime) but increases the PFP signal (more radical escape). At the highest tryptophan concentration, the RP lifetime drops below 100 ns, which limits coherent singlet-triplet mixing and consequently reduces the MFE magnitudes.

Next, we demonstrate new fluorescence-based, time-resolved MARY techniques: PP and PFP spectroscopy. The MARY spectrum reflects how the RP reaction yield, modulated by ST-mixing, varies with the magnitude of the applied magnetic field. In the absence of a low field effect (LFE), this curve is characterized by $B_{1/2}$, the field strength where the MFE reaches half saturation. While $B_{1/2}$ reflects the RP's average hyperfine interactions, dynamic relaxation processes (spin relaxation in long-lived RPs; dephasing in short-lived RPs [4]) can influence its value. Consequently, $B_{1/2}$ values from CW excitation (e.g., conventional fluorescence MFE microscopy [13,20-23]) are often ambiguous, conflating intrinsic magnetic parameters with relaxation effects. This ambiguity was previously encountered in interpreting $B_{1/2}$ from flavin-containing species in HeLa cells [13].

This limitation can be addressed by analyzing time-resolved MARY spectra. This has been previously reported for TA measurements by recording the MARY spectrum on the absorption signal at different times after photoexcitation (often referred to as TR-MARY [31,49-51] but here referred to as PP-MARY for clarity) and also by varying the duration of the applied magnetic field after photoexcitation (previously referred to as SEMF-MARY [51] but referred to here as PFP-MARY for clarity). Recently the importance of the time dependence of $B_{1/2}$ in cryptochrome has been discussed [50].

PP-MARY provides insight into how the total dark-state populations of RP reactions evolve with both time and applied magnetic field following photoexcitation. **Fig. 6i** shows the time dependence of $B_{1/2}$ for FAD with and without tryptophan (0.3−5.0 mM). In the absence of tryptophan, the $B_{1/2}$ measurement becomes unreliable at longer times as there are no long-lived species. In the presence of tryptophan, $B_{1/2}$ first increases, then decreases and stabilizes. To interpret this, we extracted time-resolved MARY spectra and fitted $B_{1/2}$ values from the simulations. The agreement between experimental and simulated $B_{1/2}$ values was also used to globally constrain kinetic parameters (**Fig. 6j**).

The simulations accurately reproduced the observed $B_{1/2}$ changes and the reduction in the $B_{1/2}$ maximum with increasing tryptophan concentration. This required two key inclusions:

1. Electron spin relaxation incorporated as singlet-triplet (ST) dephasing ($k_{std}$), caused by transient fluctuations in the intramolecular RP's exchange interaction as the conformation changes.
2. The decay of the PTS back to ground-state FAD ($k_d$). Without $k_d$, $B_{1/2}$ increased and saturated but failed to exhibit the observed maximum. This is a result of the loss of non-quenched RPs through this channel on intermediate timescales.

As for the corresponding kinetic measurements, PP-MARY reflects the cumulative MFE of all quenched and unquenched dark-state species (RPs, PTS, radicals), while PFP-MARY arises solely from long-lived radicals generated by quenching. We confirmed this by measuring time-dependent $B_{1/2}$ via PFP-MARY for a single tryptophan concentration of 300 μM (This measurement is challenging as the strength of the nanosecond-scale pulsed magnetic field cannot be directly measured and must instead be calibrated against the known static field response (see SI for details). **Fig.6k** shows that, unlike for PP-MARY, the PFP-MARY $B_{1/2}$ exhibits no peak, instead increasing then saturating. Both PFP experiments and simulations (**Fig. 6l**) revealed significant LFEs. Accurate fitting of MARY data, therefore, required a double Lorentzian model ($B_{1/2}$, saturated MFE, saturated LFE, and fixed $L_{1/2}$ as parameters, details in SI) instead of a single Lorentzian. Simulations confirmed the experimental trend (**Fig. 6l**). As expected, $B_{1/2}$ values from both PP and PFP converge at 3 μs, when only long-lived radicals remain.

As a result, the $B_{1/2}$ value in PFP-MARY gradually increases as more radicals escape, eventually saturating when all excited species have returned to the ground state or formed long-lived products. The PP-MARY signal, however, transitions from being dominated by short-lived intramolecular RP and PTS at early times to reflecting only long-lived radicals at late times. Initially, high RP/PTS concentrations result in strong MFE contributions. The subsequent $B_{1/2}$ maximum reflects the trade-off between increasing RP lifetime (which raises $B_{1/2}$ via ST-dephasing) and the simultaneous decrease in RP population.

In contrast, the PFP-MARY $B_{1/2}$ reflects a time-averaged response and shows no peak, only saturation. This is because free radicals formed from longer-lived RPs make up a progressively smaller fraction of the total, and the cumulative MFE levels off.

Collectively, the PP and PFP kinetic and MARY data provide rich, complementary details on the reaction kinetics and spin dynamics, enabling a detailed unpicking of the underlying processes. Therefore, PP and PFP measurements are invaluable for deconvoluting complex kinetic behavior in systems with multiple contributing species, offering significant applications in unraveling photochemical RP reactions in complex biological systems.

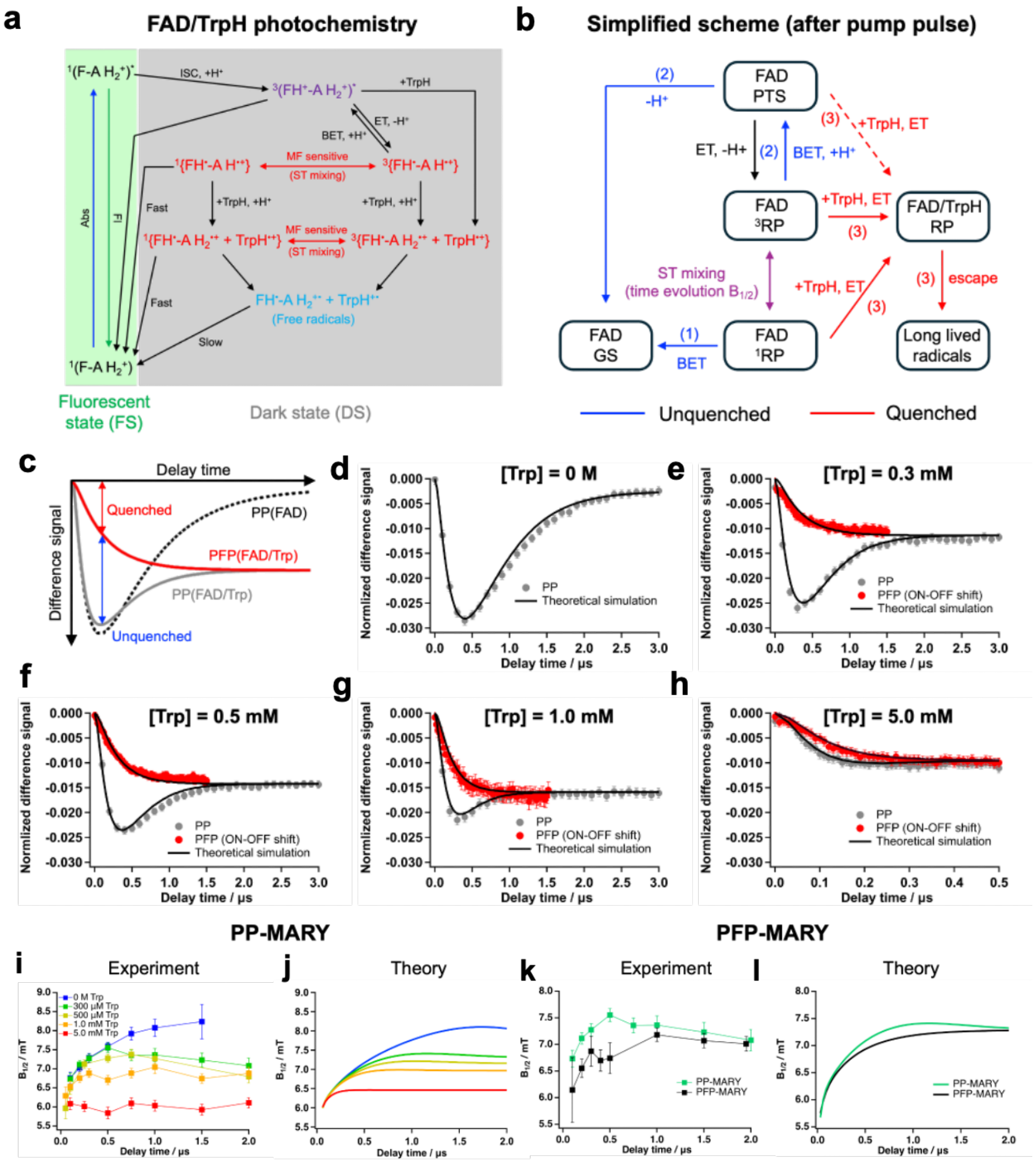

**Figure 6. The pump-field-probe measurements of 10 μM FAD and increasing tryptophan concentrations in acid solution (pH 2.3, sample thickness = 4.9 μm).** (a) Photoreaction scheme of FAD and tryptophan at low pH. Upon photoexcitation, the FAD rapidly undergoes electron transfer to form an intramolecular flavin–adenine radical pair ({$FH^{\bullet}$–$AH^{+\bullet}$}), which exists in pseudo-equilibrium with the protonated triplet state (PTS). In the presence of tryptophan, both the RP and PTS can potentially be quenched, yielding an intermolecular radical pair ($FADH^{\bullet}$ and $TrpH^{\bullet+}$). (b) Simplified scheme of the FAD/tryptophan reaction after pulse excitation. (c) Concept of discriminatory detection of quenching and non-quenching effects (d-h) The delay-time dependence of PP and PFP fluorescence signals for 10 μM FAD with increasing tryptophan concentrations. Solid lines are the results of the single parameter set kinetic / spin-dynamic simulations with no scaling. (i) PP-MARY measurements of FAD

with different tryptophan concentrations. (j) Theoretical simulation of PP-MARY using the same single parameter set simulations. (k) PP-MARY measurements of FAD with 300 μM tryptophan. For comparison, PP-MARY of 10μM FAD with 300 μM tryptophan is included. (l) Theoretical simulations (same single parameter set) of PFP-MARY and PP-MARY for 10μM FAD with 300 mM tryptophan.

## Conclusion

We developed and mathematically formalized single-color PP and PFP fluorescence detection techniques and constructed a corresponding microscope system. Under conditions mimicking living cells, these methods enable high-sensitivity monitoring of transient species and direct observation of dark-state kinetics, including photobleaching analysis and RP dynamics (with and without magnetic fields). Crucially, the techniques can discriminate between long-lived intermediates of RP and non-RP origin in complex systems. Time-resolved PP-MARY and PFP-MARY measurements monitor spin dynamics that tightly agree with quantum spin dynamics simulations. Furthermore, comparing PP and PFP measurements allows the separation of MFE contributions from LLI generating and non-generating RPs.

PP/PFP fluorescence microscopy provide a versatile platform for time-resolved studies of spin-correlated RP reactions in biological systems, offering a strong link between experiment and theory. This platform naturally extends to multi-color/multi-photon/polarized excitation systems and RYDMR/AWG spin manipulation systems. We anticipate that the technique will find important applications in mechanistic analyses of MFEs in endogenous flavins [13], optical characterization of magnetoreception candidate molecules and mechanisms [52,53], and the development of quantum sensors based on SCRPs [54-57].

## Acknowledgements

We thank Dr. Tomoaki Miura for advice on calibrating the switching magnetic field technique and Dr. Kiminori Maeda for stimulating discussions on the photochemistry of the FAD/TrpH system. This work was supported by Japan Society for the Promotion of Science (JSPS) Grants-in-Aid for Scientific Research (Grant number 20H02687 and 23H01919).

**Supporting Information**

# A fluorescence microscopy platform for time-resolved studies of spin-correlated radical pairs in biological systems

**Authors:**

Noboru Ikeya[1] and Jonathan R. Woodward[1]

1. Graduate School of Arts and Sciences, The University of Tokyo, Tokyo, Japan

**Corresponding Author:**

Jonathan R. Woodward

e-mail: jrwoodward@g.ecc.u-tokyo.ac.jp

**Contents**

## 1. Mathematical formulation of the PP/PFP fluorescence microscopy

Here, using a simplified typical flavin-based RP reaction scheme (**Fig.S1-1**) as an example model, we formulate the principle of pump probe (PP) and pump field probe (PFP) fluorescence techniques. In conclusion, we show that if the excitation pulse width is shorter than the total duration of the formation and the lifetime of the dark state species but longer than the fluorescence lifetime, then:

1. PP fluorescence detection technique can monitor total dark state population dynamics.
2. PFP fluorescence detection technique can selectively monitor radical pair (RP) dynamics.

### Model system

A simplified typical photochemical reaction scheme for a flavin-based RP reaction is shown in **Fig. S1-1** [1]. To streamline the discussion, the following photochemical processes are omitted. First. it is assumed that the majority of the molecules are photoexcited to the first excited singlet state ($S_1$) and the other excited molecules to higher singlet states ($S_n$, n≥2) and triplet states ($T_n$, n≥2) rapidly relax to their respective first excited state via internal conversion. Second, the phosphorescence is disregarded as not detected, because the phosphorescence signal is generally much lower than the fluorescence signal. Finally, non-cyclic photobleaching reactions, such as photodegradation, are omitted from the scheme for simplicity. But the quantification of these reactions is introduced after this section.

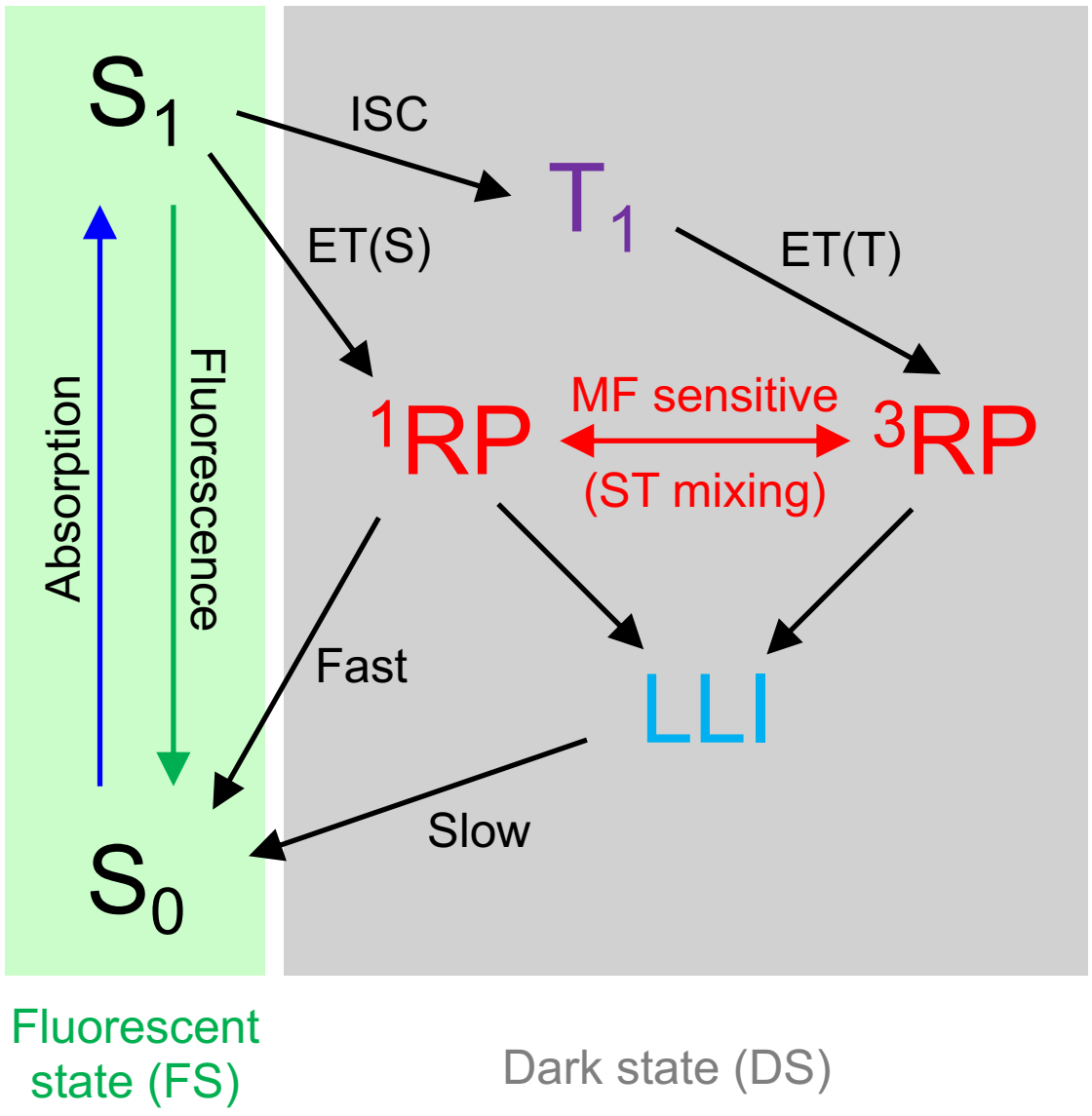

**Figure S1-1. A typical reaction scheme for flavin based RP reaction.**

$S_0$ = Singlet ground state, $S_1$ = Singlet excited state, $T_1$ = Triplet excited state, $^{1/3}RP$ = Singlet/Triplet (spin- correlated) radical pair state, LLI = Long lived intermediate state (reaction products or long lived free radicals), ISC = Intersystem crossing, ET(S/T) = Electron transfer from the singlet/triplet excited state.

## Fluorescent state and dark state

First of all, we introduce a classification of molecular states based on fluorescence emission. Under short pulse excitation on nanosecond timescales, the molecule undergoes fluorescence through repetitive transitions between the ground state ($S_0$) and the excited singlet state ($S_1$) until it undergoes transition to another excited state. On doing so, it does not return to the ground state during the pulse duration, because the pulse duration is short compared to the lifetime of that state. Therefore, a molecule that transitions to another excited state does not emit fluorescence from that point on. This leads us to classify $S_0$ and $S_1$ collectively as *fluorescent state* (FS) and $T_1$, RP, and LLI as *dark state* (DS) (**Fig. S1-1**):

Fluorescent state (FS) = {$S_0$, $S_1$}, Dark state (DS) = {$T_1$, RP, LLI}

To quantify these states, we define the total population of the fluorescent states, $[FS](t)$, and the total population of the dark states, $[DS](t)$, as follows:

$$[FS](t) = [S_0](t) + [S_1](t) \tag{S1-1}$$

$$[DS](t) = [T_1](t) + [RP](t) + [LLI](t) \tag{S1-2}$$

where $[X](t)$ denotes the population of the state $X$ at time $t$ $(X = S_0, S_1, T_1, RP, LLI)$. In principle, the total population of the molecules is conserved. At $t = 0$, the beginning of the excitation, all molecules can be regarded to be in the singlet ground state. Therefore, the following conditions are satisfied:

$$[FS](t) + [DS](t) = c_0, \qquad [S_0](0) = c_0 \tag{S1-3,4}$$

## Pump pulse and probe pulse fluorescence signals

Next, we derive the mathematical formulation of the fluorescence signals from the pump and probe pulses in terms of the total population of the fluorescent states, which is introduced in the above section. The emitted fluorescence intensity under pump excitation, $F_{pu}$, and probe excitation at a pump-probe delay time $T$, $F_{pr}(T)$, can be calculated as follows:

$$F_{pu} = \int_0^w k_F\,[S_1](t)dt, \quad F_{pr}(T) = \int_T^{T+w} k_F\,[S_1](t)dt \tag{S1-5,6}$$

where $w$ denotes the pulse width. The fluorescence intensity after pulse excitation is assumed to be negligible, as most molecules transition to the non-fluorescent dark state. If the pulse duration is shorter than the total time of the formation and lifetime of the dark states, transitions from the dark states to the ground state can be ignored. Therefore, the rate equations for the model system under pump and probe pulse excitations can be written as follows (**Fig.S1-2**):

For pump excitation ($0 \leq t \leq w$) and probe excitation ($T \leq t \leq T + w$):

$$\frac{d}{dt}[S_0](t) = -k_{ex}(t)[S_0](t) + (k_F + k_{IC})[S_1](t) \quad \text{(S1-7)}$$

$$\frac{d}{dt}[S_1](t) = k_{ex}(t)[S_0](t) - \left(k_F + k_{IC} + k_{ISC} + k_{ET(S)}\right)[S_1](t) \quad \text{(S1-8)}$$

Here, since pulses are generally not rectangular, a time dependence of the excitation rate, $k_{ex}(t)$, was introduced. Combining Equations (S1-7) and (S1-8) gives:

$$\frac{d}{dt}[FS](t) = \frac{d}{dt}\left([S_0](t) + [S_1](t)\right) = -k_D[S_1](t) \quad \text{(S1-9)}$$

where $k_D = k_{ISC} + k_{ET(S)}$. By substituting Equation (S1-9) into Equation (S1-5) and (S1-6), the emitted fluorescence intensity $F_{pu}$ and $F_{pr}(T)$ can be calculated as follows:

$$F_{pu} = -\frac{k_F}{k_D}\int_0^w \frac{d[FS](t)}{dt}dt = \frac{k_F}{k_D}\left([FS](0) - [FS](w)\right) \quad \text{(S1-10)}$$

$$F_{pr}(T) = -\frac{k_F}{k_D}\int_T^{T+w} \frac{d[FS](t)}{dt}dt = \frac{k_F}{k_D}\left([FS](T) - [FS](T+w)\right) \quad \text{(S1-11)}$$

Here we introduce the pumping ratio, $\gamma(w)$, with the pump pulse and probe pulse as follows:

$$\gamma_{pu}(w) = \frac{[FS](w)}{[FS](0)}, \gamma_{pr}(w) = \frac{[FS](T+w)}{[FS](T)} \quad \text{(S1-12, 13)}$$

where $0 < \gamma_i(w) < 1$ $(i = pu, pr)$. Then, the emitted fluorescence intensity $F_{pu}$ and $F_{pr}(T)$ can be expressed as follows:

$$F_{pu} = \frac{k_F}{k_D}\left(1 - \gamma_{pu}(w)\right)[FS](0) \propto [FS](0) \quad \text{(S1-14)}$$

$$F_{pr}(T) = \frac{k_F}{k_D}\left(1 - \gamma_{pr}(w)\right)[FS](0) \propto [FS](T) \quad \text{(S1-15)}$$

Thus, the emitted fluorescence intensities $F_{pu}$ and $F_{pr}(T)$ are proportional the total fluorescent state population at delay time 0 and $T$, respectively.

Furthermore, in this study, the pump probe delay time, $T$, is zero or much longer than the typical fluorescence lifetime, $\tau_F$:

$$T = 0 \text{ or } T \geq w \gg \tau_F \quad \text{(S1-16)}$$

Therefore, the excited singlet state at any delay time is:

$$[S_1](T) = 0 \quad \text{(S1-17)}$$

In addition, if the excitation rate (i.e., wavelength and intensity) and the pulse width of the pump and probe pulses are identical, the rate equations, (S1-7) and (S1-8), during pump excitation and probe excitation become equivalent. Consequently, the pumping ratios for the pump pulse and probe pulse are also equal:

$$\gamma_{pu}(w) = \gamma_{pr}(w) = \gamma(w) \quad \text{(S1-18)}$$

In summary, when the intensity, excitation wavelength and pulse width of the pump and probe pulses are equal and the excitation pulse width is shorter than the total duration of the formation and lifetime of the dark state species and longer than the fluorescence lifetime, the emitted fluorescence intensity $F_{pu}$ and $F_{pr}(T)$ are expressed using the same constant of proportionality:

$$F_{pu} = \alpha[FS](0) = \alpha[S_0](0)(= \alpha c_0) \qquad F_{pr}(T) = \alpha[FS](T) = \alpha[S_0](T) \quad \text{(S1-19, 20)}$$

where the constant of proportionality $\alpha$ is:

$$\alpha = \frac{k_F}{k_D}(1 - \gamma(w)) = \frac{\phi_F}{\phi_D}(1 - \gamma(w)) \quad \text{(S1-21)}$$

Where $\phi_F$ and $\phi_D$ denote the quantum yield of fluorescent and dark state, respectively. The $\alpha$ is the indicator that represents the brightness of the system.

By exploiting this information from the pump pulse and probe pulse fluorescence signals, we can develop two new detection methodologies: pump probe (PP) and pump field probe (PFP) fluorescence.

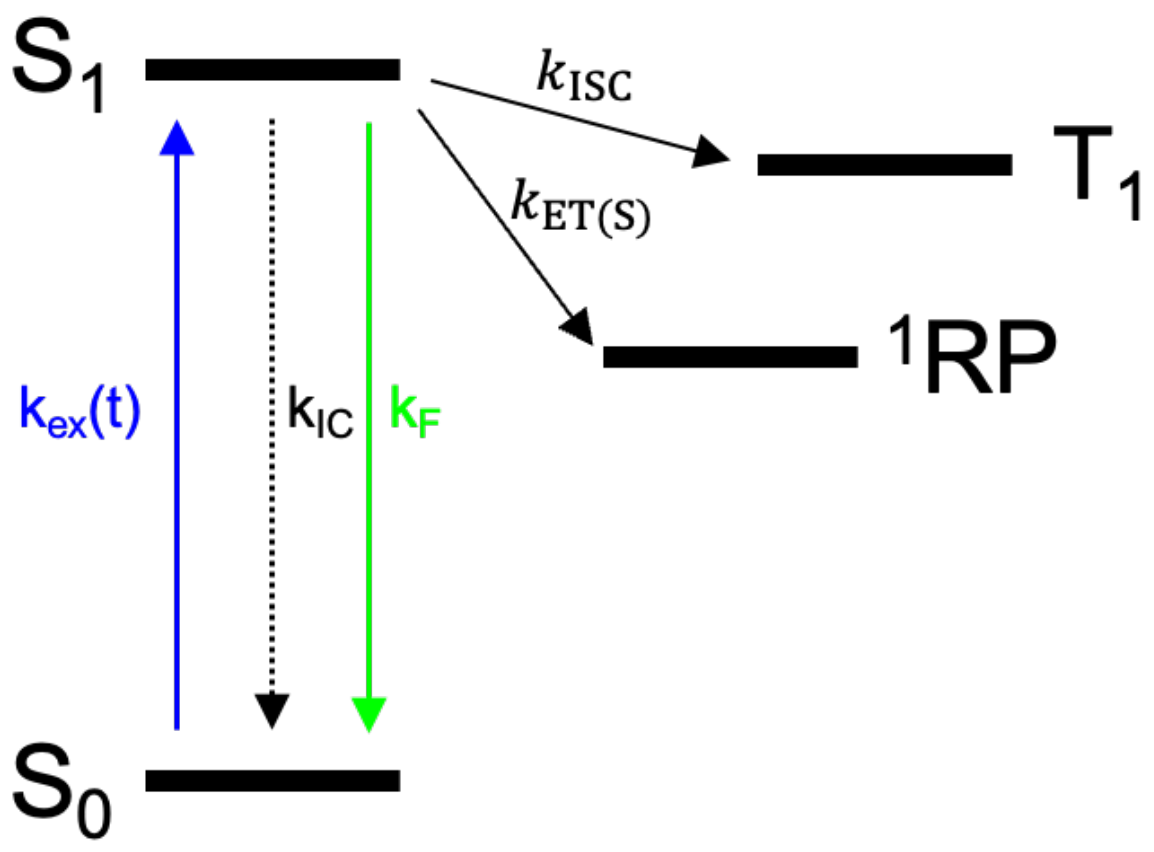

**Figure S1-2. Photoreaction scheme illustrating the processes occurring during short pulse excitation.** $k_{ex}$: Excitation rate, $k_F$: Fluorescence rate, $k_{IC}$: Internal conversion rate, $k_{ISC}$: Intersystem crossing rate, $k_{ET(S)}$: Electron transfer rate from the excited singlet state.

**Pump probe (PP) fluorescence detection**

In PP fluorescence detection, the integrated fluorescence intensity under pump-probe excitation and pump only excitation are each measured. And then, by taking the difference between these signals, the fluorescence intensity with the probe excitation only is extracted (**Fig. S1-3**). Mathematically, the fluorescence difference signal is determined as follows:

$$\Delta F_{PP}(T) = F_{pu+pr}(T) - F_{pu} = F_{pr}(T) \qquad \text{(S1-18)}$$

According to Equation (S1-15), $\Delta F_{PP}(T)$ is positively proportional to the concentration of the total fluorescent state and negatively proportional to the total dark state as follows:

$$F_{pr}(T) \propto [FS](T) = c_0 - [DS](T) \qquad \text{(S1-19)}$$

When pump and probe pulses are used with the same intensity and pulse width, the normalized fluorescence difference signal, $\overline{\Delta F}$, is expressed as follows:

$$\Delta\overline{F_{PP}}(T) = \frac{\Delta F_{PP}(T)}{F_{pu}} = \frac{\alpha\left(c_0 - [DS](T)\right)}{\alpha c_0} = 1 - \overline{[DS]}(T) \qquad \text{(S1-20)}$$

As a result, the normalized total dark state population can be written as follows:

$$\overline{[DS(T)]} = 1 - \overline{\Delta F}(T) \qquad \text{(S1-21)}$$

Therefore, the population dynamics of the total dark state can be monitored by a function of the fluorescence difference signal with respect to the pump probe delay time $T$.

In addition, the effects of an applied magnetic field ($B_0$) can be obtain by measuring the difference of the fluorescence difference, $\Delta\Delta F_{PP}$, as follows:

$$\Delta\Delta F_{PP}(T, B_0) = \Delta\bar{F}(T, B_0) - \Delta\bar{F}(T, 0) = -\left(\overline{[DS]}(T, B_0) - \overline{[DS]}(T, 0)\right) \qquad \text{(S1-22)}$$

Thus, the effect of the total dark state dynamics under an applied magnetic field appears as a negative signal in the difference of the fluorescence difference signal. Also, $\Delta\Delta F_{PP}$ can be defined from the difference between the pump probe fluorescence signals with the presence and absence of magnetic field as follows:

$$\Delta\Delta F_{PP}(T, B_0) = F_{pu+pr}(T, B_0) - F_{pu+pr}(T, 0) \propto -\left([DS](T, B_0) - [DS](T, 0)\right) \qquad \text{(S1-23)}$$

Measurement with the above definition avoids error propagation of the difference due to Equation (S23), which increases the precision of the measurement.

In the example model, the population of the excited triplet state is not changed with an application of magnetic field, therefore $\Delta\Delta F_{PP}$ monitors the time dependence of the magnetic field effects (MFEs) on the RP state and LLI state.

$$\Delta\Delta F_{PP}(T, B_0) \propto -\left(\Delta RP(T, B_0) + \Delta LLI(T, B_0)\right) \qquad \text{(S1-24)}$$

where $\Delta RP(T, B_0) = [RP](T, B_0) - [RP](T, 0)$ and $\Delta LLI(T, B_0) = [LLI](T, B_0) - [LLI](T, 0)$

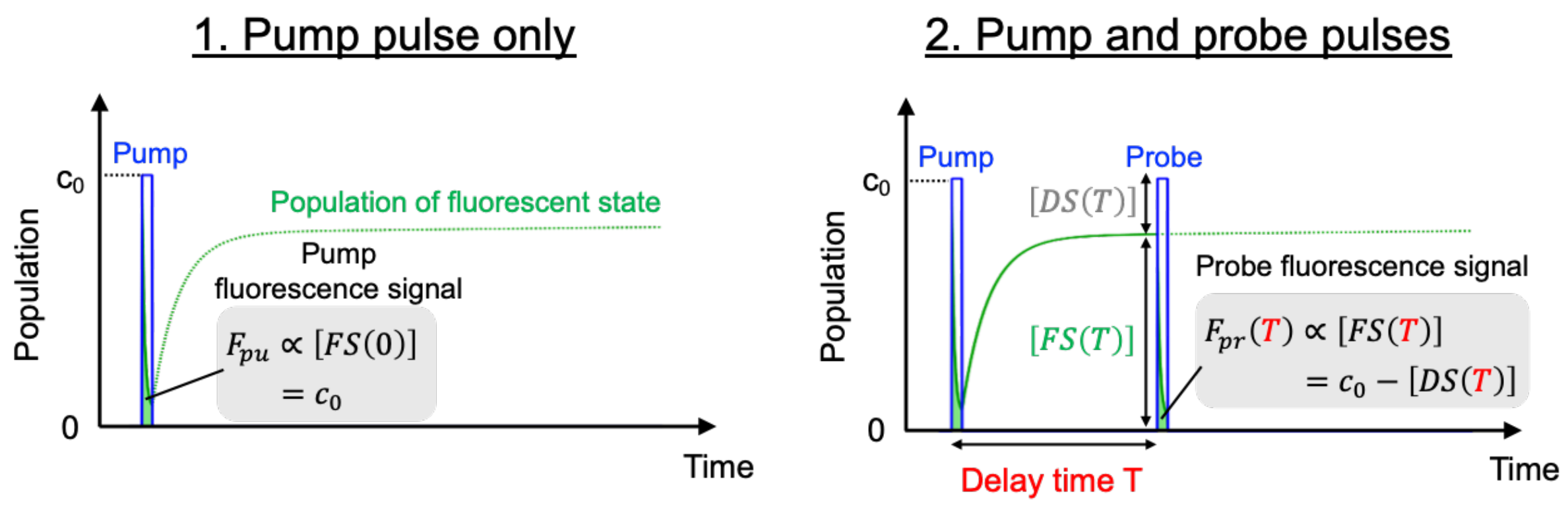

**Figure S1-3. Pump-probe fluorescence detection scheme**

**Pump field probe (PFP) fluorescence detection**

In PFP fluorescence detection, fluorescence intensities are measured with and without the application of a rapidly switched magnetic field (RSMF) at a given delay time (τ) under pump probe excitation at a fixed pump probe delay time ($T_0$). By taking the difference, the effect of the fluorescence intensity on only the probe excitation is extracted (**Fig. S1-4**).

Mathematically, the PFP fluorescence difference is defined as follows:

$$\Delta F_{PFP}(\tau, B_0) = F_{PP}(T_0, B_0; \tau) - F_{PP}(T_0, 0) \tag{S1-25}$$

Then, the $\Delta F_{PFP}$ is calculated as follows:

$$\begin{aligned} \Delta F_{PFP}(\tau, B_0) &= F_{PP}(T_0, B_0; \tau) - F_{PP}(T_0, 0) \\ &= F_{pr}(T_0, B_0; \tau) - F_{pr}(T_0, 0) \\ &\propto \left(c_0 - [DS](T_0, B_0; \tau)\right) - \left(c_0 - [DS](T_0, 0)\right) \\ &= -\left([DS](T_0, B_0; \tau) - [DS](T_0, 0)\right) \end{aligned} \tag{S1-26}$$

In the PFP detection scheme, the pump probe delay time ($T_0$) is fixed to be much longer than the lifetime of the excited triplet state and the RP state ($T_0 \gg \tau_{T_1}, \tau_{RP}$), but shorter than the lifetime of the long-lived intermediate ($T_0 < \tau_{LLI}$). As a result,

$$[DS](T_0, B_0; \tau) = [T_1](T_0) + [RP](T_0, B_0; \tau) + [LLI](T_0, B_0; \tau) \approx [LLI](T_0, B_0; \tau) \tag{S1-27}$$

$$[DS](T_0, 0) = [T_1](T_0) + [RP](T_0, 0) + [LLI](T_0, 0) \approx [LLI](T_0, 0) \tag{S1-28}$$

Substituting (S1-27) and (S1-28) for (S1-26), we obtain the following expression:

$$\Delta F_{PFP}(\tau, B_0) \propto -\left([LLI](T_0, B_0; \tau) - [LLI](T_0, 0)\right) \tag{S1-29}$$

Therefore, the difference fluorescence signal in the PFP detection ($\Delta F_{PFP}$) monitors the change of the long-lived intermediate from the RP with an application of an RSMF. The signal appears as a negative signal.

In many cases, the reaction rate equation for LLI is expressed as follows:

$$\frac{d}{dt}[LLI](t, B_0) = -k_{LLI}[LLI](t, B_0) + k_{esc}[RP](t, B_0) \tag{S1-30}$$

The general solution of a non-homogeneous 1st order differential equation is expressed as follows:

$$\frac{d}{dt}x(t) = Ax(t) + f(t) \tag{S1-31}$$

$$x(t) = x(t_0)e^{A(t-t_0)} + \left(\int_{t_0}^{t} f(t')\, e^{-A(t'-t_0)} dt'\right) e^{A(t-t_0)} \tag{S1-32}$$

Using this solution, $[LLI](T_0, B_0; \tau)$ and $[LLI](T_0, 0)$ can be expressed

$$[LLI](T_0, B_0; \tau) = [LLI](\tau, 0)e^{-k_{LLI}(T_0-\tau)} + k_{esc}\left(\int_{\tau}^{T_0} [RP](t', B_0; \tau)\, e^{k_{LLI}(t'-\tau)} dt'\right) e^{-k_{LLI}(T_0-\tau)} \tag{S1-33}$$

$$[LLI](T_0, 0) = [LLI](\tau, 0)e^{-k_{LLI}(T_0-\tau)} + k_{esc}\left(\int_{\tau}^{T_0} [RP](t', 0)\, e^{k_{LLI}(t'-\tau)} dt'\right) e^{-k_{LLI}(T_0-\tau)} \tag{S1-34}$$

The relationship between RP concentration and fluorescence detection is clarified through a more explicit formulation as follows:

$$\Delta F_{PFP}(\tau, B_0) \propto -k_{esc} \int_{\tau}^{T_0} \left( [RP](t', B_0; \tau) - [RP](t', 0) \right) \mathbf{e}^{k_{LLI}(t' - T_0)} dt' \quad \text{(S1-35)}$$

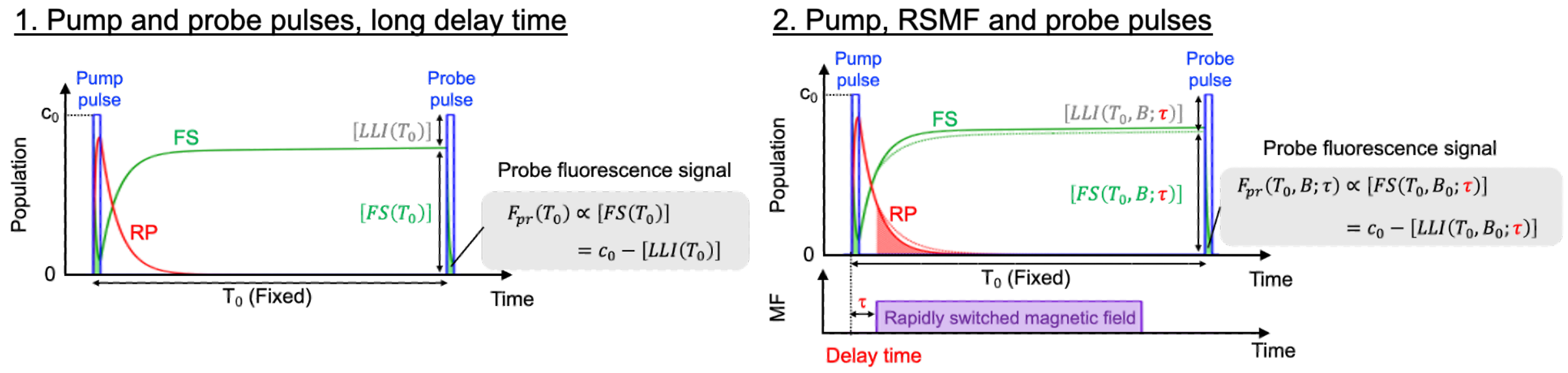

**Figure S1-4** Pump-Field-Probe detection scheme

## 2. Analytical modeling of delay-time dependence in PP/PFP fluorescence signals

While the PP/PFP fluorescence signals can be precisely analyzed using numerical simulations of corresponding rate equations, a simplified kinetic model with analytical solutions offers a more intuitive understanding of the delay time dependence of the signals. Here, we provide analytical interpretation of the observed PP/PFP fluorescence signals using a four-state model that captures the key qualitative features of MFEs in triplet born RP reactions. (Note: Singlet born RP reactions can be easily considered using a three-state model consisting of GS, RP and LLI.)

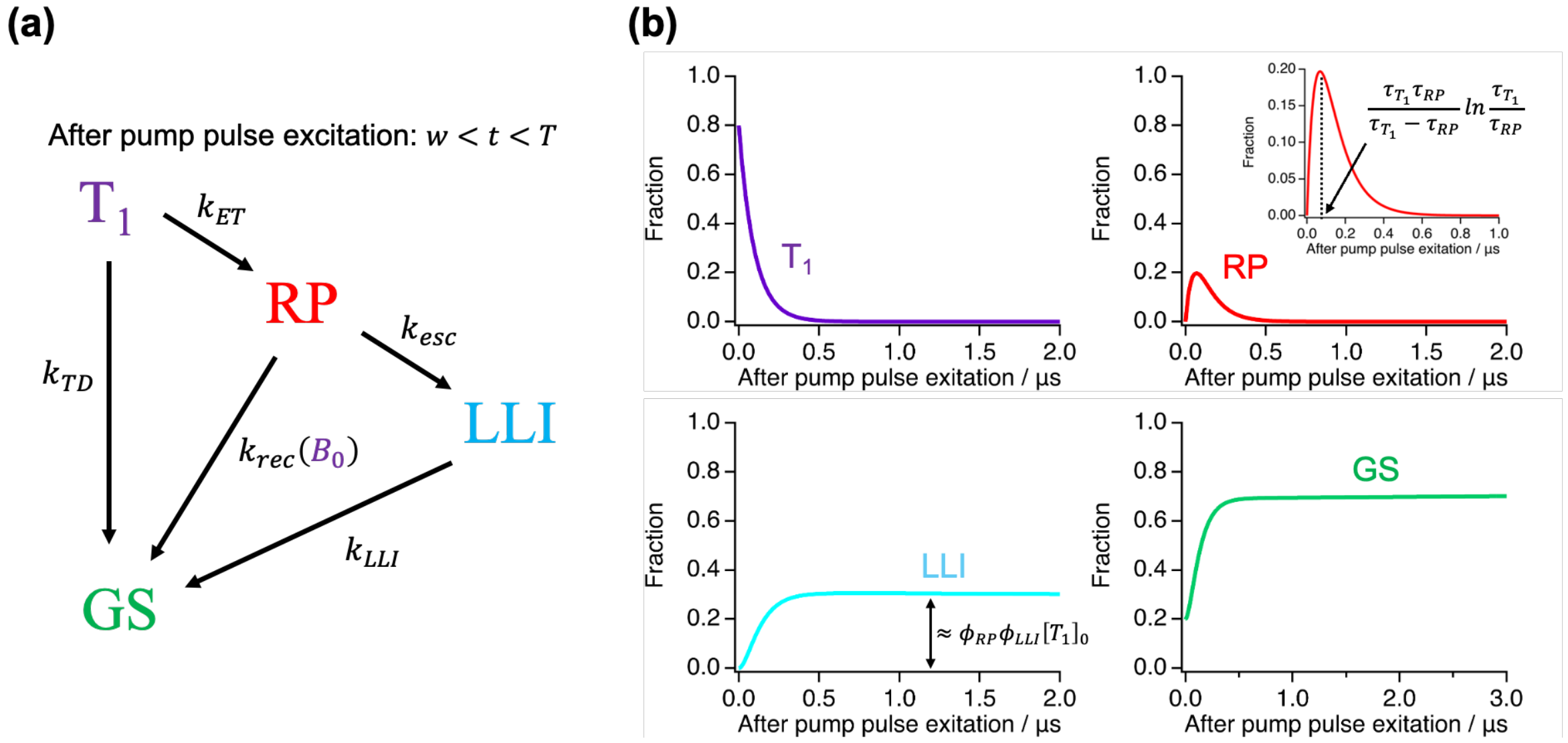

**Figure S2-1 A four-state RP reaction model after pump pulse excitation.** (a) Reaction scheme. (b) Example kinetics of excited triplet state ($T_1$), radical pair state (RP), long lived intermediate state (LLI) and ground state (GS). Rate constants are $k_{DT} = 4.0 \times 10^5$ s, $k_{ET} = k_q[Q] = 1.0 \times 10^7$ s, $k_{bet} = 1.2 \times 10^7$ s, $k_{esc} = 0.8 \times 10^7$ s, and $k_{LLI} = 1.0 \times 10^4$ s. Initial populations are $[T_1]_0 = 0.8$, $[RP]_0 = [LLI]_0 = 0.0$ and $c_0 = 1.0$.

A four-state RP reaction model after pump pulse excitation is shown in **Fig. S2-1**. This model serves as a simplified representation of MFEs originating from ST mixing. In this model, MFEs are introduced by allowing the recombination rate constant, $k_{rec}$, which governs the return of the RP to the ground state (GS)—to depend on the strength of the external magnetic field, $B_0$. A similar treatment was adopted in the Supporting Information of Ref. 2. The rate equations are given below:

$$\frac{d}{dt}[T_1](t) = -k_{T_1}[T_1](t)$$

$$\frac{d}{dt}[RP](t,B_0) = -k_{RP}(B_0)[RP](t,B_0) + k_q[T_1](t) \qquad \text{(S2-1)}$$

$$\frac{d}{dt}[LLI](t,B_0) = k_{esc}[RP](t,B_0) - k_{LLI}[LLI](t,B_0)$$

$$\frac{d}{dt}[GS](t,B_0) = k_{DT}[T_1](t) + k_{rec}(B_0)[RP](t,B_0) + k_{LLI}[LLI](t,B_0)$$

where $k_{ET} = k_q[Q]$, $k_{T_1} = k_{DT} + k_{ET}$, and $k_{RP}(B_0) = k_{rec}(B_0) + k_{esc}$.

Under the initial conditions that the total concentration is conserved and no RP and LLI are generated during the pump pulse excitation, w;

$$[T_1](t) + [RP](t, B_0) + [LLI](t, B_0) + [GS](t, B_0) = c_0 \quad \text{(S2-2)}$$

$$[T_1](w) = [T_1]_0, \qquad [RP](w) = [LLI](w) = 0 \quad \text{(S2-3)}$$

The analytical solutions are:

$$[T_1](t) = [T_1]_0 \mathbf{e}^{-k_{T_1}(t-w)} \quad \text{(S2-4)}$$

$$[RP](t, B_0) = X(B_0)[T_1]_0\left(\mathbf{e}^{-k_{RP}(B_0)(t-w)} - \mathbf{e}^{-k_{T_1}(t-w)}\right) \quad \text{(S2-5)}$$

$$[LLI](t, B_0) = X(B_0)[T_1]_0\left[Z\mathbf{e}^{-k_{T_1}(t-w)} - Y(B_0)\mathbf{e}^{-k_{RP}(B_0)(t-w)} + (Y(B_0) - Z)\mathbf{e}^{-k_{LLI}(t-w)}\right] \quad \text{(S2-6)}$$

$$[GS](t, B_0) = c_0 - [DS](t, B_0) \quad \text{(S2-7)}$$

where $X(B_0)$, $Y(B_0)$ and Z are defined as:

$$X(B_0) = \frac{k_q}{k_{T_1} - k_{RP}(B_0)}, Y(B_0) = \frac{k_{esc}}{k_{RP}(B_0) - k_{LLI}}, Z = \frac{k_{esc}}{k_{T_1} - k_{LLI}} \quad \text{(S2-8)}$$

Using these analytical solutions, (S2-4,5,6,7), the delay time dependence of the fluorescence signals on PP/PFP measurements can be analytically calculated.

1. PP fluorescence detection $\Delta F_{PP}$

**Figure S2-2** shows an example of the $\Delta F_{PP}$ signals. According to (S1-21), the normalized pump-probe fluorescence signal is given by:

$$1 - \overline{\Delta F_{PP}}(T, B_0) = \overline{[DS]}(T, B_0) \quad \text{(S2-9)}$$

Substituting the analytical solutions (S2-4 to S2-6) for $[DS](T, B_0)$, we obtain:

$$1 - \overline{\Delta F_{PP}}(T, B_0) = C_{T_1}\mathbf{e}^{-k_{T_1}(T-w)} + C_{RP(B_0)}\mathbf{e}^{-k_{RP}(B_0)(T-w)} + C_{LLI}\mathbf{e}^{-k_{LLI}(T-w)} \quad \text{(S2-10)}$$

Where the coefficients are defined as:

$$C_{T_1} = \left(1 - (1 - Z)X(B_0)\right)\overline{[T_1]_0} \quad \text{(S2-11)}$$

$$C_{RP(B_0)} = X(B_0)\left(1 - Y(B_0)\right)\overline{[T_1]_0} \quad \text{(S2-12)}$$

$$C_{LLI} = X(B_0)(Y(B_0) - Z)\overline{[T_1]_0} \quad \text{(S2-13)}$$

In many cases, the rate constants satisfy $k_{T_1}, k_{RP} \gg k_{LLI}$, allowing us to approximate

$$Y(B_0) \approx \frac{k_{esc}}{k_{RP}(B_0)}, Z \approx \frac{k_{esc}}{k_{T_1}} \quad \text{(S2-14)}$$

Therefore, the coefficient of the LLI component in (S2-13) becomes:

$$C_{LLI} \approx \frac{k_q}{k_{T_1} - k_{RP}(B_0)}\left(\frac{k_{esc}}{k_{RP}(B_0)} - \frac{k_{esc}}{k_{T_1}}\right)\overline{[T_1]_0} = \frac{k_q k_{esc}}{k_{T_1} k_{RP}(B_0)}\overline{[T_1]_0} = \phi_{RP}\phi_{LLI}\overline{[T_1]_0} \quad \text{(S2-15)}$$

Here, the quantum yields of RP state from the triplet state and long-lived intermediate state from RP state can be defined as:

$$\phi_{RP} = \frac{k_q}{k_{T_1}}, \phi_{LLI} = \frac{k_{esc}}{k_{RP}(B_0)} \tag{S2-16}$$

At long delay time where $T_{long} \gg \tau_{T_1}$, $\tau_{RP}$, the signal is dominated by the slowest decaying component

$$1 - \Delta\overline{F_{PP}}(T_{long}, B_0) \approx C_{LLI}\mathbf{e}^{-k_{LLI}(T_{long}-w)} = \phi_{RP}\phi_{LLI}(B_0)\overline{[T_1]_0}\mathbf{e}^{-k_{LLI}(T_{long}-w)} \tag{S2-17}$$

Taking the ratio of the PP fluorescence signals measured with and without an external magnetic field at long delay times provides a measure of the magnetic field-induced extension of the RP lifetime:

$$\frac{1 - \Delta\overline{F_{PP}}(T_{long}, B_0)}{1 - \Delta\overline{F_{PP}}(T_{long}, 0)} \approx \frac{\phi_{LLI}(B_0)}{\phi_{LLI}(0)} = \frac{k_{RP}(0)}{k_{RP}(B_0)} = \frac{\tau_{RP}(B_0)}{\tau_{RP}(0)} \tag{S2-18}$$

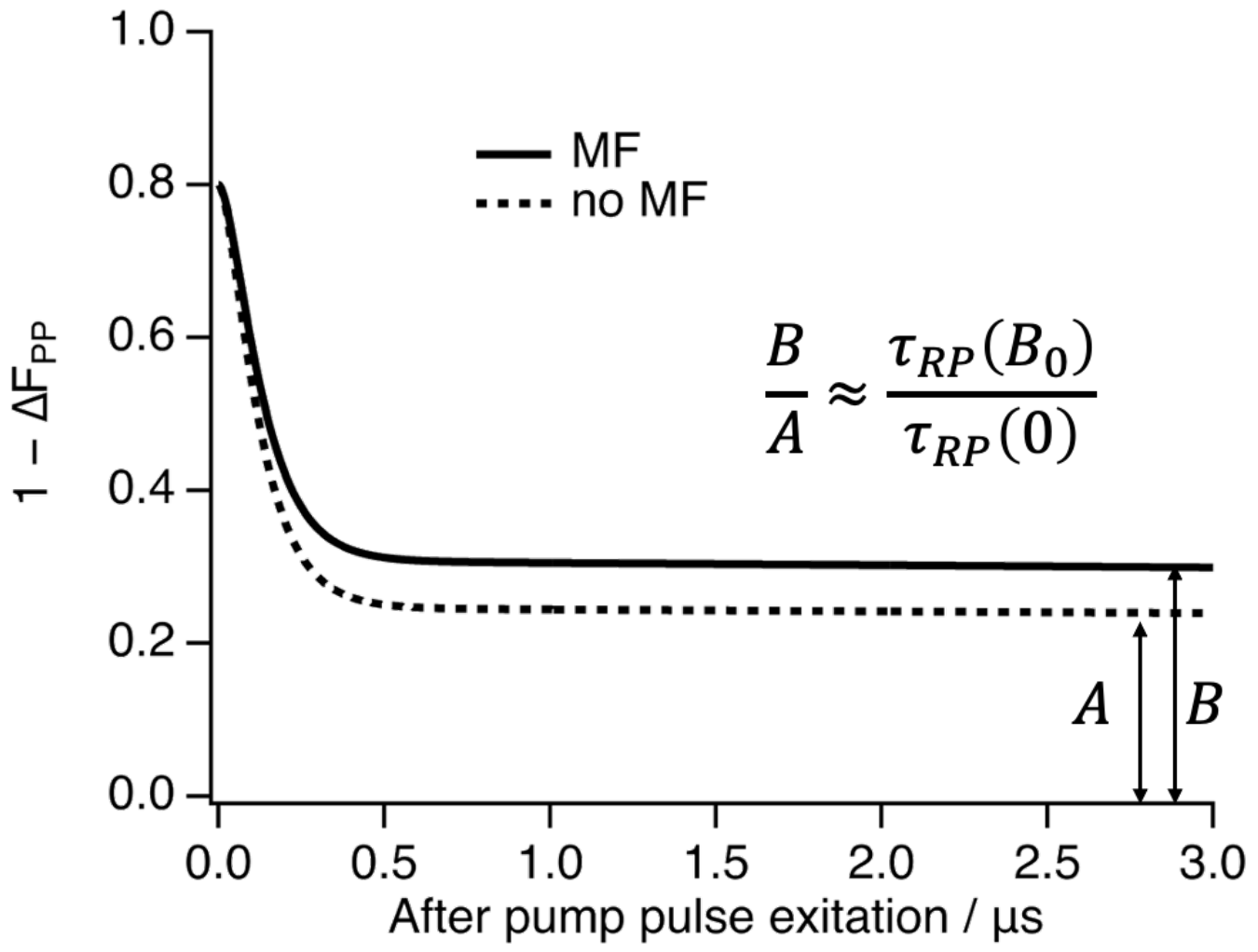

**Figure S2-2 Example of the $\Delta F_{PP}$ signals with and without magnetic field.** Rate constants are $k_{DT}$ = 4.0 ×10$^5$ s, $k_{ET}$ = $k_q$[Q] = 1.0 ×10$^7$ s, $k_{bet}$ (B) = 1.2 ×10$^7$ s, $k_{bet}$(0) = 1.7 ×10$^7$ s, $k_{esc}$ = 0.8 ×10$^7$ s, and $k_{LLI}$ =1.0 ×10$^4$ s. Initial populations are $[T_1]_0$ = 0.8, $[RP]_0$ =$[LLI]_0$= 0.0 and $c_0$ = 1.0.

2. PP fluorescence detection: $\Delta\Delta F_{PP}$

According to (S1-23), the differential signal of the pump probe fluorescence:

$$\Delta\Delta F_{PP}(T, B_0) \propto -\big([DS](T, B_0) - [DS](T, 0)\big) = -\big(\Delta RP(T, B_0) + \Delta LLI(T, B_0)\big) \tag{S2-19}$$

Substituting the general expressions (S2-5) and (S2-6) into the above equation, we obtain:

$$\Delta\Delta F_{PP}(T, B_0) = -\big[\Delta C_{T_1}\mathbf{e}^{-k_{T_1}(T-w)} + C_{RP(B_0)}\mathbf{e}^{-k_{RP}(B_0)(T-w)} - C_{RP(0)}\mathbf{e}^{-k_{RP}(0)(T-w)} + \Delta C_{LLI}\mathbf{e}^{-k_{LLI}(T-w)}\big] \tag{S2-20}$$

with coefficients defined as:

$$\Delta C_{T_1} = -(1 - Z)\big(X(B_0) - X(0)\big)[T_1]_0 \tag{S2-21}$$

$$C_{RP(B_0)} = X(B_0)\big(1 - Y(B_0)\big)[T_1]_0 \tag{S2-22}$$

$$C_{RP(0)} = X(0)\big(1 - Y(0)\big)[T_1]_0 \tag{S2-23}$$

$$\Delta C_{LLI} = \{X(B_0)(Y(B_0) - Z) - X(0)(Y(0) - Z)\}[T_1]_0 \tag{S2-24}$$

**Fig. S2-3** displays the example of the $\Delta\Delta F_{PP}$ signals.

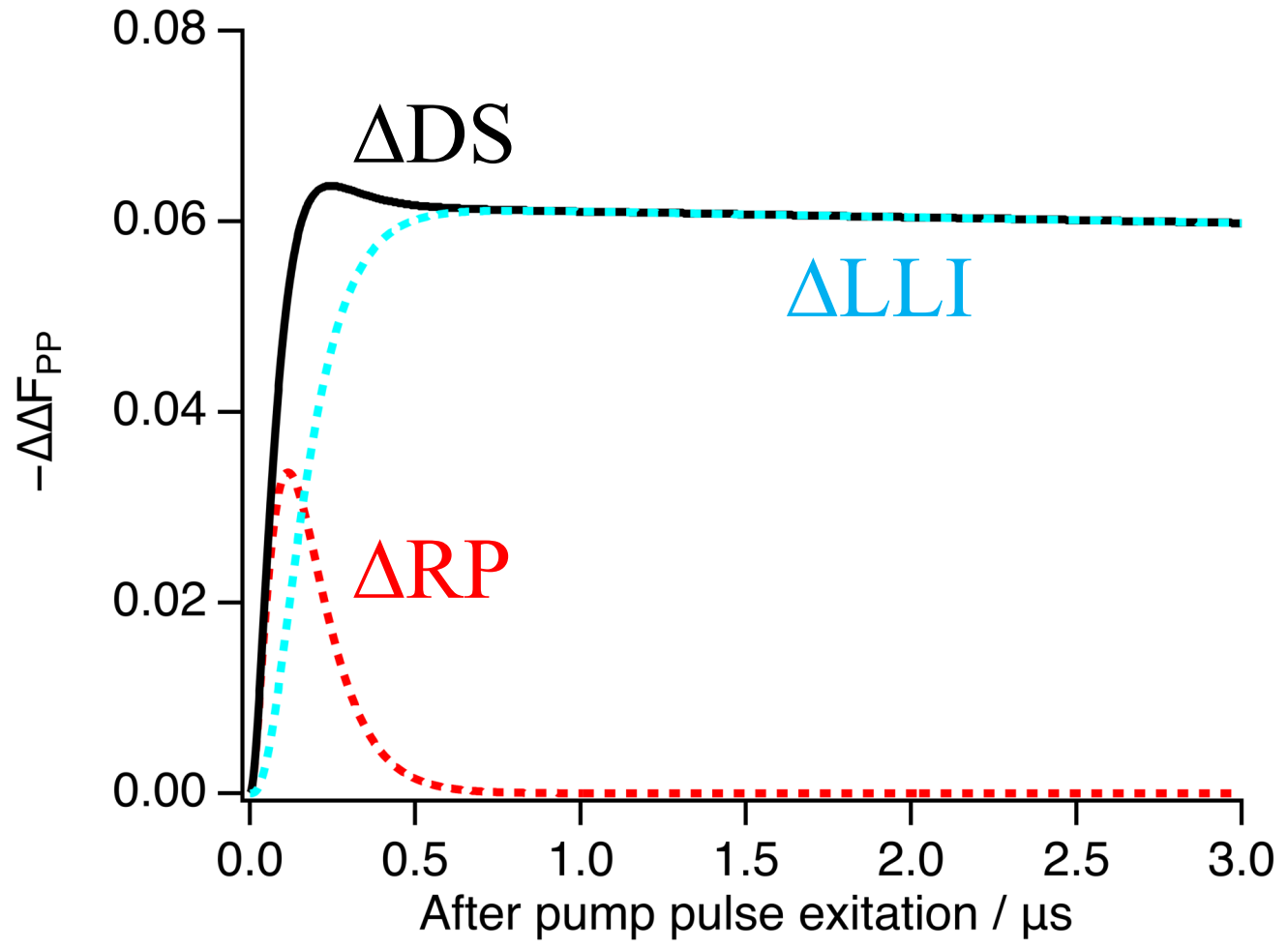

**Fig. S2-3 Example of the $\Delta\Delta F_{PP}$ signals with and without magnetic field.** Rate constants are $k_{DT}$ = 4.0 ×10$^5$ s, $k_{ET} = k_q[Q]$ = 1.0 ×10$^7$ s, $k_{bet}$ (B)= 1.2 ×10$^7$ s, $k_{bet}(0)$ = 1.7 ×10$^7$ s, $k_{esc}$ = 0.8 ×10$^7$ s, and $k_{LLI}$ =1.0 ×10$^4$ s. Initial populations are $[T_1]_0$ = 0.8, $[RP]_0$ =$[LLI]_0$= 0.0 and $c_0$ = 1.0. ΔX = [X](t,B$_0$)-[X](t,0) (X = RP, LLI, DS).

3. PFP fluorescence detection

In PFP fluorescence detection, there are two types of shifting mode for the RSMF: the Off-On shift type, where the RSMF is applied after a delay time, $\tau_{01}$, following the pump pulse excitation (**Fig. S2-4 a**), and the On-Off shift type, where the RSMF is applied during the delay time, $\tau_{10}$, after the pump pulse excitation (**Fig. S2-4 b**).

Off-On RSMF shift $\Delta F_{PFP}^{(OFF-ON)}$:

According to (S1-29), the PFP fluorescence signal, in case of Off-On shift measurement, is expressed as:

$$\Delta F_{PFP}^{(OFF-ON)}(\tau_{01}, B_0) \propto -\left([LLI](T_0, B_0; \tau_{01}) - [LLI](T_0, 0)\right) \qquad \text{(S2-25)}$$

In the Off-On shift measurement, $[LLI](T_0, B_0; \tau_{01})$ is obtained by solving the rate equations (S2-1) with the following magnetic field condition.

$$B_{RSMF}(t; \tau_{10}) = \begin{cases} 0 & (0 \le t \le \tau_{01}) \\ B_0 & (\tau_{01} \le t \le T_0) \end{cases} \qquad \text{(S2-26)}$$

In this case, a general solution of [LLI](t) on the four-state model for $\tau_{01} \le t \le T_0$:

$$[LLI](t, B_0; \tau_{01}) = \left[[LLI](\tau_{01}, 0) + Y(B_0)[RP](\tau_{01}, 0) + (Y(B_0) - Z)X(B_0)[T_1](\tau_{01})\right]\mathbf{e}^{-k_{LLI}(t-\tau_{01})}$$
$$- Z\left([RP](\tau_{01}, 0) + X(B_0)[T_1](\tau_{01})\right)\mathbf{e}^{-k_{RP}(B_0)(t-\tau_{01})} + ZX(B_0)[T_1](\tau_{01})\mathbf{e}^{-k_{T_1}(t-\tau_{01})} \qquad \text{(S2-27)}$$

Since $T_0 \gg \tau_{T_1}, \tau_{RP}$, $[LLI](T_0, B_0; \tau_{01})$ is obtained:

$$[LLI](T_0, B_0; \tau_{01}) = \left[[LLI](\tau_{01}, 0) + Y(B_0)[RP](\tau_{01}, 0) + (Y(B_0) - Z)X(B_0)[T_1](\tau_{01})\right]\mathbf{e}^{-k_{LLI}(T_0-\tau_{01})} \qquad \text{(S2-28)}$$

Also, the population of the long-lived intermediate states without the application of RSMF $[LLI](T_0, 0)$ is:

$$[LLI](T_0,0)=\left[[LLI](\tau_{01},0)+Y(0)[RP](\tau_{01},0)+(Y(0)-Z)X(0)[T_1](\tau_{01})\right]\mathbf{e}^{-k_{LLI}(T_0-\tau_{01})} \quad \text{(S2-29)}$$

Substituting (S2-28) and (S2-29) into (S2-25), and then analytical solutions (S2-4,5,6) into that, we obtain:

$$\Delta F_{PFP}^{(OFF-ON)}(\tau_{01},B_0)\propto-\left[C_{RP(0)}\mathbf{e}^{-(k_{RP}(0)-k_{LLI})(\tau_{01}-w)}+C_{T_1}\mathbf{e}^{-(k_{T_1}-k_{LLI})(\tau_{01}-w)}\right] \quad \text{(S2-30)}$$

with coefficients defined as:

$$C_{RP(0)}=X(0)\Delta Y(B_0)[T_1]_0C(T_0) \quad \text{(S2-31)}$$

$$C_{T_1}=\Delta X(B_0)(Y(B_0)-Z)[T_1]_0\,C(T_0) \quad \text{(S2-32)}$$

$$C(T_0)=\mathbf{e}^{-k_{LLI}(T_0-w)} \quad \text{(S2-33)}$$

where $\Delta X(B_0)=X(B_0)-(0),\Delta Y(B_0)=Y(B_0)-Y(0)$. Assuming $k_{T_1},k_{RP}\gg k_{LLI}$:

$$\Delta F_{PFP}^{(OFF-ON)}(\tau_{01},B_0)\propto-\left[C_{RP(0)}\mathbf{e}^{-k_{RP}(0)(\tau_{01}-w)}+C_{T_1}\mathbf{e}^{-k_{T_1}(\tau_{01}-w)}\right] \quad \text{(S2-34)}$$

Therefore, in the Off-On shift type, the MFEs with the RSMF, as a function of delay time, reflects the RP dynamics in zero magnetic field. When RPs are formed quickly, this dependence directly reflects the lifetimes of the RPs in zero field.

<u>On-Off RSMF shift $\Delta F_{PFP}^{(ON-OFF)}$:</u>

Similarly, in case of the On-Off shift type, the PFP fluorescence signal is expressed as follow:

$$\Delta F_{PFP}^{(ON-OFF)}(\tau_{10},B_0)\propto-\left([LLI](T_0,B_0;\tau_{10})-[LLI](T_0,0)\right) \quad \text{(S2-35)}$$

In the On-Off shift measurement, the$[LLI](T_0,B_0;\tau_{10})$ is obtained by solving the rate equations (S2-1) with the following magnetic field condition.

$$B_{RSMF}(t;\tau_{10})=\begin{cases}B_0 & (0\le t\le\tau_{10})\\ 0 & (\tau_{10}\le t\le T_0)\end{cases} \quad \text{(S2-36)}$$

A general solution of [LLI](t) on the four-state model for $\tau_{10}<t<T_0$ is:

$$\begin{aligned}[LLI](t,B_0;\tau_{10})&=\left[[LLI](\tau_{10},B_0)+Y(0)[RP](\tau_{10},B_0)+(Y(0)-Z)X(0)[T_1](\tau_{10})\right]\mathbf{e}^{-k_{LLI}(t-\tau_{10})}\\&-Z\left([RP](\tau_{10},B_0)+X(0)[T_1](\tau_{10})\right)\mathbf{e}^{-k_{RP}(0)(t-\tau_{10})}+ZX(0)[T_1](\tau_{10})\mathbf{e}^{-k_{T_1}(t-\tau_{10})}\end{aligned} \quad \text{(S2-37)}$$

Since $T_0\gg\tau_{T_1},\tau_{RP}$, $[LLI](T_0,B_0;\tau_{10})$ is obtained:

$$[LLI](T_0,B_0;\tau_{10})=\left[[LLI](\tau_{10},B_0)+Y(0)[RP](\tau_{10},B_0)+(Y(0)-Z)X(0)[T_1](\tau_{10})\right]\mathbf{e}^{-k_{LLI}(T_0-\tau_{10})} \quad \text{(S2-38)}$$

Substituting (S2-37) and (S2-38) into (S2-35), and then analytical solutions (S2-4,5,6) into that, we obtain:

$$\Delta F_{PFP}^{(ON-OFF)}(\tau_{10},B_0)\propto-\left[C_{RP(B_0)}\left(1-\mathbf{e}^{-(k_{RP}(B_0)-k_{LLI})(\tau_{01}-w)}\right)+C_{T_1}\left(1-\mathbf{e}^{-(k_{T_1}-k_{LLI})(\tau_{01}-w)}\right)\right] \quad \text{(S2-39)}$$

with coefficients defined as:

$$C_{RP(B_0)}=X(B_0)\Delta Y(B_0)[T_1]_0C(T_0) \quad \text{(S2-40)}$$

$$C_{T_1}=\Delta X(B_0)(Y(0)-Z)[T_1]_0C(T_0) \quad \text{(S2-41)}$$

$$C(T_0)=\mathbf{e}^{-k_{LLI}(T_0-w)} \quad \text{(S2-42)}$$

Assuming $k_{T_1}, k_{RP} \gg k_{LLI}$:

$$\Delta F_{PFP}^{(ON-OFF)}(\tau_{10}, B_0) \propto -\left[C_{RP(B_0)}\left(1-\mathbf{e}^{-k_{RP}(B_0)(\tau_{10}-w)}\right) + C_{T_1}\left(1-\mathbf{e}^{-k_{T_1}(\tau_{10}-w)}\right)\right] \tag{S2-43}$$

Therefore, in the On-Off shift type, the delay time dependence of the MFEs with the RSMF reflects the RP dynamics in a magnetic field. When RPs are formed quickly, this dependence directly reflects the lifetimes of the RPs in a magnetic field.

As a side note, defining the differential signal of PFP signals as follows, we can obtain the MFEs of only RP dynamics.

$$\Delta\Delta F_{PFP}(\tau, B_0) = \left(\Delta F_{PFP}^{(ON-OFF)}(T_0, B_0) - \Delta F_{PFP}^{(ON-OFF)}(\tau, B_0)\right) - \Delta F_{PFP}^{(OFF-ON)}(\tau, B_0) \propto -\Delta RP(\tau, B_0) \tag{S2-44}$$

In summary, we employed a four-state RP reaction model to analytically describe the delay-time dependence of fluorescence signals observed in both PP and PFP measurements. This model provides a simple framework for linking the parameters obtained from exponential curve fitting to the reaction rate constants. In PFP, the delay-time dependence appears as a double-exponential form, while in PP it is represented by a triple-exponential expression. Indeed, PP can obtain decay rate of the triplet state and the RP state in the case that the decay of the LLI is much slower than in other states, but this simplification in PFP enhances the reliability and accuracy of kinetic analysis, enabling more precise characterization of RP dynamics.

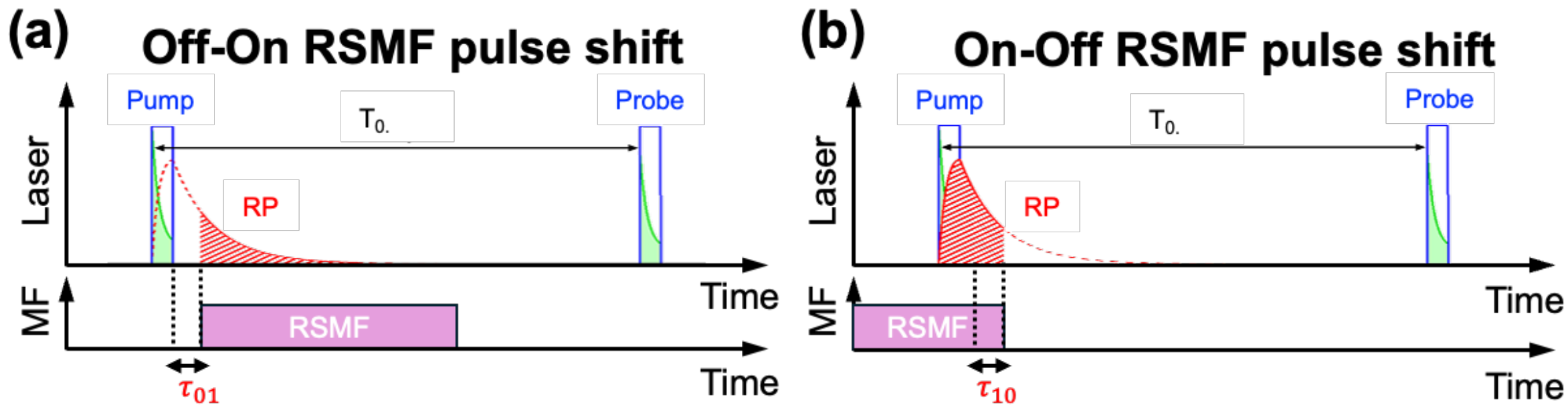

**Fig. S2-4 PFP fluorescence detection scheme.** (a) Off-On RSMF pulse shift. (b)On-Off RSMF pulse shift.

## 3. Experimental setup

### 3.1 Materials

FMN, FAD, and Trp were purchased from Sigma-Aldrich. Samples at pH 2.3 and pH 7.4 were prepared using citrate/phosphate buffer and PBS buffer (Sigma-Aldrich), respectively. Sample thicknesses of 2.9 μm and 4.9 μm were achieved by adding polymer microbeads (2.0–2.9 μm and 4.5–4.95 μm in diameter, respectively; Spherotech Inc.) to the sample solution and then sandwiching 1 μL of the solution between glass cover slips (No.1, 24 × 60 mm, 0.13–0.17 mm thick; Matsunami) [3]. To prevent evaporation, silicon grease was applied to the edges of the slide. For 250 μm thickness samples, 5 μL of the solution was placed in a chamber sealed with adhesive spacers (SLF0201, Bio-Rad) [4].

### 3.2 Microscope principle

**Figure S3-1** shows the schematic of a new custom-built fluorescence microscope for PP and PFP fluorescence detection measurements. To achieve arbitrary delay times between pump and probe pulses, two independent, identical 450 nm nanosecond pulse lasers (NPL45C, Thorlabs) are used for the single-color pump-probe excitation system. These laser pulses are combined into a multimode fiber (M42L02, Thorlabs) using a knife-edge prism mirror (MRAK25-P01, Thorlabs) and collimated through an aspheric lens (CFC11P-A, Thorlabs), allowing optimal spatial overlap on the sample. In PP fluorescence detection measurements, the static magnetic field is generated by a projected field electromagnet (GMW5204, GMW Associates). In PFP fluorescence detection measurements, the rapidly switching magnetic field (RSMF) is generated by a capacitor bank-based custom pulser circuit and a homemade solenoid coil (5 turns, 4 mm diameter). This setup allows sub-10 ns rise-time switching and provides a flat magnetic field output on the microsecond timescale [5]. The fluorescence signals are captured using an sCMOS camera (ORCA Flash 4.0 V3, Hamamatsu) through a 100× oil objective lens with a numerical aperture of 1.49 (UAPON100XOTIRF, Olympus), a dichroic mirror (T470lpxr, Chroma), a reflection mirror (PFR10-P01, Thorlabs), a long-pass filter (ET500lp, Chroma), and a tube lens (AC254-200-A, Thorlabs). The timing for the two laser pulses, the static magnetic field, the RSMF, and the camera is controlled by a custom controller circuit based on the Raspberry Pi Pico microcontroller with data acquisition programs written in Micropython / PIO assembly (Pi Pico) and LabVIEW code (control PC).

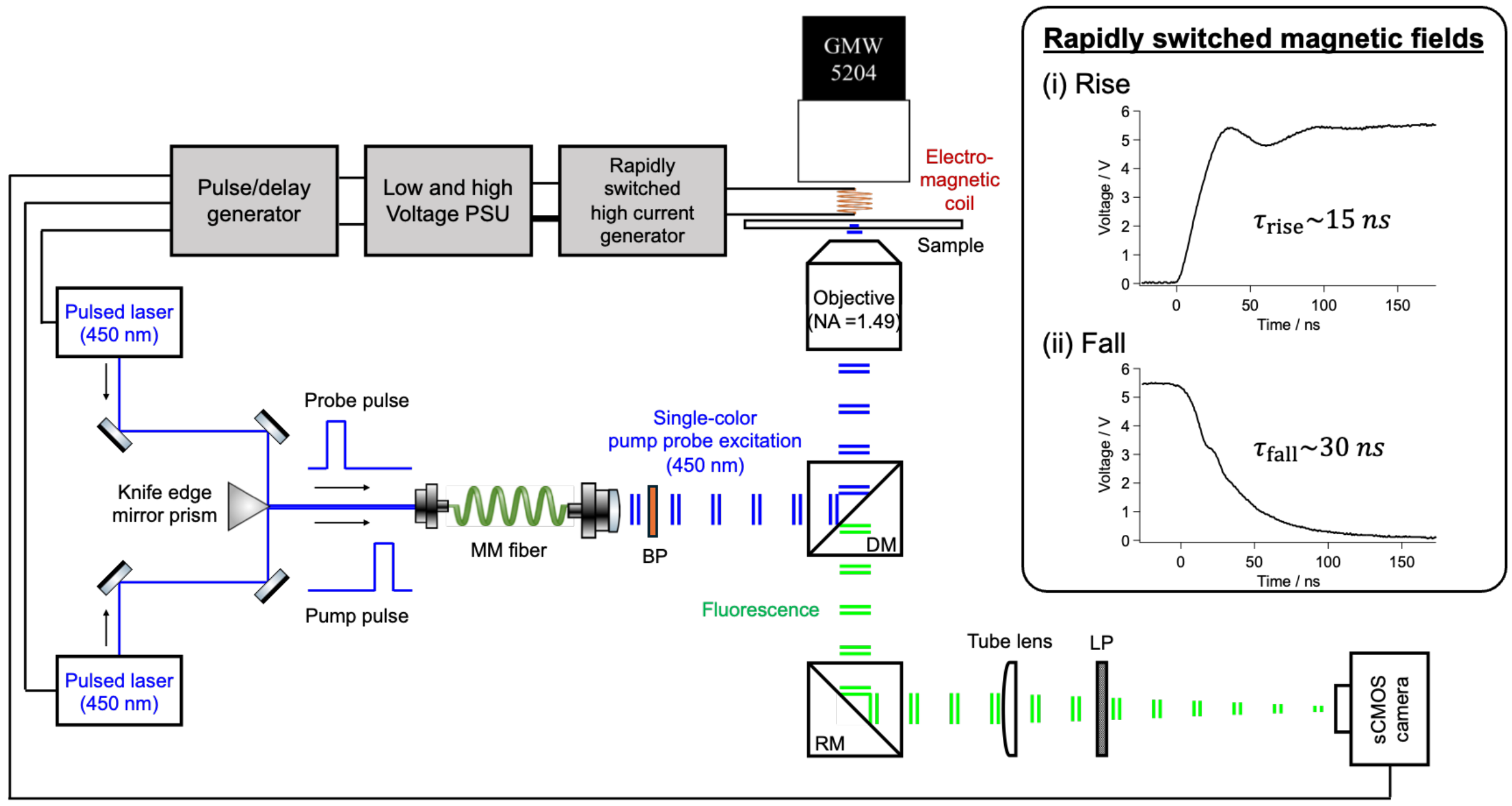

**Figure S3-1. Microscope setup.** The rise and fall times of the rapidly switched magnetic field were estimated using a mono-exponential fit.

## 4. Microscope calibration

### 4.1 RSMF magnitude calibration

Direct measurement of short, microsecond-scale magnetic field pulses with low repetition rates is technically difficult using conventional magnetic sensors such as Hall probes. Therefore, before starting measurements, the magnitude of the rapidly switched magnetic field (RSMF) was calibrated indirectly using MFEs observed in a chemical reaction system. To perform the calibration, an external static magnetic field (DC field) of known magnitude was applied in the opposite direction to the RSMF (**Fig. S4-1a**). By sweeping the magnitude of the DC field to cancel the MFE induced by the RSMF, the magnitude of the RSMF was estimated.

The MFE induced by the RSMF was measured by setting the pump–probe laser delay time after the rise and before the fall of the RSMF pulse, so that the flat part of the pulse was probed (**Fig. S4-1b**). During this period, a DC field in the opposite direction to the RSMF was applied, and its magnitude was swept to determine the value that canceled the MFE.

The magnitude of the DC field that cancels the MFE is half that of the RSMF. This is because when the combined RSMF and DC field is equal in magnitude but opposite in direction to the DC field alone, the resulting fluorescence intensities are equal and the MFE is zero. This can be derived mathematically.

When the MFE due to the combined field is zero, the fluorescence signals under the combined field of RSMF and DC field and DC field alone:

$$MFE = \frac{F(B_{RSMF} - B_{DC}) - F(-B_{DC})}{F(-B_{DC})} = 0 \Leftrightarrow F(B_{RSMF} - B_{DC}) = F(-B_{DC}) \quad \text{(S4-1)}$$

Fluorescence under a magnetic field, B, is expressed using the MARY curve.

$$\frac{F(B) - F(0)}{F(0)} = -MFE_{sat}\frac{B^2}{B^2 + \left(B_{1/2}\right)^2} \Leftrightarrow F(B) = F(0)\left(1 - MFE_{sat}\frac{B^2}{B^2 + \left(B_{1/2}\right)^2}\right) \quad \text{(S4-2)}$$

Substituting (S2-2) into (S2-1) and dividing both sides by $F(0)$.

$$1 - MFE_{sat}\frac{(B_{RSMF} - B_{DC})^2}{(B_{RSMF} - B_{DC})^2 + \left(B_{1/2}\right)^2} = 1 - MFE_{sat}\frac{(B_{DC})^2}{(B_{DC})^2 + \left(B_{1/2}\right)^2} \quad \text{(S4-3)}$$

Solving this equation:

$$(B_{RSMF} - B_{DC})^2 = (B_{DC})^2 \Leftrightarrow B_{DC} = \frac{1}{2}B_{RSMF} \quad (\because B_{DC} \neq 0) \quad \text{(S4-4)}$$

Thus, by measuring the DC field magnitude that cancels the MFE and multiplying it by two, the magnitude of the RSMF can be indirectly measured. **Fig. S4-1c** shows an example of this calibration procedure using FAD and tryptophan in acidic solution. The magnetic-field step (ΔB) in the PFP-MARY experiment was determined from the applied current–voltage relationship (**Fig. S4-1d**) and the maximum RSMF field estimated in **Fig. S4-1c**, yielding a calibration B(I) to convert any applied current to its corresponding field.

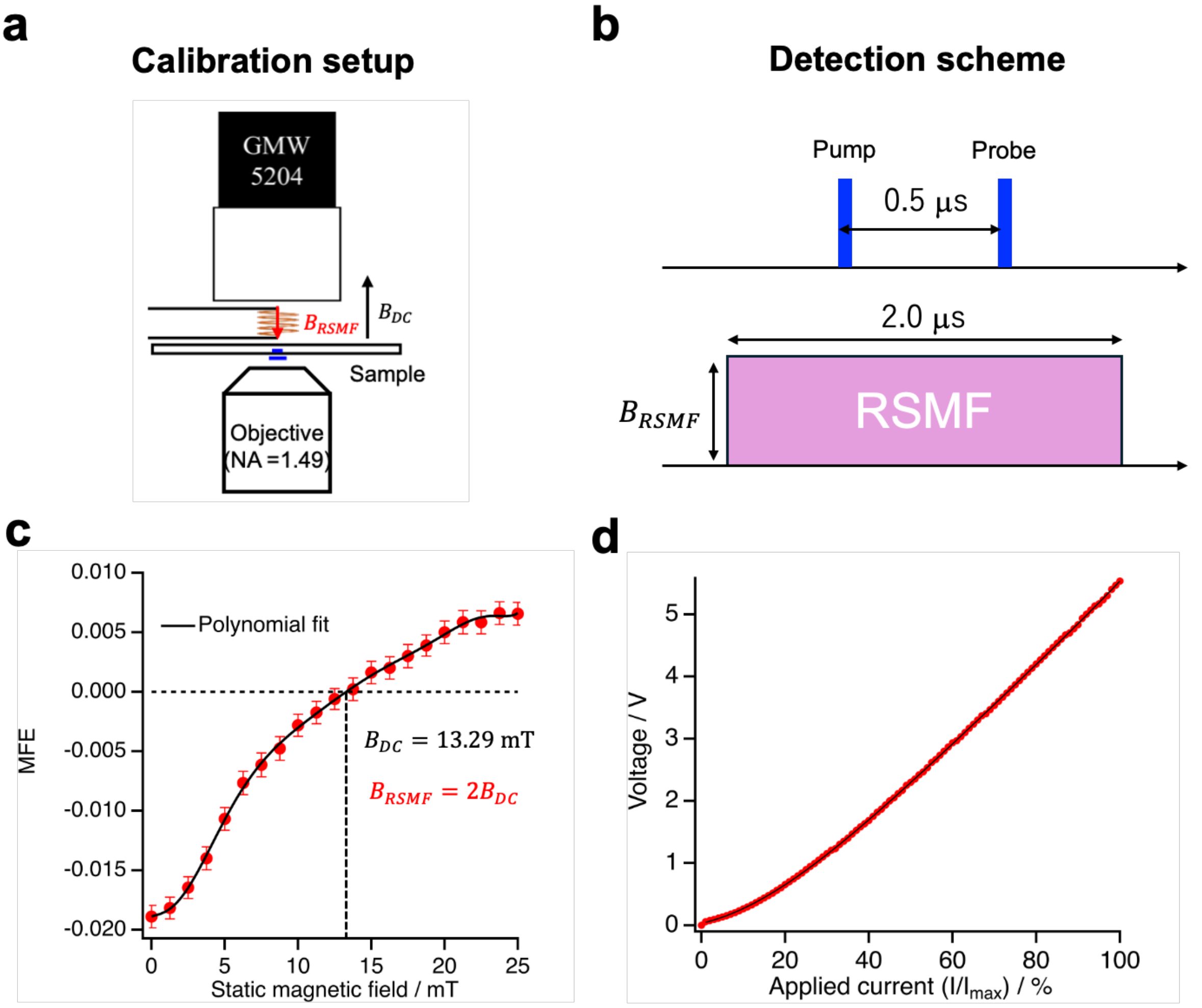

**Figure S4-1. RSMF magnitude calibration.** (a) Schematics of the calibration setup. (b) Detection scheme of MFEs for the calibration. (c) Example data of RSMF calibration using FAD (10 μM) and tryptophan (0.3 mM) in acidic solution (pH 2.3) for the measurements shown in **Fig.6**. (d) Relationship between applied current and applied voltage. From the current–voltage curve and the maximum RSMF magnetic field estimated in **Fig. S4-1 c**, the applied-current sweep was converted to a magnetic-field sweep.

**4.2 Optimization of pump probe pulse excitation**

To ensure that the observed MFEs under repetitive pump-probe excitation arise only from those occurring between the pump and probe pulses, we optimize the pump-probe pulse width and laser repetition rate to meet the following conditions:

1. **Pulse width:** MFEs must not be established during the pump pulse (i.e., not MFEs typically observed under CW or pseudo-CW excitation).
2. **Repetition rate:** MFEs must not be established by repeated excitation (i.e., not MFEs between the probe pulse and the next pump pulse).

**Figure S4-2** shows the dependence of MFEs on pulse width at different laser repetition rates in FAD systems. Although, at high laser repetition rates (400 Hz), short pulse width excitation produced unexpected positive MFEs, typical triplet-born RP MFEs appeared as the pulse width increased, becoming clearly observable when the pulse width exceeded 40 ns. Therefore, in the FAD system, the MFE is not detected below 200 Hz because the MFE is not established by repetition rates below 200 Hz in single-pulse excitation, so pump-probe excitation below 100 Hz can be used to monitor the MFE of the pump-probe interactions. dynamics MFE can be monitored.

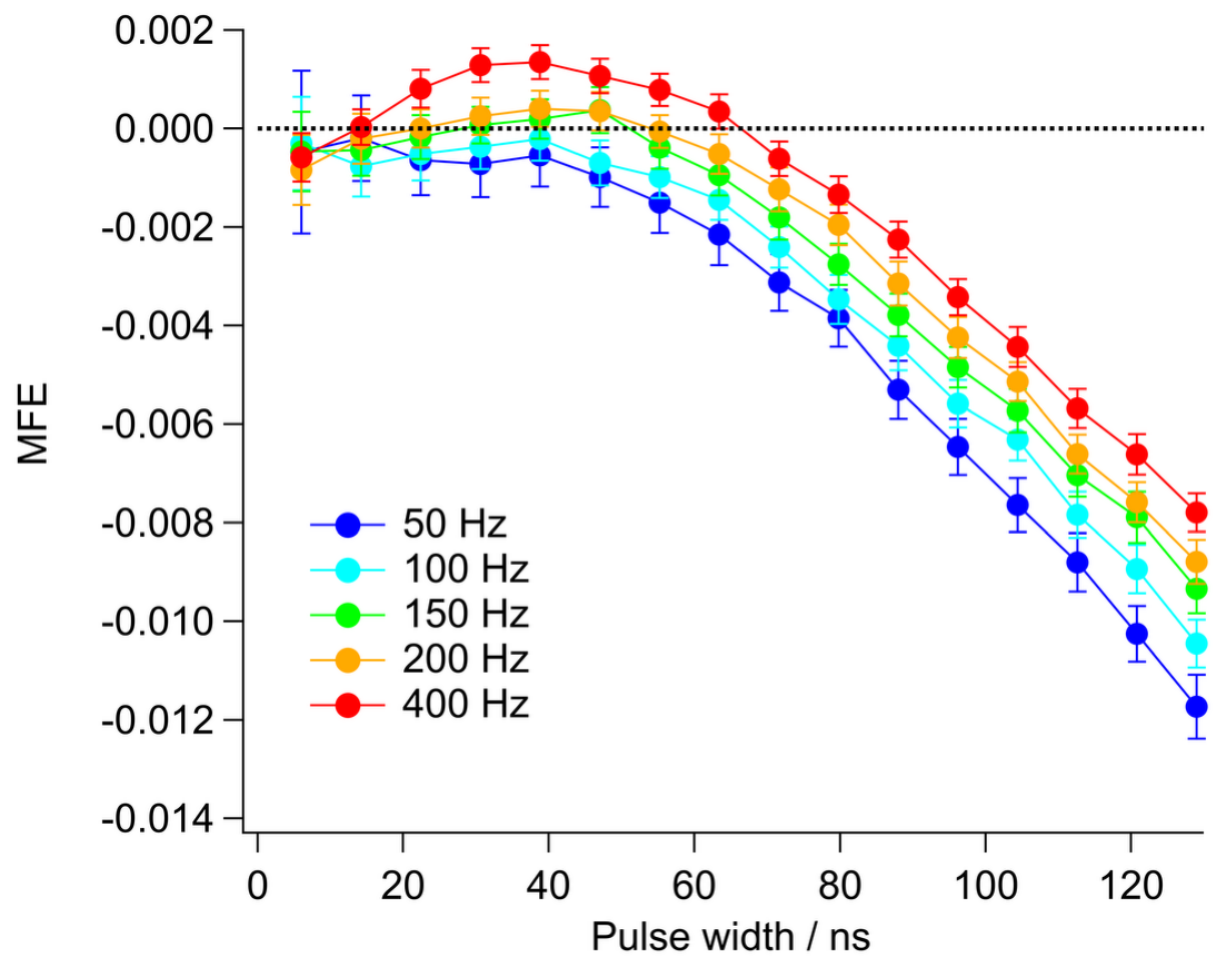

**Figure S4-2 The pulse width dependence of MFEs on 100 μM FAD at pH 2.3.**

## 5. Data acquisition sequence and analysis

This section describes the data acquisition sequence and the analytical methods used for the kinetics and MARY measurements exploiting PP and PFP fluorescence detection.

In all measurements, fluorescence was generated by the pump-probe laser excitation system at a repetition rate, $f_{rep}$, which was set sufficiently low to eliminate the residual LLIs formed via RPs and their MFEs. The fluorescence was captured as image data by camera-based detection over an exposure time, Δt, corresponding to $f_{rep}\Delta t$ excitations per frame. To evaluate the fluorescence response relative to the excitation intensity – which is assumed to be spatially uniform across the illuminated area - the integrated fluorescence intensity was defined as the average fluorescence intensity per pixel within the region of interest (ROI) within illumination area. This averaging reduces the pixel-to-pixel fluctuations due to excitation intensity variation and improves the signal-to-noise ratio. Therefore, the integrated fluorescence intensity was calculated as follows:

$$F = \frac{1}{N_{ROI}} \sum_{(i,j)\in ROI} F(i,j) \tag{S5-1}$$

where $N_{ROI}$ is the number of pixels within the ROI region, $F(i,j)$ is the fluorescence intensity at a pixel at position $(i,j)$. This calculation was performed the open-source imaging processing software ImageJ.

For each delay time point, the measurement was performed by repeatedly cycling the Probe, static MF or RSMF between Off and On states over a certain time ($T_{off/on}$) and acquiring fluorescence images accordingly. The outline of data analysis is:

1. **Data acquisition**: Acquire fluorescence image data by performing Probe or MF Off/On cycle measurements at each delay time or MF strength and calculated the averaged intensity.
2. **Signal extraction**: Extract difference signals between On and Off measurements using the Off-only data or through residual analysis of curve fitting.
3. **Mean and error calculation**: Calculate the overall mean and error of the signals based on the extracted signals.
4. **Plotting:** Plot the final mean signals and their errors are plotted as functions of either delay time or magnetic field strength.

Details for each measurement are presented in Sections 5.1 to 5.4.

## 5.1 PP kinetics measurements

*$\Delta F_{PP}$ measurements*

**Table S5-1** displays the data acquisition sequence to measure $\Delta F_{PP}$ shown in **Fig. 3**. In this sequence, $\Delta F_{PP}$ is measured by performing M repetitions (typically 8 times) of integrated fluorescence acquisition under each Probe off and on condition (corresponding to $F_{pu}$ and $F_{pu+pr}$, respectively) with $T_{off}/T_{on}$ cycles (typically 5-seconds each). Following these cycles, a recovery (non-irradiation) period (typically 60 or 80 seconds) is applied, followed by an additional Probe Off ($F_{pu}$) measurement with the same acquisition time. This additional measurement is used to correct for fluorescence decay due to photobleaching. After this, the same recovery period is applied before proceeding to the next measurement again. This entire procedure is repeated for each pump-probe delay time to determine the corresponding $\Delta F_{PP}$ signal.

**Figure S5-1** outlines the data analysis procedure for $\Delta F_{PP}$ measurement. First, a series of integrated fluorescence signals is measured at each Pump-Probe delay time step by calculating the average intensity from the image data acquired according to the sequence shown in **Table S5-1** (**Figure S5-1 a**). From the data, the $\Delta\overline{F_{PP}}(T,B)$ signals are extracted at each Pump-Probe delay time step by performing residuals analysis using Probe Off fluorescence signals (**Figure S5-1 b**), followed by calculation of their normalized mean and error (**Figure S5-1 c**). Finally, these values are plotted against Pump-Probe delay time (**Figure S5-1 d**).

$\Delta\overline{F_{PP}}(T)$ was calculated as the difference between the normalized residuals of the fluorescence signals acquired under Probe On and Probe Off conditions.

$$\Delta\overline{F_{PP}}(T) = \overline{Res}(probe_\text{on}) - \overline{Res}(probe_\text{off}) \tag{S5-2}$$

Here, the normalized residuals are given as:

$$\overline{Res} = \frac{F_{\text{off/on}} - F_{\text{off}}}{F_{\text{off}}} \tag{S5-3}$$

Where $F_{\text{off/on}}$ and $F_{\text{off}}$ denote the integrated fluorescence signal under Probe Off/On and under Off-only measurement, respectively.

The mean and standard deviation of $\Delta\overline{F_{PP}}$ at each step kth off/on step were calculated as:

$$Mean\left(\Delta\overline{F_{PP}^{(k)}}(T)\right) = Mean\left(\overline{Res^{(k)}}(probe_\text{on})\right) - Mean\left(\overline{Res^{(k)}}(probe_\text{off})\right) \tag{S5-4}$$

$$SD\left(\Delta\overline{F_{PP}^{(k)}}(T)\right) = \sqrt{\left[SD\left(\overline{Res^{(k)}}(probe_\text{on})\right)\right]^2 + \left[SD\left(\overline{Res^{(k)}}(probe_\text{off})\right)\right]^2} \tag{S5-5}$$

Here, SD denotes the standard deviation. To obtain accurate values for the mean and SD, a few data points at the beginning and end of each Off/On step were excluded from the analysis. Since each Off/On step measurement is independent, the mean of the $\Delta F_{PP}$ was calculated as:

$$Mean\left(\Delta\overline{F_{PP}}(T)\right) = \frac{1}{M-1}\sum_{k=2}^{M} Mean\left(\Delta\overline{F_{PP}^{(k)}}(T)\right) \tag{S5-6}$$

The error of $\Delta\overline{F_{PP}}$ was calculated using the standard deviation of mean:

$$Error\left(\Delta\overline{F_{PP}}(T)\right) = \frac{1}{\sqrt{M-1}}\sum_{k=2}^{M} SD\left(\Delta\overline{F_{PP}^{(k)}}(T)\right) \quad \text{(S5-7)}$$

Here, to obtain accurate values for the overall mean and error, data from k = 2 onward are used. Therefore, the effective number of repetitions is M – 1.

**Table S5-1** Data acquisition sequence for $\Delta F_{PP}$ measurement shown in **Fig. 3**.

| | Repeats over Pump-Probe Delay Steps ΔT | | | | |
|---|---|---|---|---|---|
| | M Cycles (M = 8) | | Recovery ($T_{rec}$) | M Cycles (M = 8) | Recovery ($T_{rec}$) |
| Pump-Probe Laser System | Probe Off ($T_{off}$) | Probe On ($T_{on}$) | | Probe Off ($2T_{off}$) | |
| Camera | 200 ms Camera Exposure | | | | |

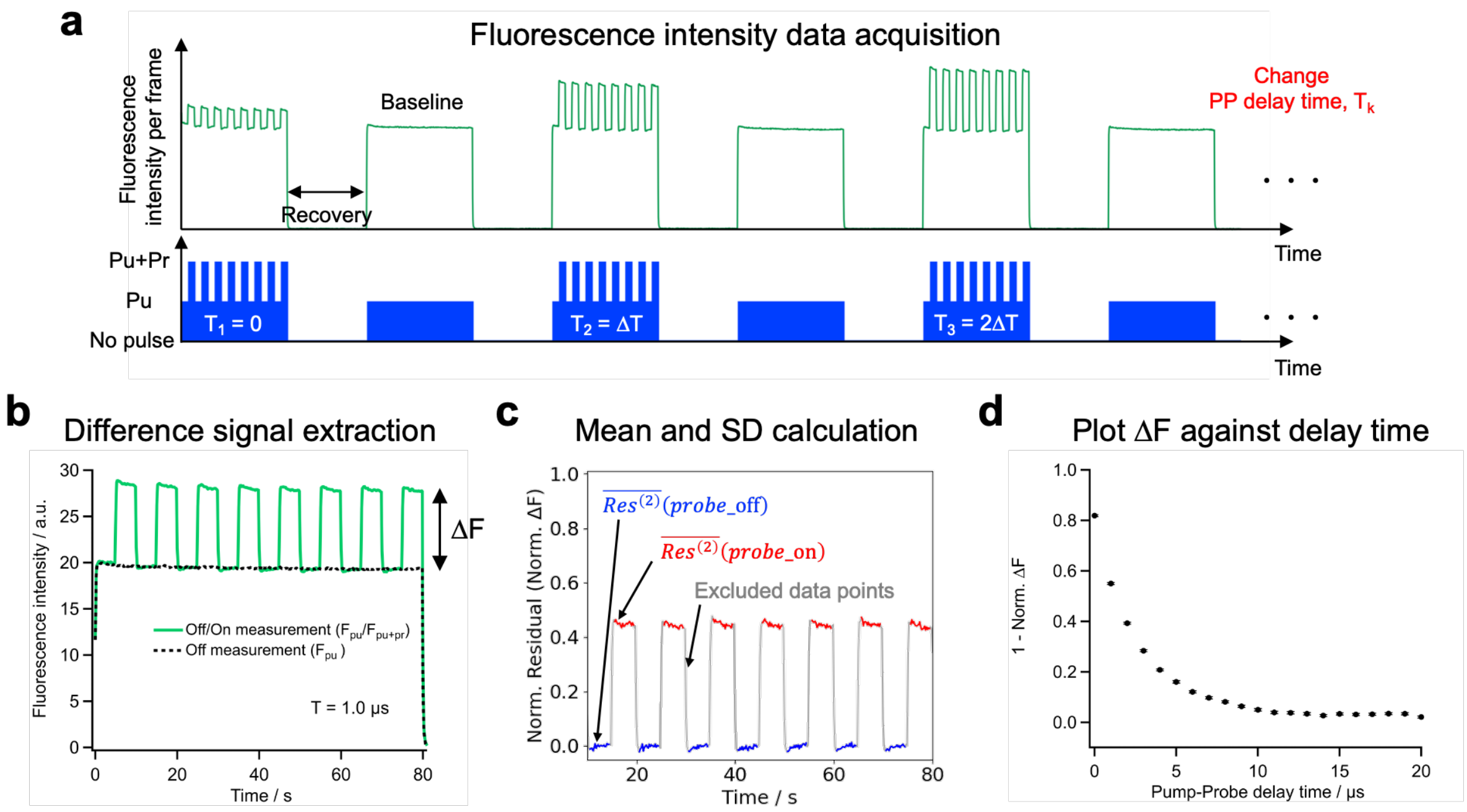

**Figure S5-1 Data analysis procedure for $\Delta F_{PP}$ measurement shown in Fig. 3b and 3c.** (a) Fluorescence intensity acquisition timeline. (b). Integrated fluorescence signals of Probe Off/On and Off measurement at a fixed Pump-Probe delay time. (c) Normalized residuals corresponding to normalized $\Delta F_{PP}$. (d) Normalized $\Delta F_{PP}$ against delay time.

*$\Delta\Delta F_{PP}$ measurements (obtained from MFEs of $\Delta F_{PP}$)*

**Table S5-2** shows the data acquisition sequence used to measure the $\Delta\Delta\overline{F_{PP}}(T)$ measurements obtained from MFEs of the $\Delta F_{PP}$ signals. In this sequence, $\Delta\overline{F_{PP}}(0,0)$ is first measured, following the procedure outlined in **Table S5-1**, and then the same measurement is repeated under a magnetic field to obtain $\Delta\overline{F_{PP}}(0,B_0)$. This process is performed for each pump–probe delay time. Since it has been shown that the fluorescence intensity doesn't change with pump pulse only (i.e. during the Probe Off period measurement), the Probe Off signal can be assumed to be identical with and without the magnetic field. Accordingly, the same analytical procedure can be used to extract $\Delta\overline{F_{PP}}(T)$ values under both magnetic and non-magnetic conditions.

**Figure S5-2** outlines the analysis procedure for $\Delta\Delta\overline{F_{PP}}$ measurement derived from $\Delta F_{PP}$. First, a series of integrated fluorescence signals is measured at each Pump-Probe delay time step under magnetic field Off and On by calculating the average intensity from the image data acquired according to the sequence shown in **Table S5-2** (**Figure S5-2 a**). From the data, the $\Delta\overline{F_{PP}}(T,B_0)$ were obtained by performing residuals analysis on the integrated fluorescence signals at each Pump-Probe delay time, comparing them under magnetic field Off and On condition, using Probe Off fluorescence signals (**Figure S5-5 b**). The normalized mean and error under magnetic field Off and On condition were then calculated from these residuals (**Figure S5-5 c**). Finally, the $\Delta\Delta\overline{F_{PP}}(T,B_0)$ were plotted against Pump-Probe delay time (**Figure S5-2 d**).

As in the previous section, $\Delta\overline{F_{PP}}(T,0)$ and $\Delta\overline{F_{PP}}(T,B_0)$ were calculated as the difference between the normalized residuals of the fluorescence signals acquired under Probe On and Probe Off conditions, under magnetic field off and on respectively:

$$\Delta\overline{F_{PP}}(T,0) = \overline{Res}(MF_\mathrm{off}, probe_\mathrm{on}) - \overline{Res}(MF_\mathrm{off}, probe_\mathrm{off}) \tag{S5-8}$$

$$\Delta\overline{F_{PP}}(T,B_0) = \overline{Res}(MF_\mathrm{on}, probe_\mathrm{on}) - \overline{Res}(MF_\mathrm{on}, probe_\mathrm{off}) \tag{S5-9}$$

Here, the normalized residuals are given as:

$$\overline{Res} = \frac{F_{\mathrm{off/on}} - F_{\mathrm{off}}}{F_{\mathrm{off}}} \tag{S5-10}$$

Where $F_{\mathrm{off/on}}$ and $F_{\mathrm{off}}$ denotes the integrated fluorescence signal under Probe Off/On and under Off-only measurement, respectively.

The mean and error of $\Delta\Delta\overline{F_{PP}}(T,B_0)$ were calculated from the overall mean and error values of $\Delta\overline{F_{PP}}(T,B_0)$ and $\Delta\overline{F_{PP}}(T,0)$ obtained by following the previous discussion. Since $\Delta\overline{F_{PP}}(T,B_0)$ and $\Delta\overline{F_{PP}}(T,0)$ measurements are independent, the mean and error of $\Delta\Delta\overline{F_{PP}}(T,B_0)$ were calculated as follow:

$$Ave\left(\Delta\Delta\overline{F_{PP}}(T,B_0)\right) = Ave\left(\Delta\Delta\overline{F_{PP}}(T,B_0)\right) - Ave\left(\Delta\Delta\overline{F_{PP}}(T,0)\right) \tag{S5-11}$$

$$Error\left(\Delta\Delta\overline{F_{PP}}(T,B_0)\right) = \sqrt{\left[Error\left(\Delta\overline{F_{PP}}(T,B_0)\right)\right]^2 + \left[Error\left(\Delta\overline{F_{PP}}(T,0)\right)\right]^2} \tag{S5-12}$$

**Table S5-2** Data acquisition sequence of $\Delta\Delta F_{PP}$ measurement shown in **Fig. 4b**.

| | Repeats over Pump-Probe Delay Steps ΔT | | | | | | | | | |
|---|---|---|---|---|---|---|---|---|---|---|
| | M Cycles (M = 8) | | Recovery ($T_{rec}$) | M Cycles (M = 8) | Recovery ($T_{rec}$) | M Cycles (M = 8) | | Recovery ($T_{rec}$) | M Cycles (M = 8) | Recovery ($T_{rec}$) |
| Pump-Probe Laser System | Probe Off ($T_{off}$) | Probe On ($T_{on}$) | | Probe Off ($2T_{off}$) | | Probe Off ($T_{off}$) | Probe On ($T_{on}$) | | Probe Off ($2T_{off}$) | |
| Static MF | MF off | | | | | MF on | | | | |
| Camera | 200 ms Camera Exposure | | | | | | | | | |

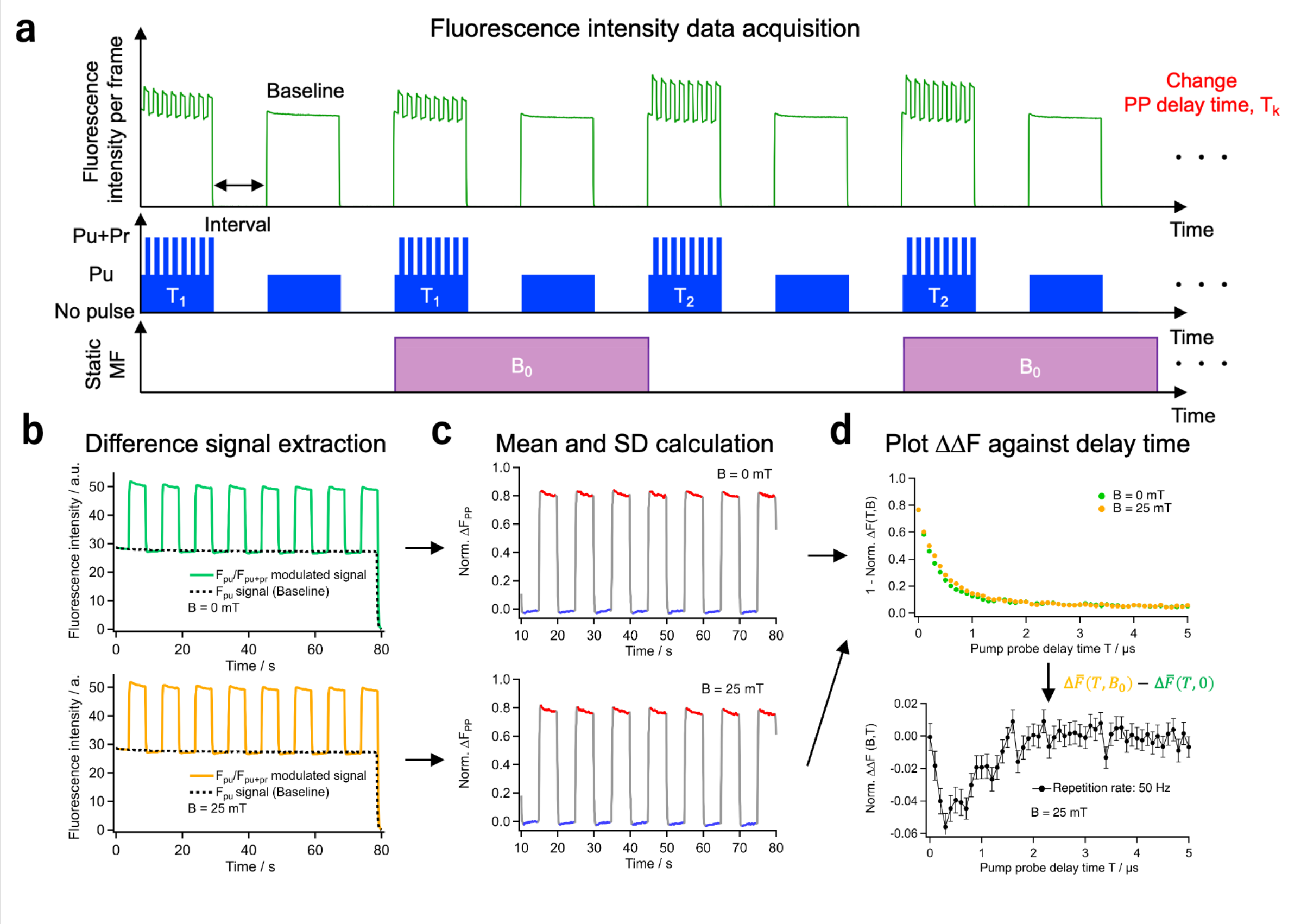

**Figure S5-2 Data acquisition sequence of $\Delta\Delta F_{PP}$ measurement shown in Fig. 4b.** (a). Fluorescence intensity acquisition timeline. (b) Integrated fluorescence signals of Probe-Off/On and Off measurement at a fixed Pump-Probe delay time with and without a magnetic field ($B_0$). (c). Normalized residuals corresponding to the $\Delta\overline{F_{PP}}(T,B_0)$ and $\Delta\overline{F_{PP}}(T,0)$. (d).Plots of $\Delta\overline{F_{PP}}(T,B_0)$, $\Delta\overline{F_{PP}}(T,0)$ and $\Delta\Delta\overline{F_{PP}}(T,B_0)$.

*$\Delta\Delta F_{PP}$ measurements (obtained from MFEs of pump-probe fluorescence signals)*

**Table S5-3** displays the data acquisition sequence for $\Delta\Delta F_{PP}$ measurement, obtained from MFEs of pump-probe fluorescence signals. In this sequence, under Pump-Probe excitation at a given delay time, fluorescence signals under alternating magnetic field Off and On conditions (corresponding to $F_{pu+pr}(T,0)$ and $F_{pu+pr}(T,B)$, respectively) are acquired **M** times with $T_{off}/T_{on}$ cycles. Following these cycles, a recovery period ($T_{rec}$) is applied. Here, photobleaching correction measurement is not performed, as the photobleaching trend can be estimated through curve fitting. Therefore, this acquisition process is simply repeated for different Pump–Probe delay times.

**Figure S5-3** outlines the data analysis procedure for this $\Delta\Delta F_{PP}$ measurement. First, a series of integrated fluorescence signals was obtained at each Pump-Probe delay time step by calculating the average intensity from the image data acquired according to the sequence shown in **Table S5-3** (**Figure S5-3 a**). From the data, the $\Delta\Delta F_{PP}$ signals were extracted at each Pump-Probe delay time step by performing residual analysis through curve fitting of the MF Off/On cycle measurements (**Figure S5-3 b**), followed by calculation of their mean and error (**Figure S5-3 c**), Finally, these values are plotted against Pump–Probe delay (**Figure S5-3 d**).

The corrected fluorescence intensity corresponding to $F_{pu+pr}(T,0)$ was estimated by curve fitting using the following biexponential function.

$$BiExp(x|x_0, y_0, A_1, \tau_1, A_2, \tau_2) = y_0 + A_1\mathbf{e}^{-\frac{x-x_0}{\tau_1}} + A_2\mathbf{e}^{-\frac{x-x_0}{\tau_2}} \tag{S5-13}$$

To accurately estimate the photobleaching decay, the fitting was performed after excluding several early-time data points (typically, from 0 to 8s).

The $\Delta\Delta F_{PP}(T,B_0)$ was calculated as the difference between the residuals of the fluorescence signals acquired under MF On and Off conditions.

$$\Delta\Delta F_{PP}(T,B_0) = Res(MF_\mathrm{on}) - Res(MF_\mathrm{off}) \tag{S5-14}$$

Here, the residuals denote as:

$$Res = F_{\mathrm{off/on}} - fitted\ value \tag{S5-15}$$

Where $F_{\mathrm{off/on}}$ denotes the integrated fluorescence signal under MF Off/On measurement.

The mean and error of the $\Delta\Delta F_{PP}(T,B_0)$ signal were calculated from the means and standard deviations of the residuals obtained from each MF Off/On cycle, as described below.

The mean and standard deviations of $\Delta\Delta F_{PP}(T,B_0)$ for the kth Off/On cycle were calculated as:

$$Ave\left(\Delta\Delta F_{PP}^{(k)}(T,B_0)\right) = Mean\left(Res^{(k)}(MF_\mathrm{on})\right) - Mean(Res^{(k)}(MF_\mathrm{off})) \tag{S5-16}$$

$$SD\left(\Delta\Delta F_{PP}^{(k)}(T,B_0)\right) = \sqrt{\left[SD\left(Res^{(k)}(MF_\mathrm{on})\right)\right]^2 + \left[SD\left(Res^{(k)}(MF_\mathrm{off})\right)\right]^2} \tag{S5-17}$$

Then, the overall mean and error of the $\Delta\Delta F_{PP}(T,B_0)$ were then calculated by averaging over M-1 Off/On cycles (excluding the first cycle, k = 1):

$$Mean(\Delta\Delta F_{PP}(T,B_0)) = \frac{1}{M-1}\sum_{k=2}^{M} Ave\left(\Delta\Delta F_{PP}^{(k)}(T,B_0)\right) \tag{S5-18}$$

$$Error\left(\Delta\Delta F_{PP}(T,B_0)\right) = \frac{1}{\sqrt{M-1}}\sum_{k=2}^{M} SD\left(\Delta\Delta F_{PP}^{(k)}(T,B_0)\right) \tag{S5-19}$$

In addition, to eliminate the variation of the fluorescence intensity, the normalized value of $\Delta\Delta F_{PP}$, $\Delta\Delta\widetilde{F_{PP}}$, show in **Fig. 6** was introduced and calculated as follows.

$$\Delta\Delta\widetilde{F_{PP}}(T,B_0) \overset{\text{def}}{=} \frac{F_{pu+pr}(T,B_0) - F_{pu+pr}(T,0)}{F_{pu+pr}(T,0)} \tag{S5-20}$$

$$\Delta\Delta\widetilde{F_{PP}}(T,B_0) = \overline{Res}(MF_\text{on}) - \overline{Res}(MF_\text{off}) \tag{S5-21}$$

$$\overline{Res} = \frac{F_{\text{off/on}} - (fitted\ value + \delta)}{fitted\ value + \delta} \tag{S5-22}$$

where the collection of value, $\delta$, was calculated as the overall mean of the residual for the off-cycle measurement.

$$\delta = Mean(Res(MF_\text{off})) \tag{S5-23}$$

The mean and error of $\Delta\Delta\widetilde{F_{PP}}$ were calculated in the same way for $\Delta\Delta F_{PP}$. In addition, $\Delta\Delta\widetilde{F_{PP}}$ can be collected using the $\Delta\overline{F_{PP}}(T,0)$, although the $\Delta F_{PP}$ measurement is necessary.

$$\Delta\Delta\overline{F_{PP}}(T,B_0) = \left(1 + \Delta\overline{F_{PP}}(T,0)\right)\Delta\Delta\widetilde{F_{PP}}(T,B_0) \tag{S5-24}$$

**Table S5-3** Data acquisition sequence of $\Delta\Delta F_{PP}$ measurement shown in **Fig. 4c, 4d** and **6d-h**

| | | | |
|---|---|---|---|
| | Repeats over Pump-Probe Delay Steps $\Delta T$ | | |
| | M Cycles (M = 8) | | Recovery ($T_{rec}$) |
| Pump-Probe Laser System | Probe On | | |
| Static MF | MF off ($T_{off}$) | MF on ($T_{on}$) | |
| Camera | 200 ms Camera Exposure | | |

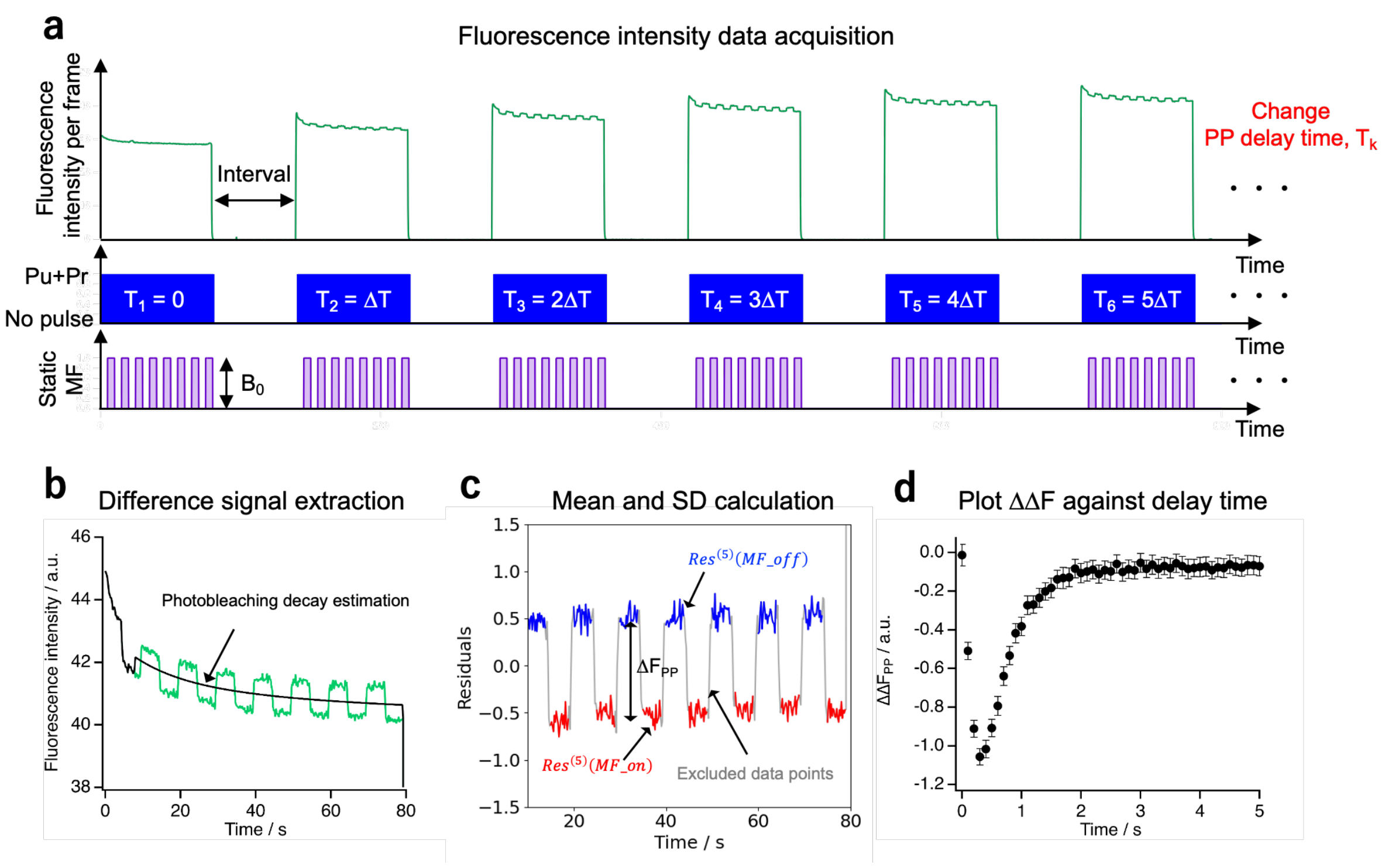

**Figure S5-3 Data acquisition sequence of $\Delta\Delta F_{PP}$ measurement shown in Fig. 4c, 4d and 6d-h.** (a). Fluorescence intensity acquisition timeline. (b) Integrated fluorescence signals of MF-Off/On and Off measurement at a fixed Pump-Probe delay time. Photobleaching decay is modeled with a bi-exponential fit. (c) Residuals corresponding to the $\Delta F_{PP}(T, B_0)$. (d) Plot of $\Delta F_{PP}(T, B_0)$.

## 5.2 PFP kinetics measurement

**Table S5-4** displays the data acquisition sequence for $\Delta F_{PFP}$ measurement. In this sequence, under Pump-Probe excitation with a fixed long Pump-Probe delay time ($T_0$), fluorescence signals under alternating RSMF Off and On conditions (corresponding to $F_{pu+pr}(T_0, 0)$ and $F_{pu+pr}(T_0, B_0; \tau)$, respectively) are acquired M times with $T_{off}/T_{on}$ cycles (typically 5-seconds each). Following these cycles, a recovery period (typically 60 or 80 seconds) is applied. This entire procedure is repeated for each RSMF delay time to generate $\Delta F_{PFP}$ as a function of the RSMF delay time.

**Figure S5-4** outlines the data analysis procedure for the $\Delta F_{PFP}$ measurement. First, a series of integrated fluorescence signals was obtained at each RSMF delay time step by calculating the average intensity from the image data acquired according to the sequence shown in **Table S5-4** (**Figure S5-4 a**). From the data, the $\Delta F_{PFP}$ signals were extracted at each RSMF delay time step by performing residual analysis through curve fitting of the RSMF Off/On cycle measurements (**Figure S5-3 b**), followed by calculation of their mean and error (**Figure S5-3 c**), Finally, these values are plotted against RSMF delay (**Figure S5-3 d**).

The residual analysis to obtain $\Delta F_{PFP}$ was performed by curve fitting using biexponential function (eq.S5-13).To eliminate the variation of fluorescence intensity, the following normalized value of $\Delta F_{PFP}$, $\widetilde{\Delta F_{PFP}}$, was defined and calculated as:

$$\widetilde{\Delta F_{PFP}}(\tau, B_0) = \frac{F_{pu+pr}(T_0, B_0; \tau_{10}) - F_{pu+pr}(T_0, 0)}{F_{pu+pr}(T_0, 0)} \tag{S5-25}$$

$$\widetilde{\Delta F_{PFP}}(\tau, B_0) = \overline{Res}(RSMF_on) - \overline{Res}(RSMF_off) \tag{S5-26}$$

$$Res = \frac{F_{\mathrm{off/on}} - (fitted\ value + \delta)}{fitted\ value + \delta} \tag{S5-27}$$

where the collection of value, $\delta$, was calculated as the overall mean of the residual for the off-cycle measurement.

The mean and error of the $\widetilde{\Delta F_{PFP}}$ signals were calculated from the means and standard deviations of the residuals obtained from each RSMF Off/On cycle, as described below.

The mean and standard deviations of $\widetilde{\Delta F_{PFP}}(\tau, B_0)$ for the kth Off/On cycle were calculated as:

$$Ave\left(\widetilde{\Delta F^{(k)}_{PFP}}(\tau, B_0)\right) = Mean\left(res^{(k)}(on)\right) - Mean(res^{(k)}(off)) \tag{S5-28}$$

$$Error\left(\widetilde{\Delta F^{(k)}_{PFP}}(\tau, B_0)\right) = \sqrt{\left[SD\left(res^{(k)}(on)\right)\right]^2 + \left[SD\left(res^{(k)}(off)\right)\right]^2} \tag{S5-29}$$

Then, the overall mean and error of the $\Delta\Delta F_{PP}(T, B_0)$ were then calculated by averaging over M-1 Off/On cycles (excluding the first cycle, k = 1):

$$Ave\left(\widetilde{\Delta F_{PFP}}(\tau, B_0)\right) = \frac{1}{N_M}\sum_{k=1}^{N_M} Ave\left(\widetilde{\Delta F^{(k)}_{PFP}}(\tau, B_0)\right) \tag{S5-30}$$

$$Error\left(\widetilde{\Delta F_{PFP}}(\tau, B_0)\right) = \frac{1}{\sqrt{N_M}}\sum_{k=1}^{N_M} Error\left(\widetilde{\Delta F^{(k)}_{PFP}}(\tau, B_0)\right) \tag{S5-31}$$

As a supplement, $\Delta\widetilde{F_{PFP}}$ can be collected using the $\Delta\overline{F_{PP}}(T,0)$, as $F_{pu+pr}(T_0,0) = (1+\Delta\overline{F_{PP}})F_{pu}$. But the $\Delta F_{PP}$ measurement is necessary.

$$\Delta\overline{F_{PFP}}(\tau,B_0) = (1+\Delta\overline{F_{PP}})\Delta\widetilde{F_{PFP}}(\tau,B_0) \quad \text{(S5-32)}$$

**Table S5-4** Data acquisition sequence for $\Delta F_{PFP}$ measurement shown in **Fig. 5** and **Fig. 6**.

| | | | |
|---|---|---|---|
| | Repeats over RSMF Delay Steps Δτ | | |
| | M Cycles (M = 8) | | Recovery ($T_{rec}$) |
| Pump-Probe Laser System | Probe On [$T_{fixed}$] | | |
| RSMF | RSMF off ($T_{off}$) | RSMF on ($T_{on}$) | |
| Camera | 200 ms Camera Exposure | | |

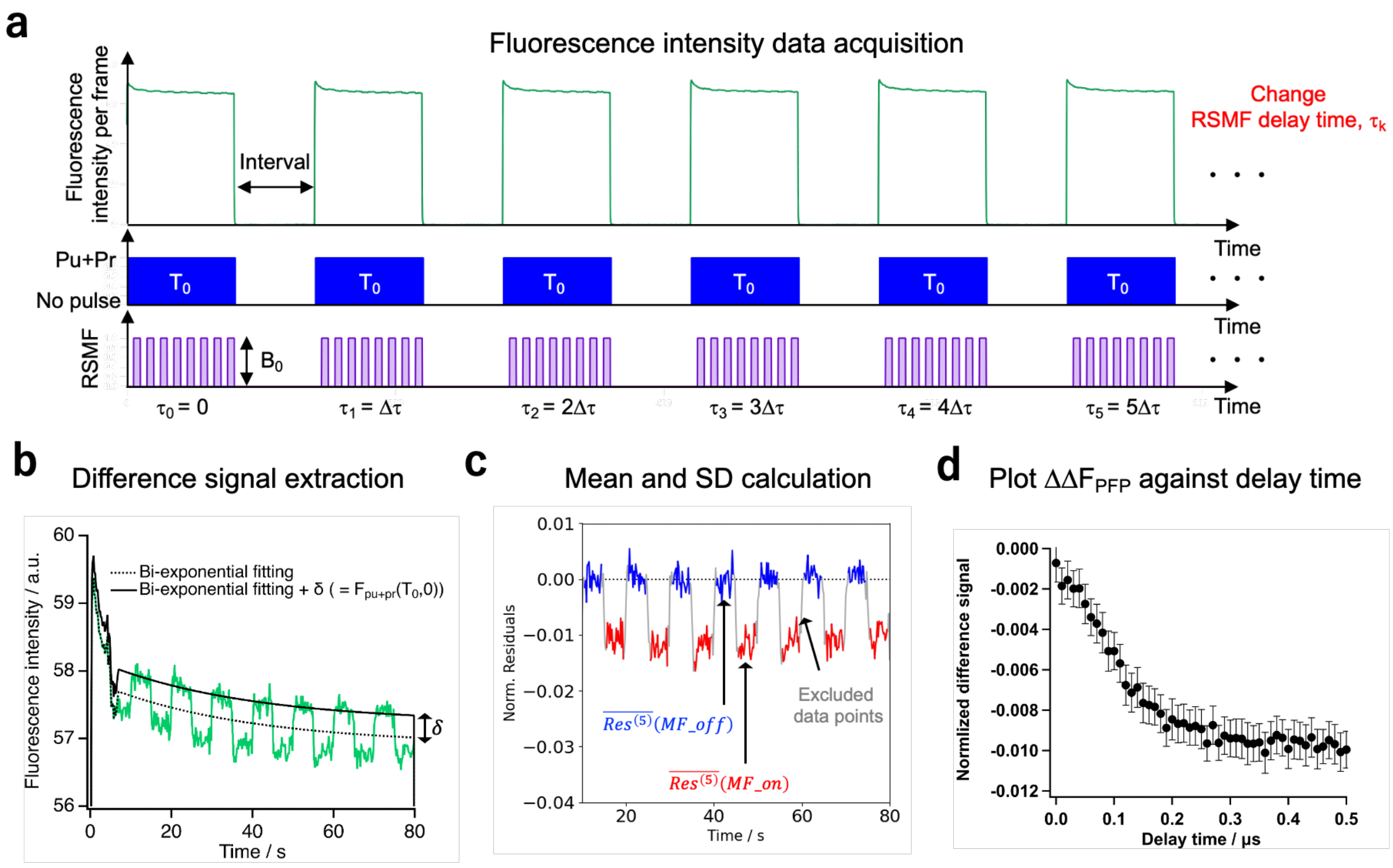

**Figure S5-4 Data acquisition sequence of $\Delta\Delta F_{PFP}$ measurement shown in Fig. 5 and Fig. 6** (a). Fluorescence intensity acquisition timeline. (b) Integrated fluorescence signals of RSMF-Off/On and Off measurement at a fixed RSMF delay time. $F_{pu+pr}(T_0, 0)$ is estimated with a bi-exponential fit. (c) Normalized residuals corresponding to the $\Delta\widetilde{F_{PFP}}(\tau, B_0)$. (d) Plot of $\Delta\widetilde{F_{PFP}}(\tau, B_0)$.

### 5.3 PP-MARY measurement

**Table S5-5** displays the data acquisition sequence for PP-MARY measurement. In this sequence, under fixed Pump-Probe delay time, fluorescence signals under alternating magnetic field Off and On conditions (corresponding to $F_{pu+pr}(T_0, 0)$ and $F_{pu+pr}(T_0, B)$, respectively) are acquired M times with $T_{off}/T_{on}$ cycles. Following these cycles, a recovery period ($T_{rec}$) is applied. This process is repeated for each magnetic field step to obtain a PP-MARY curve at the given Pump-Probe delay time. Finally, the entire sequence is repeated with different fixed Pump-Probe delay times to generate PP-MARY curves as a function of the Pump-Probe delay time.

**Figure S5-5** outlines the data analysis procedure for PP-MARY measurement. First, a series of integrated fluorescence signals is obtained at each magnetic field strength step by calculating the average intensity from the image data acquired according to the sequence shown in **Table S5-5** (**Figure S5-5**). From the data, the $\Delta\Delta F_{PP}(T, B)$ signals are extracted at each magnetic field strength step by performing residuals analysis through curve fitting of the MF Off/On cycle measurements (**Figure S5-5 b**), followed by calculation of their mean and error (**Figure S5-5 c**), Finally, these values are plotted against magnetic field strength to generate the PP-MARY curve at a given Pump-Probe delay time (**Figure S5-5 d**).

The fluorescence intensity corresponding to $F_{pu+pr}(T_0, 0)$ was estimated by curve fitting using biexponential functions (eq. S5-13) for the 0 M, 0.3 mM, 0.5 mM, and 1 mM Trp samples. For the 5 mM Trp samples, where the fluorescence signal is relatively flat and exhibits small, slow fluctuations, a 5th-order polynomial function was used instead:

$$Poly_5(x|x_0, K_0, K_1, K_2, K_3, K_4, K_5) = \sum_{i=0}^{5} K_i (x - x_0)^i \tag{S5-33}$$

Based on this, the normalized residual corresponding to $\Delta\widetilde{F_{PP}}(T, B)$ was determined as (S5-26). The mean and error of the $\Delta\widetilde{F_{PP}}(T, B)$ signal were calculated from the means and standard deviations of the normalized residuals obtained from each Off/On cycle, as described below. The mean and standard deviations of $\Delta\widetilde{F_{PP}}(T, B)$ for the kth MF Off/On cycle were calculated as:

$$Ave\left(\Delta\Delta\widetilde{F_{PP}^{(k)}}(T_{\text{fixed}}, B)\right) = Mean\left(\overline{Res^{(k)}}(\text{MF_on})\right) - Mean\left(\overline{Res^{(k)}}(\text{MF_off})\right) \tag{S5-34}$$

$$Error\left(\Delta\Delta\widetilde{F_{PP}^{(k)}}(T_{\text{fixed}}, B)\right) = \sqrt{\left[SD\left(\overline{Res^{(k)}}(\text{MF_on})\right)\right]^2 + \left[SD\left(\overline{Res^{(k)}}(\text{MF_off})\right)\right]^2} \tag{S5-35}$$

Then, the overall mean and error of the $\Delta\widetilde{F_{PP}}(T, B)$ were then calculated by averaging over M-1 Off/On cycles (excluding the first cycle, k = 1):

$$Ave\left(\Delta\Delta\widetilde{F_{PP}}(T_{\text{fixed}}, B)\right) = \frac{1}{N_M} \sum_{k=1}^{N_M} Ave\left(\Delta\Delta\widetilde{F_{PP}^{(k)}}(T_{\text{fixed}}, B)\right) \tag{S5-36}$$

$$Error\left(\Delta\Delta\widetilde{F_{PP}}(T_{\text{fixed}}, B)\right) = \frac{1}{\sqrt{N_M}} \sum_{k=1}^{N_M} Error\left(\Delta\Delta\widetilde{F_{PP}^{(k)}}(T_{\text{fixed}}, B)\right) \tag{S5-37}$$

**Table S5-5** Data acquisition sequence of PP-MARY measurement shown in **Fig. 6**

| | | | |
|---|---|---|---|
| | Repeats over Static MF Steps $\Delta B_S$ | | |
| | M Cycles (M = 8) | | Recovery ($T_{rec}$) |
| Pump-Probe Laser System | Probe On [$T_{fixed}$] | | |
| Static MF | MF off ($T_{off}$) | MF on ($T_{on}$) | |
| Camera | 200 ms Camera Exposure | | |

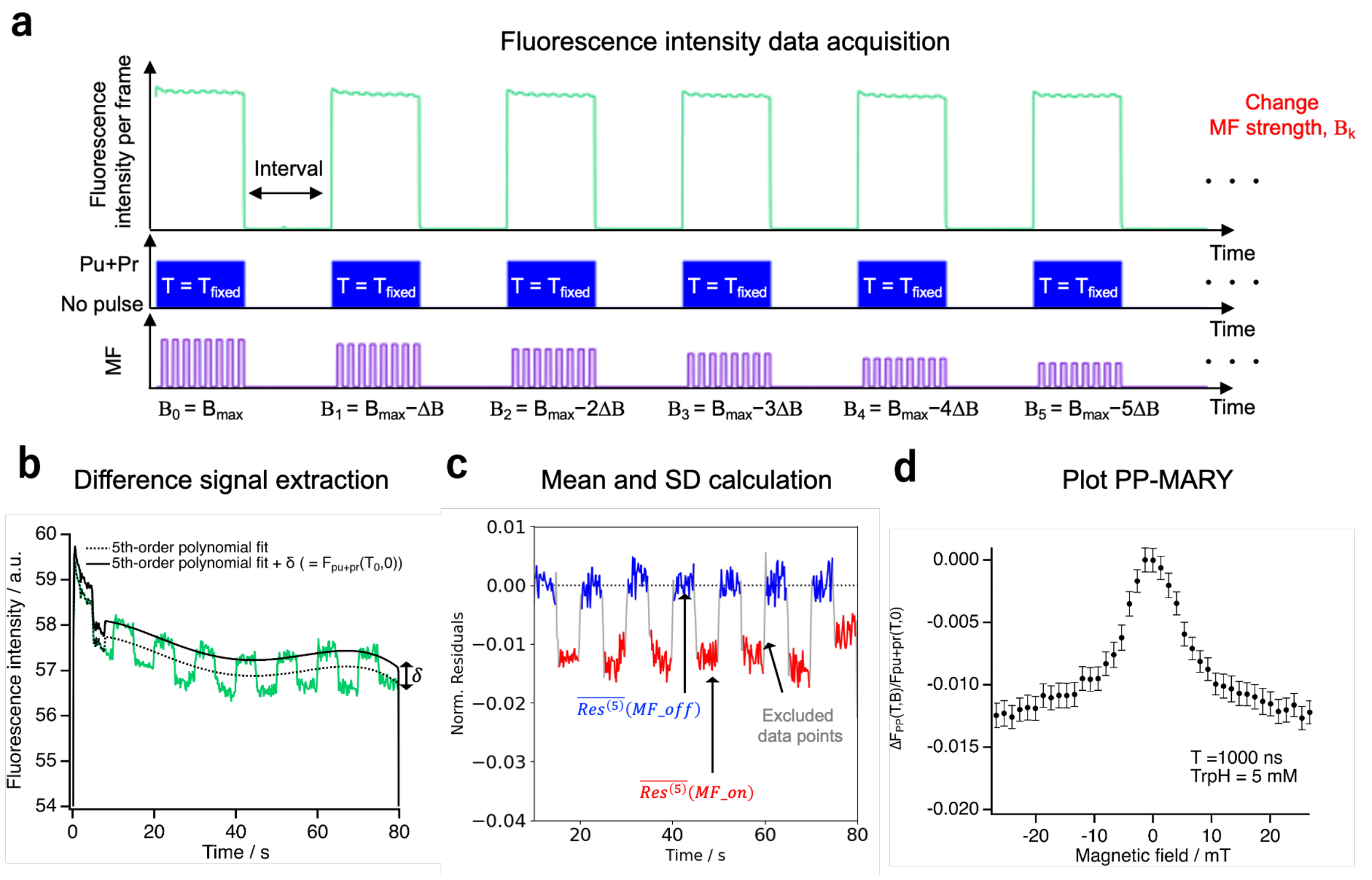

**Figure S5-5** Data analysis procedure of PP-MARY measurements shown in **Fig. 6**. (a). Fluorescence intensity acquisition timeline. (b) Integrated fluorescence signals of MF-Off/On and Off measurement at a fixed pump-probe delay time. $F_{pu+pr}(T_0, 0)$ is estimated with either a bi-exponential fit or 5th-order polynomial fit(the 5th-order polynomial fit is shown here). (c) Normalized residuals corresponding to the $\Delta\widetilde{F_{PP}}(T, B)$. (d) PP-MARY plot.

**5.4 PFP-MARY measurement**

**Table S5-6** displays the data acquisition sequence for PFP-MARY measurements. In this sequence, under fixed Pump-Probe and RSMF delay times, fluorescence signals under variable magnetic field RSMF Off and On conditions (corresponding to $F_{pu+pr}(T_0, 0)$ and $F_{pu+pr}(T_0, B; \tau)$, respectively) are acquired M times with $T_{off}/T_{on}$ cycles. Following these cycles, a recovery period ($T_{rec}$) is applied. This process is repeated for each RSMF field step to obtain a PFP-MARY curve at the given RSMF delay time. Finally, the entire sequence is repeated with different fixed RSMF delay times to generate a PFP-MARY curves as a function of the RSMF delay.

**Figure S5-6** outlines the data analysis procedure for PFP-MARY measurement. First, a series of integrated fluorescence signals is obtained at each magnetic field step by calculating the average intensity from the image data acquired according to the sequence shown in **Table S5-6** (**Figure S5-6 a)**. From the data, the $\Delta F_{PFP}(\tau, B)$ signal is extracted at each magnetic field step by performing residuals analysis through curve fitting of the RSMF Off/On cycle measurements (**Figure S5-6 b**). The extracted signals from each cycle are then used to calculate the mean and error values (**Figure S5-6 c**). Finally, the mean and error values are plotted as a function of magnetic field strength to generate the PFP-MARY curve at a given RSMF delay time (**Figure S5-6 d**).

As described in Section 5.2, the fluorescence intensity corresponding to $F_{pu+pr}(T_0, 0)$ was estimated by curve fitting using biexponential functions (eq. S5-13). Based on this, the normalized residual corresponding to $\widetilde{\Delta F_{PFP}}(\tau, B)$ was determined. The mean and error of the $\widetilde{\Delta F_{PFP}}(\tau, B)$ signal were calculated from the means and standard deviations of the normalized residuals obtained from each Off/On cycle, as described below. The mean and standard deviations of $\widetilde{\Delta F_{PFP}}(\tau, B)$ for the kth RSMD Off cycle were calculated as:

$$Mean\left(\widetilde{\Delta F^{(k)}_{PFP}}(\tau_{\text{fixed}}, B)\right) = Mean\left(\overline{Res^{(k)}}(\text{MF_on})\right) - Mean\left(\overline{Res^{(k)}}(\text{MF_off})\right) \quad \text{(S5-38)}$$

| | |
|---|---|
| $SD\left(\widetilde{\Delta F^{(k)}_{PFP}}(\tau_{\text{fixed}}, B)\right) = \sqrt{\left[SD\left(\overline{Res^{(k)}}(\text{MF_on})\right)\right]^2 + \left[SD\left(\overline{Res^{(k)}}(\text{MF_off})\right)\right]^2}$ | (S5-39) |

Then, the overall mean and error of the $\widetilde{\Delta F_{PFP}}(\tau, B)$ were then calculated by averaging over M-1 Off/On cycles (excluding the first cycle, k = 1):

$$Ave\left(\widetilde{\Delta F_{PFP}}(\tau_{\text{fixed}}, B)\right) = \frac{1}{M-1}\sum_{k=2}^{M} Mean\left(\widetilde{\Delta F^{(k)}_{PFP}}(\tau_{\text{fixed}}, B)\right) \quad \text{(S5-40)}$$

$$Error\left(\widetilde{\Delta F_{PFP}}(\tau_{\text{fixed}}, B)\right) = \frac{1}{\sqrt{M-1}}\sum_{k=2}^{M} SD\left(\widetilde{\Delta F^{(k)}_{PFP}}(\tau_{\text{fixed}}, B)\right) \quad \text{(S5-41)}$$

**Table S5-6** Data acquisition sequence of PFP-MARY measurement shown in **Fig. 6**.

| | | | |
|---|---|---|---|
| | Repeats over RSMF Field Steps $\Delta B_{RSMF}$ | | |
| | M Cycles (M = 8) | | Recovery ($T_{rec}$) |
| Pump-Probe Laser System | Probe On [$T_{fixed}$] | | |
| RSMF | RSMF off ($T_{off}$) | RSMF on [$\tau_{fixed}$] ($T_{on}$) | |
| Camera | 200 ms Camera Exposure | | |

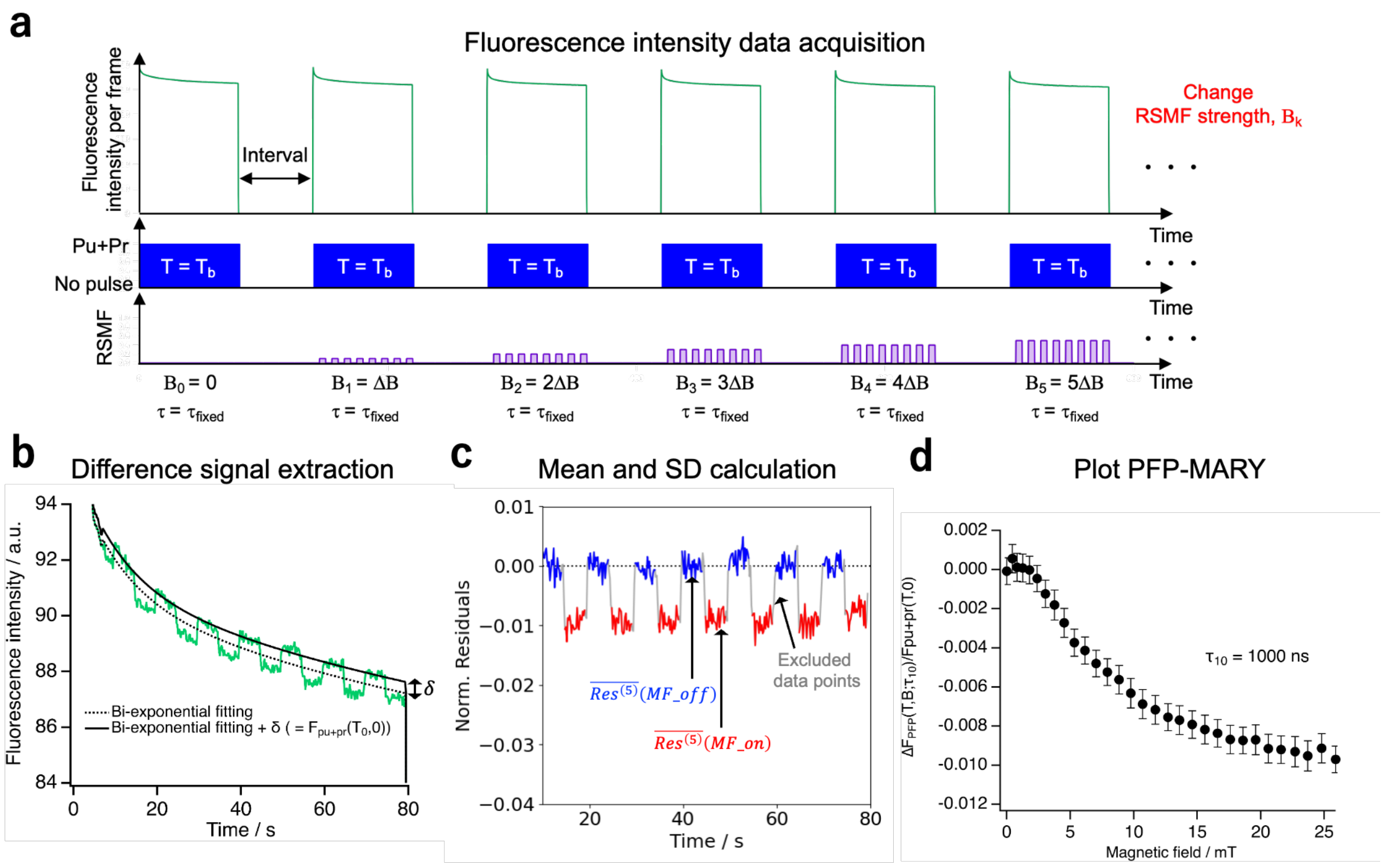

**Figure S5-6** Data analysis procedure of PFP-MARY measurements shown in **Fig. 6**. (a). Fluorescence intensity acquisition timeline. (b) Integrated fluorescence signals of RSMF-Off/On and Off measurement at a fixed pump-probe delay time. $F_{pu+pr}(T_0, 0)$ is estimated with a bi-exponential fit. (c) Normalized residuals corresponding to the $\Delta\widetilde{F_{PFP}}(\tau, B)$. (d) PFP-MARY plot.

## 5.5 Experimental parameters

**Table S5-7** Experimental parameters common to all measurements

| Excitation energy[a] | Irradiation spot diameter | Excitation intensity[b] | Camera exposure time (Δt) |
|---|---|---|---|
| 960 mW | 8.1 μm | 1.78 MW/cm² | 200 ms |

a: The value was calculated as the manufacturer-specified laser power (1400 mW) multiplied by the ratio of the power output at the laser source to that after the objective lens measured using a power meter (LP-1, Sanwa).

b: The value was calculated from the excitation energy and the diameter of the irradiation spot.

**Table S5-8** Experimental parameters for $\Delta F_{PP}$ measurement

| Figure | Pulse width (w) | Delay time step (ΔT) | Laser repetition rate ($f_{rep}$) | Probe On-Off time ($T_{on}$ and $T_{off}$) | On-Off captures (M) | Recovery period ($T_{rec}$) |
|---|---|---|---|---|---|---|
| 3b | 38.8 ns | 1 μs | 30, 50, and 100 Hz | 5 s each | 8 times | 60 s |
| 3c | 38.8 ns | 30 μs | Shown in the figure | 5s each | 8 times | None |
| 4b | 38.8 ns | 100 ns | 50 and 100 Hz | 5s each | 8 times | 60 s |

**Table S5-9** Experimental parameters for $\Delta\Delta F_{PP}$ measurement

| Figure | Pulse width (w) | Delay time step (ΔT) | Laser repetition rate ($f_{rep}$) | MF On-Off time ($T_{on}$ and $T_{off}$) | On-Off captures (M) | Recovery period ($T_{rec}$) |
|---|---|---|---|---|---|---|
| 4c,4d | 38.8 ns | 100 ns | 50 and 100 Hz | 5 s each | 8 times | 60 s |
| 6d-h | 30.6 ns | 100 ns | 100 Hz | 5 s each | 8 times | 80 s |

**Table S5-9** Experimental parameters for $\Delta F_{PFP}$ measurement

| Figure | Pulse width (w) | Delay time step (Δτ) | Laser repetition rate ($f_{rep}$) | RSMF On-Off time ($T_{on}$ and $T_{off}$) | On-Off captures (M) | Recovery period ($T_{rec}$) |
|---|---|---|---|---|---|---|
| 5b,5c | 30.6 ns | 100 ns | 100 Hz | 5 s each | 8 times | 60 s |
| 6e-h | 30.6 ns | 100 ns | 100 Hz | 5 s each | 8 times | 80 s |

**Table S5-9** Experimental parameters for PP-MARY measurement

| Figure | Pulse width (w) | MF step (ΔB) | Laser repetition rate ($f_{rep}$) | MF On-Off time ($T_{on}$ and $T_{off}$) | On-Off captures (M) | Recovery period ($T_{rec}$) |
|---|---|---|---|---|---|---|
| 6i | 30.6 ns | 100 ns | 100 Hz | 5 s each | 8 times | 80 s |

**Table S5-9** Experimental parameters for PFP-MARY measurement

| Figure | Pulse width (w) | MF step (ΔB) | Laser repetition rate ($f_{rep}$) | RSMF On-Off time ($T_{on}$ and $T_{off}$) | On-Off captures (M) | Recovery period ($T_{rec}$) |
|---|---|---|---|---|---|---|
| 6k | 30.6 ns | See the Fig S4-2 | 100 Hz | 5 s each | 8 times | 80 s |

## 6. Supporting information for FMN system

### 6.1 Photobleaching effects on pump-probe fluorescence signal

Many molecules, when photoexcited, undergo largely cyclic processes and primarily regenerate the original ground state chromophore. This is an important fact for molecules used as, for example, fluorescent probes. Indeed even when photoexcitation leads to the formation of RPs, it is possible for the reaction to be cyclic, with a magnetic field only influencing the rate at which the ground state chromophore is regenerated. In practice, however, photocycles are almost never 100% efficient and some fraction of the photoexcited molecules undergo photoreactions to form other chemical species, and do not regenerate the original ground state chromophore. This process is typically referred to as photobleaching. In general, this process depends on the intensity and duration of the irradiation. Here, we describe how the photobleaching affects the observations of the present techniques.

Note that, for simplicity, we assume here that the photobleaching rate per pulse is constant. In reality, when the pulse excitation is repeated, the concentration of photobleached molecules within the observation area increases, and the concentration gradient with respect to fresh molecules outside the observation area changes, thus causing the photobleaching rate to vary between pulses.

(i) Single Pulse Excitation.

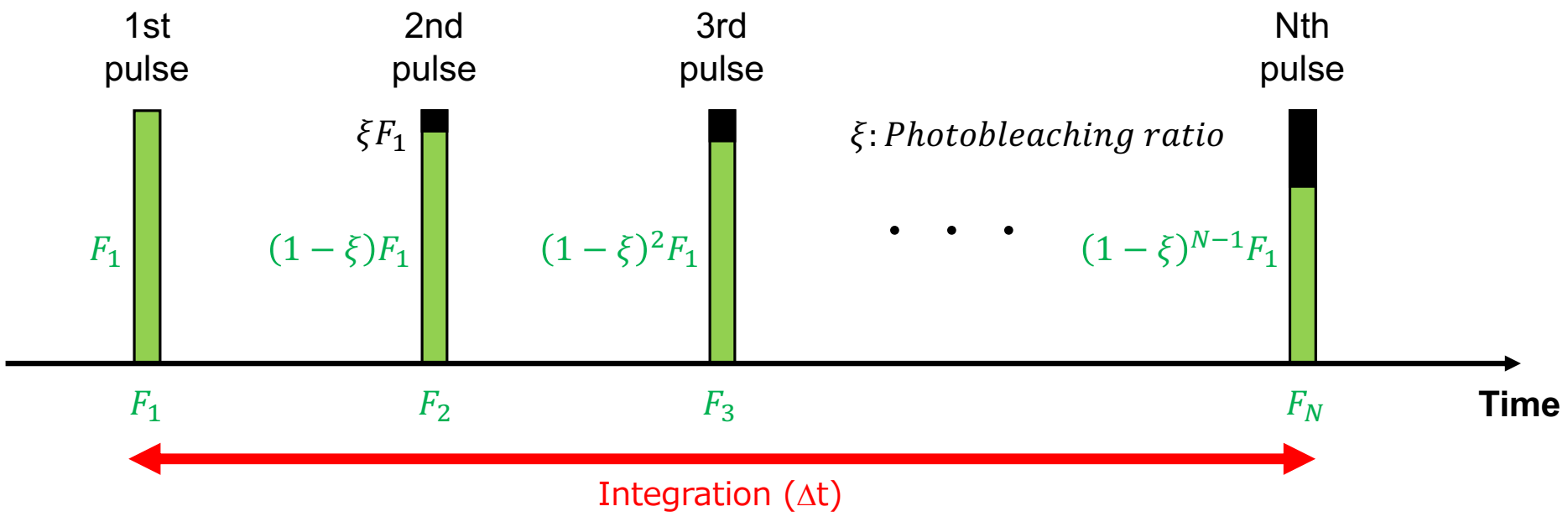

**Figure S6-1 The definition of the photobleaching rate on single pulse excitation.** Green and black colors represent fluorescent state and photobleaching population, respectively.

First, we consider a scenario where the fluorescence signal is detected using an integration time of $\Delta t$ from an excitation with the same intensity and pulse width at a repetition rate $f$ (see **Figure S6-1**).

We assume that the repetition rate of the excitation is shorter than the time required for the photobleached molecules to diffuse out of the excitation region. The rate of photobleaching, $\xi$, is assumed to be constant per pulse. Under these conditions, the fluorescence signal generated by the kth pulse excitation is expressed as follows:

$$F_k = \gamma^{k-1} F_1 \qquad \text{(eq. 6-1)}$$

Where $\gamma = 1 - \xi$. The integrated fluorescence signal with $N\left(= f_{rep}\Delta t\right)$ pulse excitations can be expressed as:

$$I_N = \sum_{k=1}^{N} F_k = \sum_{k=1}^{N} \gamma^{k-1} F_1 = \frac{1-\gamma^N}{1-\gamma} F_1 \qquad (\gamma \neq 1) \qquad \text{(eq. 6-2)}$$

(ii) Pump-Probe Excitation

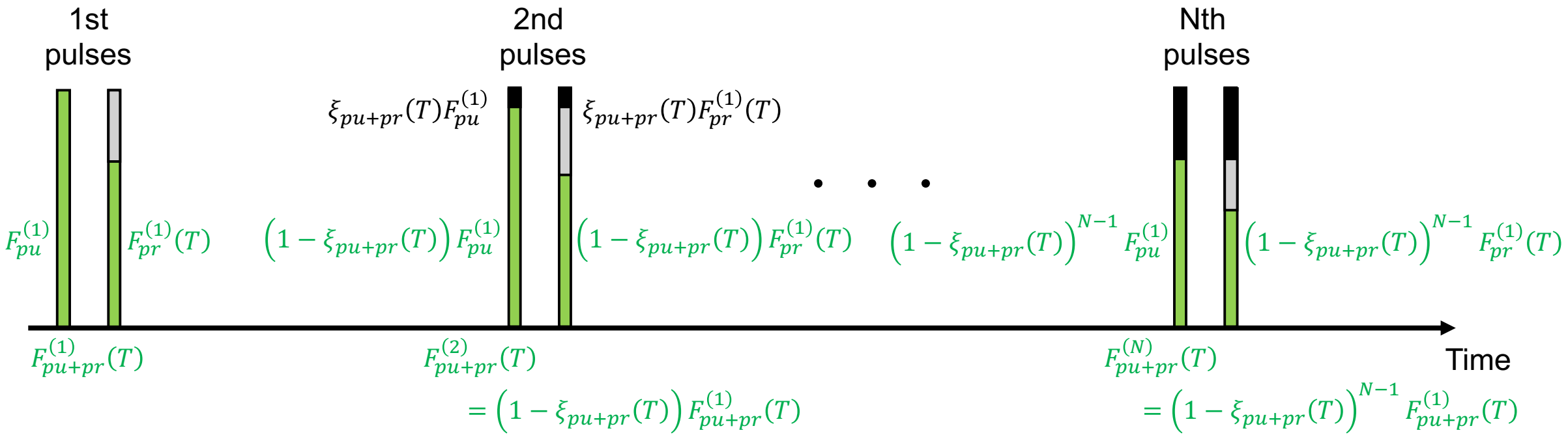

**Figure S6-2 The definition of the photobleaching rate on pump-probe excitation.** Green, gray, black colors represent fluorescent state, dark state and photobleaching population, respectively.

In a similar manner to the case of the single excitation, we consider the photobleaching rate per unit pulse, $\xi$, when the fluorescence signal is detected from pump-probe excitation with a repetition frequency $f$ and an integration time $\Delta t$ (see **Fig S6-2**). The fluorescence signal generated by the kth pump-probe pulse excitation is written as follows:

$$F_{pu+pr}^{(k)}(T) = F_{pu}^{(k)} + F_{pr}^{(k)}(T) \qquad \text{(eq. 6-3)}$$

In a similar manner to **eq.2**, by introducing the photobleaching rate per pump-probe excitation at a pump probe delay time T, $\xi_{pu+pr}(T)$, the integrated fluorescence signal by pump-probe and only pump excitation can be expressed as follows:

$$I_{pu+pr}(T) = \sum_{k=1}^{N} F_{pu+pr}^{(k)}(T) = \sum_{k=1}^{N} \gamma_{pu+pr}^{k-1}(T) F_{pu+pr}^{(1)}(T) = \frac{1-\gamma_{pu+pr}^{N}(T)}{1-\gamma_{pu+pr}(T)} F_{pu+pr}^{(1)}(T) \qquad \text{(eq. 6-4)}$$

$$I_{pu} = \sum_{k=1}^{N} F_{pu}^{(k)} = \sum_{k=1}^{N} \gamma_{pu}^{k-1} F_{pu}^{(1)} = \frac{1-\gamma_{pu}^{N}}{1-\gamma_{pu}} F_{pu}^{(1)} \qquad \text{(eq. 6-5)}$$

By taking the above expressions, the differences in fluorescence signals using pump-probe detection can be derived.

$$\Delta I_{PP}(T) = I_{pu+pr}(T) - I_{pu}$$
$$= \left(\frac{1-\gamma_{pu+pr}^{N}(T)}{1-\gamma_{pu+pr}(T)} - \frac{1-\gamma_{pu}^{N}}{1-\gamma_{pu}}\right) F_{pu}^{(1)} + \frac{1-\gamma_{pu+pr}^{N}(T)}{1-\gamma_{pu+pr}(T)} F_{pr}^{(1)}(T) \qquad \text{(eq. 6-6)}$$

Here, if the pump-probe delay time is long enough (longer than the time it takes for the cyclic photoreaction to fully finish), the fluorescence signal from pump-probe excitation is equivalent to that from single pulse excitation at twice the repetition rate. Therefore,

$$I_{pu+pr}(T_{long}) = \frac{1-\gamma_{pu+pr}^{N}(T_{long})}{1-\gamma_{pu+pr}(T_{long})} F_{pu+pr}^{(1)}(T_{long}) \approx \frac{1-\gamma_{pu}(2f)^{2N}}{1-\gamma_{pu}(2f)^{2}} F_{pu+pr}^{(1)}(T_{long}) \qquad \text{(eq. 6-7)}$$

The rate of photobleaching depends on the repetition rate, so we introduce $\gamma_{pu}(2f) = \alpha\gamma_{pu}(f)$. The normalized differences in fluorescence signals can be derived as follows:

$$\Delta\overline{I_{PP}}(T_{long}) = \frac{\Delta I_{PP}(T_{long})}{I_{pu}}$$

$$\approx \frac{\left(\frac{1-\alpha^{2N}\gamma^{2N}}{1-\alpha^2\gamma^2} - \frac{1-\gamma^N}{1-\gamma}\right)F_{pu}^{(1)} + \frac{1-\alpha^{2N}\gamma^{2N}}{1-\alpha^2\gamma^2}F_{pr}^{(1)}(T_{long})}{\frac{1-\gamma^N}{1-\gamma}F_{pu}^{(1)}} \quad \text{(eq. 6-8)}$$

$$\approx \left(\frac{1-\alpha^{2N}\gamma^{2N}}{1-\alpha^2\gamma^2}\frac{1-\gamma}{1-\gamma^N} - 1\right) + \frac{1-\alpha^{2N}\gamma^{2N}}{1-\alpha^2\gamma^2}\frac{1-\gamma}{1-\gamma^N}\frac{F_{pr}^{(1)}(T_{long})}{F_{pu}^{(1)}}$$

Where $\gamma_{pu}(f) = \gamma$. By taking $F_{pr}^{(1)}(T_{long}) = \gamma_{pu}(2f)F_{pu}^{(1)}$, the normalized differential signal can be expressed as:

$$\Delta\overline{I_{PP}}(T_{long}) \approx \frac{1-\alpha^{2N}\gamma^{2N}}{1-\gamma^N}\frac{1-\gamma}{1-\alpha\gamma} - 1 \quad \text{(eq. 6-9)}$$

At very low repetition rates, when the photo recovery following pump pulse photobleaching varies negligibly, $\gamma_{pu}(2f) \approx \gamma_{pu}(f)$, i.e. $\alpha \approx 1$. This gives:

$$\Delta\overline{I_{PP}}(T_{long}) \approx \gamma^N = \gamma^{f_{rep}\Delta t} \quad \text{(eq. 6-10)}$$

This behavior qualitatively corresponds to the repetition rate dependence of $\Delta F_{PP}$ measurements shown in **Fig.3b**. Note that this approximation does not hold when the repetition rate is high.

### 6.2 Corrections to photobleaching quantum yield estimation

In practice, it is difficult to adjust the excitation intensities of the pump and probe pulses so as to fully satisfy (eq.S1–21). Therefore, a correction formula must be applied to obtain accurate quantification. If the excitation intensities of the pump and probe pulses are not perfectly matched, $\Delta F_{PP}$ can be expressed as follows.

$$\Delta\overline{F_{PP}}(T) = \frac{\Delta F_{PP}(T)}{F_{pu}} = \frac{\alpha_{pr}\left(c_0 - [DS](T)\right)}{\alpha_{pu}c_0} = \frac{\alpha_{pr}}{\alpha_{pu}}\left(1 - \overline{[DS]}(T)\right) \quad \text{(S6-11)}$$

Therefore, in practice, the photobleaching quantum yield shown in eq.15 is corrected as follows:

$$\phi_B = 1 - \frac{\alpha_{pu}}{\alpha_{pr}}\Delta\overline{F_{PP}}\left(T_{long}\right) \quad \text{(S6-12)}$$

The proportionality constants $\alpha_{pu}$ and $\alpha_{pr}$ can be estimated from the integrated fluorescence recorded under pump-only and probe-only excitation, respectively. Following careful pump and probe pulse intensity configuration, as shown in **Fig. S6-3,** we measured the pump-only and probe-only fluorescence intensities and, after averaging over 50.0–79.8 s, obtained the following values:

$$\frac{\alpha_{pu}}{\alpha_{pr}} = 1.00693589 \quad \text{(S6-13)}$$

Using the above value, we obtained **Fig.3c**.

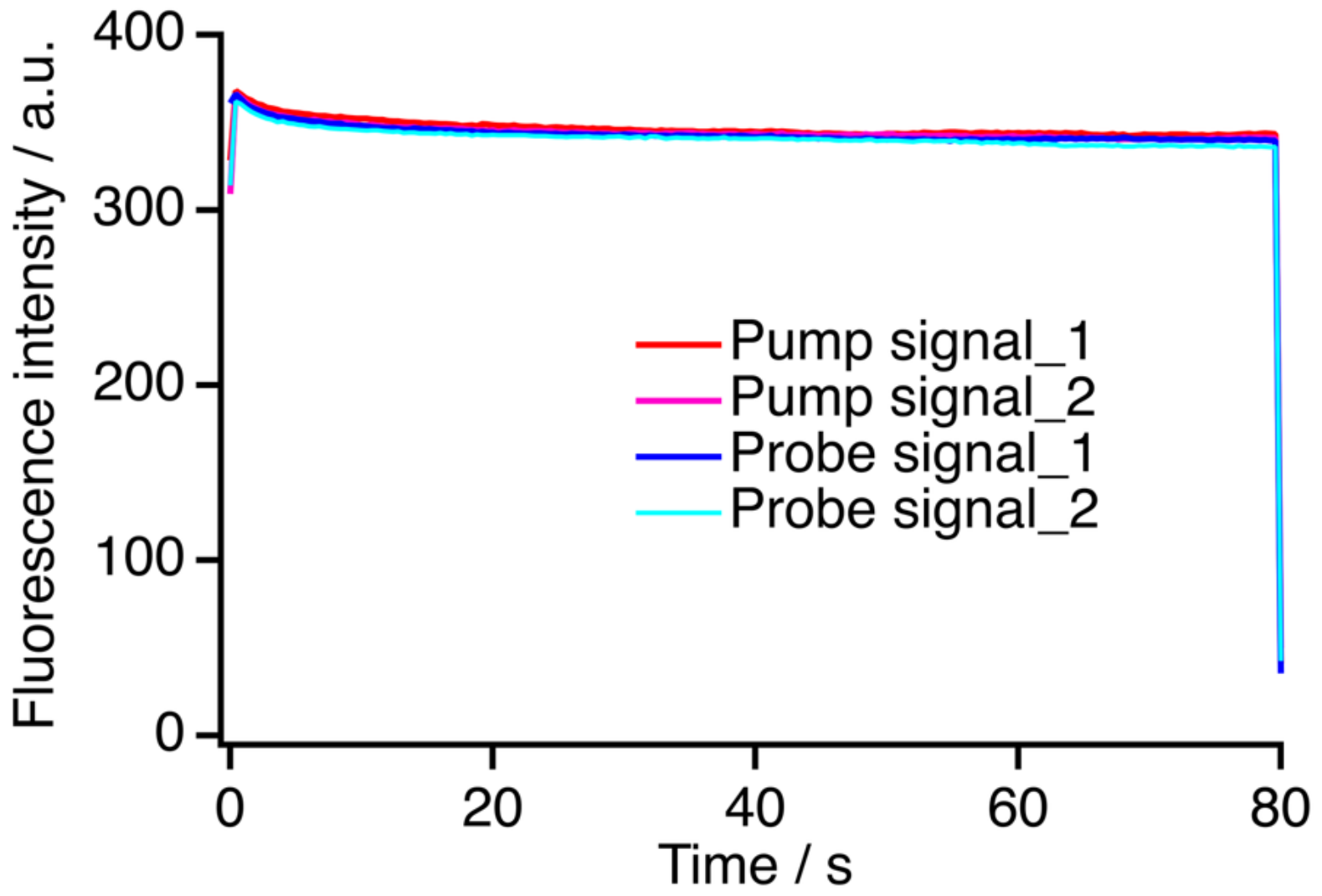

**Figure S6-3. Pump and probe excitation intensity calibration.** Fluorescence signals obtained from the 10 μM FMN sample under the same measurement conditions as **Fig. 3c**, using pump-only and probe-only excitation with 100 Hz of laser repetition.

## 7. Supporting information for FAD system

### 7.1 Repetition rate dependence

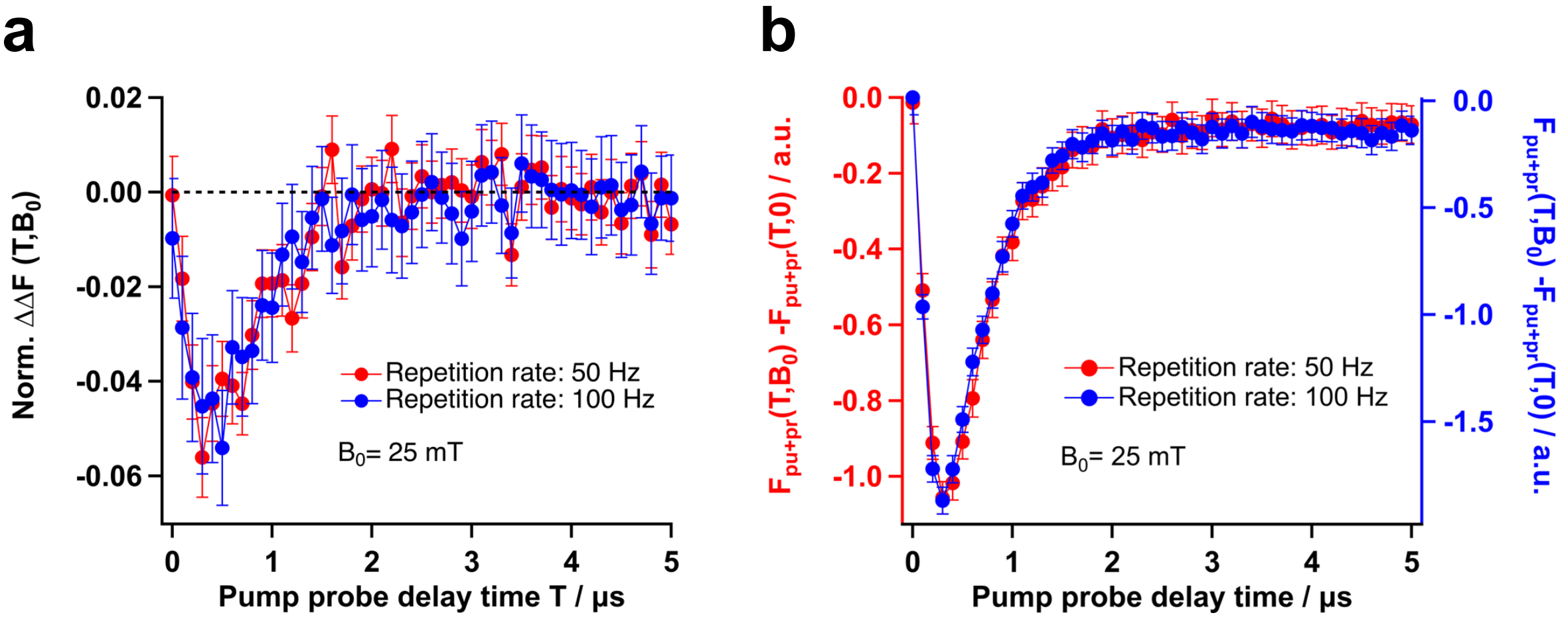

**Figure S7-1. Repetition rate dependence of $\Delta\Delta F_{PP}$ measurements.** (a)$\Delta\Delta\overline{F_{PP}}(T,B_0) = \Delta\overline{F_{PP}}(T,B_0) - \Delta\overline{F_{PP}}(T,0)$, at 50 Hz and 100 Hz laser repetition rates. $B_0$ = 25 mT. (b)$\Delta\Delta F_{PP}(T,B_0) = F_{pu+pr}(T,B_0) - F_{pu+pr}(T,0)$, at 50 Hz and 100 Hz laser repetition rates. $B_0$ = 25 mT.

## 7.2. Sample thickness effects of FAD photochemistry

Comparing the ΔΔF signals in **Fig. 4b and 4c** with ΔΔA signals from conventional cuvette TA [6] and our TA-based microscope (TOAD) [4] revealed differing time dependences. After careful pH calibration and matching solute concentrations, the discrepancy persisted. Remarkably, matching the sample thickness—the sole remaining difference—eliminated the discrepancy (**Fig. S7-2**). We attribute this to interfacial pH shifts near borosilicate glass: buffer pH increases by ~2 units within a few micrometres of the glass surface[7], consistent with our geometry where the maximum distance from the surface is ~2.5 μm. Thus, in thin films the minute volume places essentially all FAD molecules within the interfacial zone, fully explaining the altered kinetics. Indeed, using the same thickness (~3 μm) in TOAD reproduced the PP-fluorescence time dependence in ΔΔA (Fig. 5e). Fitting the 5 μm sample's kinetics with the Murakami pH-dependent model estimated an average local pH of ~2.8 (vs. bulk pH 2.3). We conclude that the observed time-dependence differences arise from sample thickness (interfacial pH effects), not from detection modality.

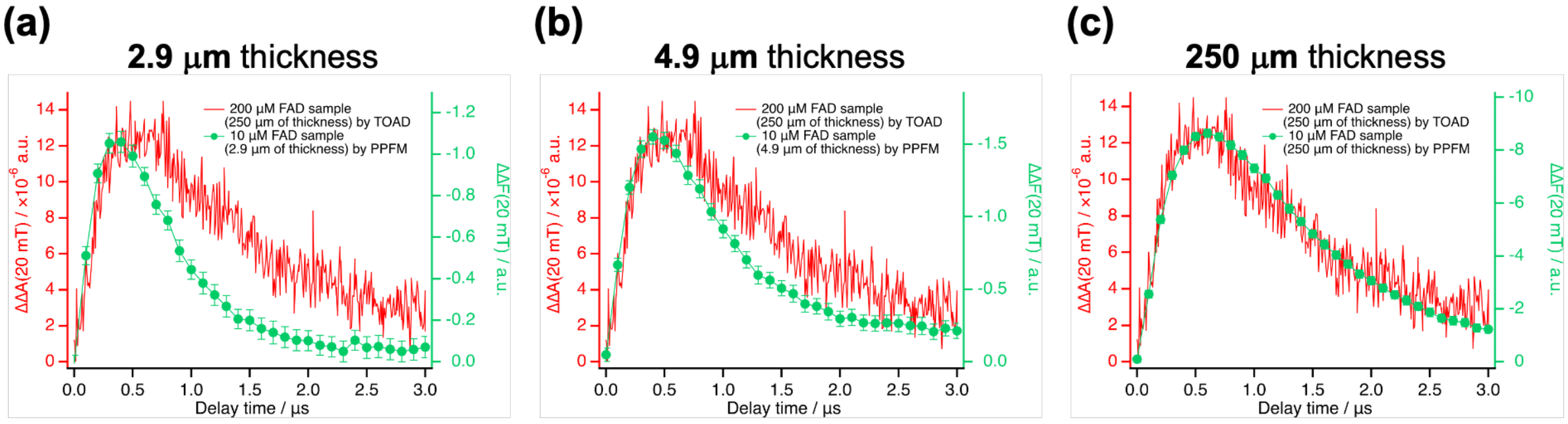

**Figure S7-2. Time-dependent MFE signals ($\Delta\Delta F_{PP}$) of 10 μM FAD in pH 2.3 buffer measured at different sample thicknesses under an external magnetic field of 20 mT.** (a) 2.9 μm, (b) 4.95 μm, and (c) 250 μm. For reference, transient absorption data (ΔΔA) of 200 μM FAD in pH 2.3 buffer at a sample thickness of 250 μm measured by TOAD are also shown in Ref.5.

## 8. Supporting information for FMN/Trp system

### 8.1 ΔΔ$F_{PP}$ at pH 2.3 and pH6.4

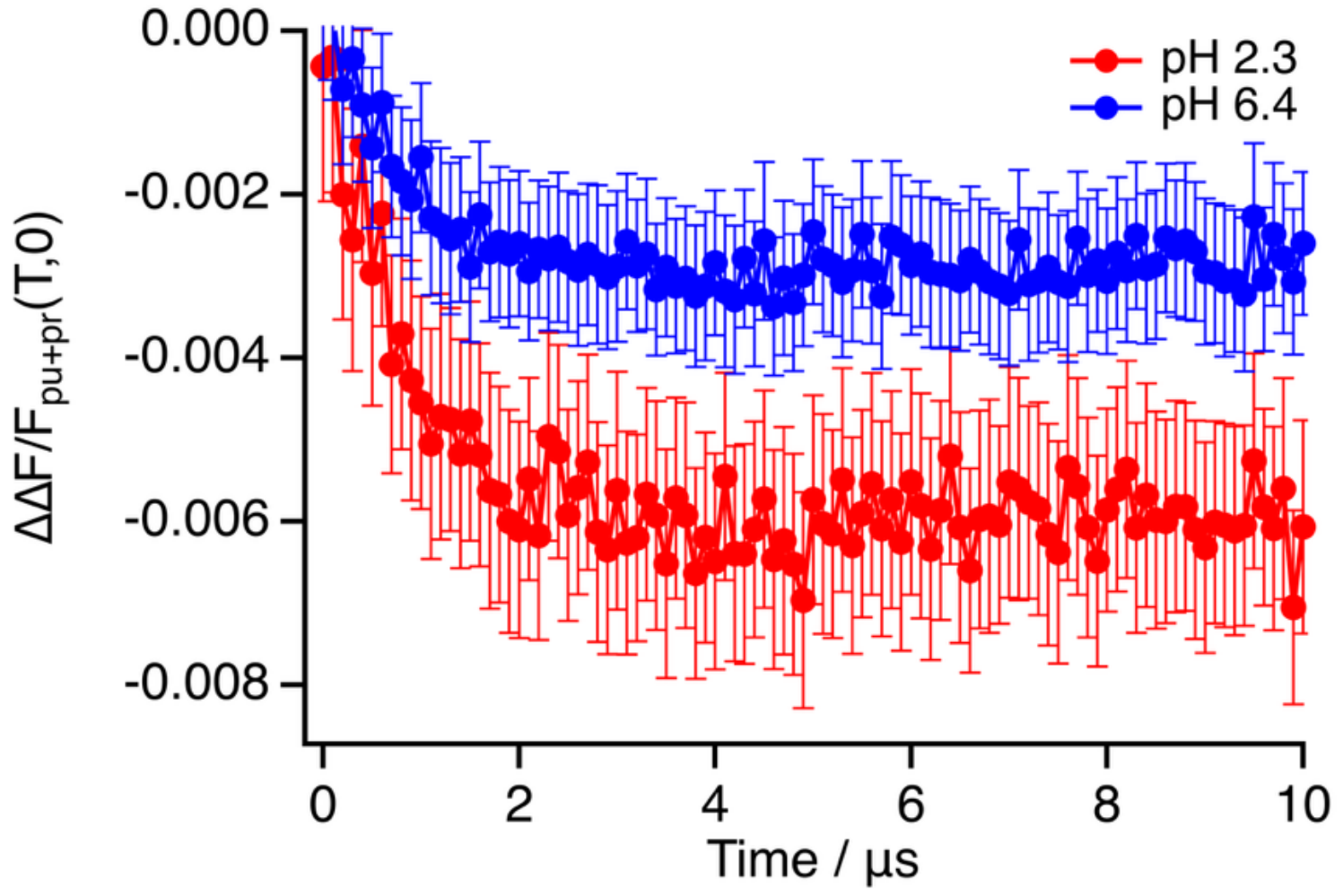

**Figure S8-1. ΔΔ$F_{PP}$ of 10 µM FMN + 1.0 mM tryptophan in buffer at pH 2.3 and pH 6.4.** $B_0$ = 25 mT. $f_{rep}$= 100 Hz. Sample thickness = 5.0 µm

## 8.2 Biexponential fitting of the 10 mM Trp Off-On data

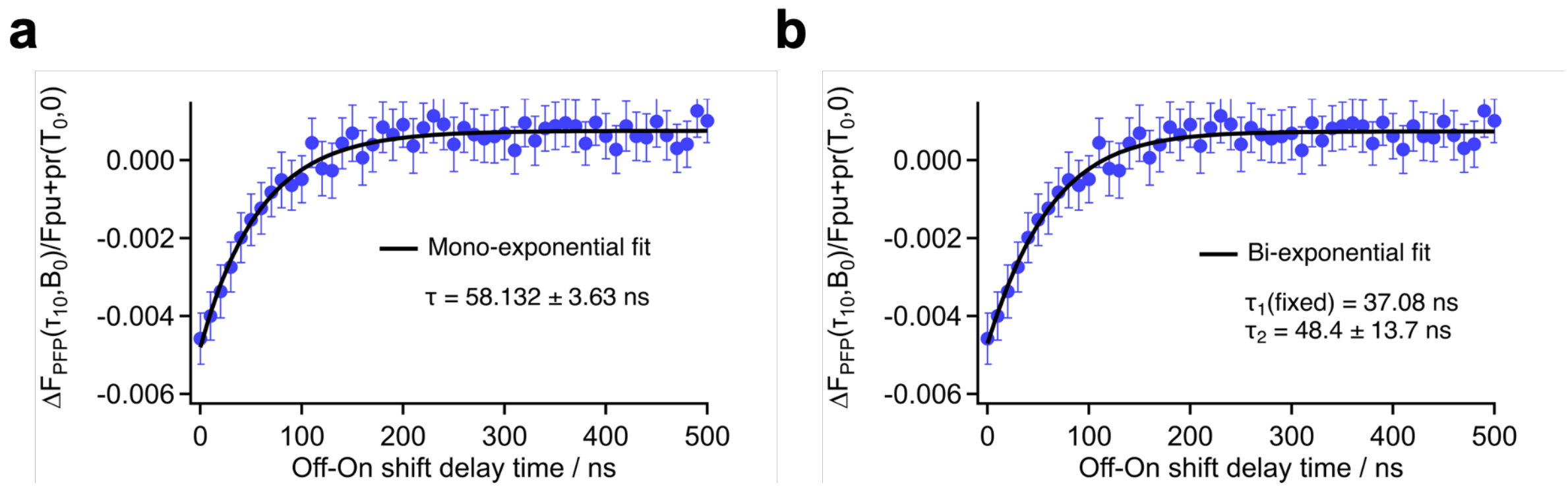

**Figure S8-2. Off-On RSMF shift measurement for 10 µM FMN with10 mM tryptophan (Delay time step = 10 ns)**. (a) Mono-exponential fit. $f(x|x_0, y_0, A, \tau) = y_0 + A\mathbf{e}^{-\frac{x-x_0}{\tau}}$ (b)Bi-exponential fit. $f(x|x_0, y_0, A_1, \tau_1, A_2, \tau_2) = y_0 + A_1\mathbf{e}^{-\frac{x-x_0}{\tau_1}} + A_2\mathbf{e}^{-\frac{x-x_0}{\tau_2}}$

## 9. Supporting information for FAD/Trp system

### 9.1 Spin dynamics simulation

Simultaneous simulations of both the PP and PFP reaction kinetics and the associated MFE and its dependence on magnetic field strength (MARY) were performed using a modified version of the RadicalPy Python based simulation library [8]. RadicalPy was selected for its open source, object oriented code and convenient implementation of matrix generation for complex kinetic schemes. Furthermore, we prioritized using an existing framework over developing a custom program to facilitate reproducibility by other groups, even though substantial modifications were ultimately required. Reasons for the modifications were twofold:

1) The original MARY code as described in ref.[8] and as released on GitHub [https://github.com/Spin-Chemistry-Labs/radicalpy] as version 0.81 (still the current version as of writing) contains a number of minor and more major errors. These are presented below. These errors lead to a number of problems with the simulations presented in ref.[8]. In particular for the MARY curves presented in Figure 4 of ref.[8] are erroneous for two main reasons:

    1) Despite these being transient experiments, the main simulation code contains a non-negligible value for kex (1 x $10^4$ $s^{-1}$) which means that the FAD ground state is continuously excited throughout the observation period of the reaction.
    2) More significantly, however, are two significant errors in the underlying library code, specifically the functions that perform the kinetic quantum simulation (kine_quantum_mary in experiments.py) and the function that performs the semiclassical simulation (SemiclassicalSimulation(LiouvilleSimulation) in simulation.py). The details of these errors and corrected code are provided below. The result is that the simulated MARY curves have $B_{1/2}$ values which are incorrect and in particular are too large as a result of the error in the kinetic quantum simulation code which causes the MFE value to have a component which increases linearly with each field step. The details of the two errors are as follows:

        a) Error in the kinetic quantum simulation code. The problem here lies in an incorrect positioning of a variable assignment in the main loop which results in accumulation of data throughout the simulation rather than a necessary reset for each new field value. The original and corrected versions are presented below.

        b) Errors in the semiclassical simulation code based on the Schulten-Wolynes approach[9]. The algorithm used to calculate the semiclassical hyperfine vectors (based on [9]) is problematic and so the code was completely rewritten (see below). There are two major issues:

i) instead of calculating a random hyperfine vector length for each electron spin, the code takes all the hyperfine couplings for both radicals and creates a single random hyperfine vector length which is applied to only one radical. The other radical is treated as having no hyperfine coupling, which can lead to the generation of non-negligible low field effects.
ii) For each radical, the random vector length should point along a random direction in 3D space. In the original code, the random vector length is assigned to the x,y and z components of the hyperfine vector, which generates a longer vector which always points in the same direction.

2) Additional functions were created for the following purposes:
   1) In the reaction of FAD with Trp, there are two RPs generated (intramolecular and intermolecular). Therefore a new experiment type was created in experiments.py to handle this. Strictly these two RPs have different hyperfine couplings in one of the radicals (adenine vs tryptophan) but in practice, the contribution to the MFE from the intermolecular RP is small and so the same hyperfine couplings were used for simplicity and code efficiency (this was tested against code that uses the correct hyperfine couplings for each radical and no appreciable differences were observed).
   2) The new experiment type was also rewritten to allow the calculation of both PP (possible using the original time simulation code) and PFP (new code needed) measurement schemes. This was combined with the code for the inclusion of two RPs described above into a new function called kine_quantum_mary2rp_onoff. The code is included below.
   3) It was discovered that even when the full set of hyperfine couplings for FAD and adenine radicals are included, a non-negligible LFE was observed in both the PFP MARY experiments and simulations for field-off switching at early times. This is an interesting phenomenon and something we intend to explore further in a future publication. The code determines the $B_{1/2}$ values for the simulated MFE curves by fitting with a two parameter (zero offset) Lorentzian function. Where the LFEs were non-negligible, this led to time-dependent distortion of the fitted $B_{1/2}$ value (**Fig. S8**). Therefore a new four parameter double Lorentzian function (DblLorentzian) and a new curve fitting function (Bhalf_withLFE_fit) were added to utils.py to allow inclusion of the LFE in the fit. In order to ensure reliable fits, the value of Lhalf (the LFE equivalent of Bhalf) was assigned a fixed value of 2mT based on manual fitting of the simulated MARY spectra and a three-parameter fit was performed. This produced consistent and reliable fitting of the simulated PFP MARY curves. On the same basis, a three-parameter fit was performed for the experimental PFP MARY data (see details below).

Double Lorentzian function:

$$\mathcal{L}_2(B|L_{1/2}, LFE_{sat}, B_{1/2}, MFE_{sat}) = LFE_{sat}\frac{B^2}{B^2 + L_{1/2}^2} - MFE_{sat}\frac{B^2}{B^2 + B_{1/2}^2} \quad \text{(S9-1)}$$

Finally, the main simulation code was written to simulate the reaction scheme presented in **Fig.6** for the various tryptophan concentrations. The full code is provided below and if the necessary functions are added to the base RadicalPy code, then this single simulation will produce all the simulation data included in the main text. In the case of FAD alone, it is clear from the data, that there is a small amount of quenching due to dissolved oxygen. This is accounted for in the simulations by including a small quenching concentration which provides the best fit. It should be noted that in Ref. 9 the semiclassical simulation was employed with 400 samples (individual Monte-Carlo trajectories). Repeat simulations shows that this number is insufficient and substantial changes in the simulated MARY spectra are produced from run to run. Therefore, for the included simulation data, the number of samples was increased to 20000 which led to consistent simulations with negligible differences between runs.

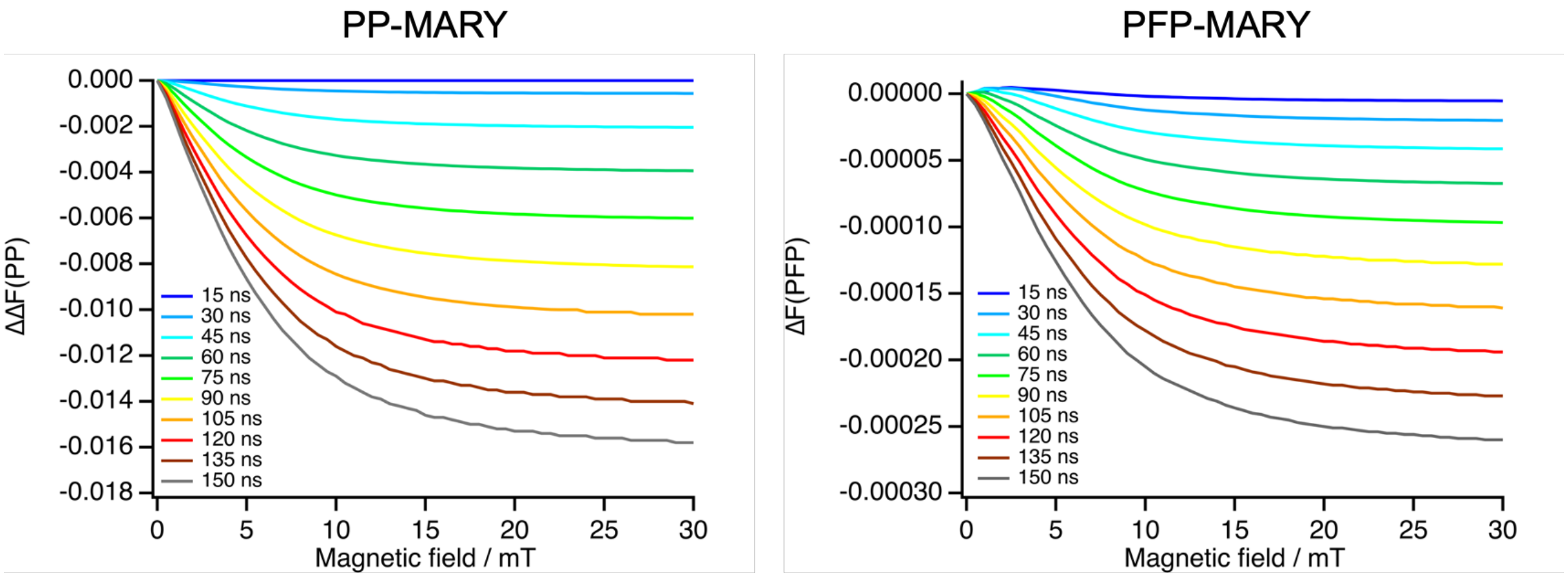

**Figure S9-1. Simulated PP/PFP-MARY curves at early delay time.** Non-negligible LFEs are observed in the PFP-MARY curves.

The rate coefficients used for the simulations in **Fig.6** were as follows:

| | |
|---|---|
| $k_{ex} = 0$ | (ground state excitation rate) zero as pulsed excitation |
| $k_{fl} = 3.55 \times 10^{8}\ s^{-1}$ | (fluorescence rate) Ref.10 |
| $k_{ic} = 1.28 \times 10^{9}\ s^{-1}$ | (internal conversion rate) Ref.8 |
| $k_{isc} = 3.64 \times 10^{8}\ s^{-1}$ | (intersystem crossing rate) Ref.8 |
| $k_{d} = 1 \times 10^{7}\ s^{-1}$ | (protonated triplet to ground state) - fitting parameter |
| $k_{1} = 7 \times 10^{6}\ s^{-1}$ | (protonated triplet to RP) Ref.8 |
| $k_{-1} = 2.7 \times 10^{9} s^{-1}$ | (RP to protonated triplet) Ref.8 |
| $k_{rT} = 0$ | (triplet state relaxation rate) - no measurable effect |
| $k_{bet} = 6.5 \times 10^{6}\ s^{-1}$ | (singlet recombination rate) fitting parameter |
| pH = 2.8 | (pH of the solution) estimated based on kinetic fitting using Ref.6 |
| $k_{esc} = 1.5 \times 10^{8}\ s^{-1}$ | (intermolecular RP escape rate) fitting parameter |
| $k_{STD} = 1.6 \times 10^{7}\ s^{-1}$ | (ST dephasing rate) RP relaxation fitting parameter |
| $k_{qt} = 0$ | (triplet quenching rate) fitting parameter |
| $k_{qr} = 3.4 \times 10^{8}\ mol^{-1}\ dm^{3}\ s^{-1}$ | (RP quenching rate) fitting parameter |
| $k_{p} = 0$ | (free radical recombination rate) - no significant recombination on this observation timescale |

Modified RadicalPy code:

1) kine_quantum_mary function

Original code:

```
def kine_quantum_mary(
    sim: SemiclassicalSimulation,
    num_samples: int,
    init_state: ArrayLike,
    radical_pair: list,
    ts: NDArray[float],
    Bs: ArrayLike,
    D: float,
    J: float,
    kinetics: ArrayLike,
    relaxations: list[ArrayLike],
):
    dt = ts[1] - ts[0]
    total_yield = np.zeros((len(ts), len(init_state), len(Bs)),
dtype=complex)
    kinetic_matrix = np.zeros((len(kinetics), len(kinetics)),
dtype=complex)
    loop_rho = np.zeros((len(ts), len(init_state)), dtype=complex)
    loop_yield = np.zeros((len(ts), len(init_state)), dtype=complex)
    HHs = sim.semiclassical_HHs(num_samples)
    HJ = sim.exchange_hamiltonian(J)
    HD = sim.dipolar_hamiltonian(D)

    for i, B0 in enumerate(tqdm(Bs)):
        Hz = sim.zeeman_hamiltonian(B0)
        for HH in HHs:
            Ht = Hz + HH + HJ + HD
            L = sim.convert(Ht)
            sim.apply_liouville_hamiltonian_modifiers(L, relaxations)
            kinetic_matrix[
                radical_pair[0] : radical_pair[1], radical_pair[0] :
radical_pair[1]
            ] = L
            kinetic = kinetics + kinetic_matrix
            rho0 = init_state
```

```
        propagator = sp.sparse.linalg.expm(kinetic * dt)

        for k in range(0, len(ts)):
            loop_rho[k, :] = rho0
            rho0 = propagator @ rho0

        loop_yield = loop_yield + loop_rho
    total_yield[:, :, i] = loop_yield / num_samples

return {"ts": ts, "Bs": Bs, "yield": total_yield}
```

Corrected code:

```
def kine_quantum_mary_fix(
    sim: SemiclassicalSimulationJ,
    num_samples: int,
    init_state: ArrayLike,
    radical_pair: list,
    ts: NDArray[float],
    Bs: ArrayLike,
    D: float,
    J: float,
    kinetics: ArrayLike,
    relaxations: list[ArrayLike],
):
    dt = ts[1] - ts[0]
    total_yield = np.zeros((len(ts), len(init_state), len(Bs)),
dtype=complex)
    kinetic_matrix = np.zeros((len(kinetics), len(kinetics)),
dtype=complex)
    loop_rho = np.zeros((len(ts), len(init_state)), dtype=complex)
    HHs = sim.semiclassical_HHs(num_samples)
    HJ = sim.exchange_hamiltonian(J)
    HD = sim.dipolar_hamiltonian(D)

    for i, B0 in enumerate(tqdm(Bs)):
        Hz = sim.zeeman_hamiltonian(B0)
        loop_yield = np.zeros((len(ts), len(init_state)), dtype=complex)
        for HH in HHs:
            Ht = Hz + HH + HJ + HD
            L = sim.convert(Ht)
            sim.apply_liouville_hamiltonian_modifiers(L, relaxations)
            kinetic_matrix[
                radical_pair[0] : radical_pair[1], radical_pair[0] :
radical_pair[1]
            ] = L
            kinetic = kinetics + kinetic_matrix
            rho0 = init_state
            propagator = sp.sparse.linalg.expm(kinetic * dt)
```

```
        for k in range(0, len(ts)):
            loop_rho[k, :] = rho0
            rho0 = propagator @ rho0

        loop_yield = loop_yield + loop_rho
    total_yield[:, :, i] = loop_yield / num_samples

return {"ts": ts, "Bs": Bs, "yield": total_yield}
```

2) SemiclassicalSimulation(LiouvilleSimulation) class:

Original:

```
class SemiclassicalSimulation(LiouvilleSimulation):
    def semiclassical_HHs(
        self,
        num_samples: int,
    ) -> np.ndarray:
        assert len(self.radicals) == 2
        assert self.radicals[0].multiplicity == 2
        assert self.radicals[1].multiplicity == 2

        spinops = np.array([self.spin_operator(0, ax) for ax in "xyz"])
        cov = np.diag([m.semiclassical_std for m in self.molecules])
        samples = np.random.multivariate_normal(
            mean=[0, 0],
            cov=cov,
            size=(num_samples, 3),
        )
        result = np.einsum("nam,axy->nxy", samples, spinops) * 2
        return result * self.radicals[0].gamma_mT

    @property
    def nuclei(self):
        return []
```

Rewritten:

```
class SemiclassicalSimulation_fix(LiouvilleSimulation):
    def semiclassical_HHs(self, num_samples: int,) -> np.ndarray:

        spinops = [[self.spin_operator(ri, ax) for ax in "xyz"] for ri in
range(len(self.radicals))]
        result = np.zeros((num_samples,4,4),dtype=complex)

        for i in range(num_samples):
            for ri, m in enumerate(self.molecules):
                tau = np.sqrt(2)/m.semiclassical_std
                #maxfI=((tau**2)/4*np.pi)**(3/2) * np.exp(-
(1/4)*(1/2)*(1/2)*tau**2)



                #Randomly sample the length of the composite nuclear spin

                I = (tau**2/4*np.pi)*np.random.normal(0, m.semiclassical_std,
size=1)

                #Randomly sample the direction of the composite nuclear spin
                theta = np.arccos(1-2*np.random.rand())
                phi = 2*np.random.rand()*np.pi
                gamma = m.radical.gamma_mT

                result[i,:,:] +=
gamma*I*np.sin(theta)*np.cos(phi)*spinops[ri][0] +
gamma*I*np.sin(theta)*np.sin(phi)*spinops[ri][1] +
gamma*I*np.cos(theta)*spinops[ri][2]

        return result

    @property
    def nuclei(self):
        return []
```

New code for these simulations.

Library functions:

kine_quantum_mary2rp_onoff function

```
def kine_quantum_mary2rp_onoff(
    sim: SemiclassicalSimulation_fix,
    num_samples: int,
    init_state: ArrayLike,
    radical_pair: list,
    radical_pair2: list,
    ts: NDArray[float],
    switchpoint: int,
    pfpdecimate:int,
    Bs: ArrayLike,
    D: float,
    J: float,
    kinetics: ArrayLike,
    relaxations: list[ArrayLike],
):
    dt = ts[1] - ts[0]
    dtpts=int(1+(len(ts)-1)/pfpdecimate)
    total_yield = np.zeros((len(ts), len(init_state), dtpts , len(Bs)),
dtype=complex)
    kinetic_matrix = np.zeros((len(kinetics), len(kinetics)),
dtype=complex)
    kinetic_matrix0 = np.zeros((len(kinetics), len(kinetics)),
dtype=complex)
    loop_rho = np.zeros((len(ts), len(init_state)), dtype=complex)
    HHs = sim.semiclassical_HHs(num_samples)
    HJ = sim.exchange_hamiltonian(J)
    HD = sim.dipolar_hamiltonian(D)

    for i, B0 in enumerate(tqdm(Bs)):
        Hz = sim.zeeman_hamiltonian(B0)
        loop_yield = np.zeros((len(ts), len(init_state), dtpts),
dtype=complex)
        for HH in HHs:
            H0 = HH + HJ + HD
            Ht = Hz + H0
```

```
            L0 = sim.convert(H0)
            L = sim.convert(Ht)
            sim.apply_liouville_hamiltonian_modifiers(L0, relaxations)
            sim.apply_liouville_hamiltonian_modifiers(L, relaxations)
            kinetic_matrix0[
                radical_pair[0] : radical_pair[1], radical_pair[0] :
radical_pair[1]
            ] = L0
            kinetic_matrix0[
                radical_pair2[0] : radical_pair2[1], radical_pair2[0] :
radical_pair2[1]
            ] = L0
            kinetic_matrix[
                radical_pair[0] : radical_pair[1], radical_pair[0] :
radical_pair[1]
            ] = L
            kinetic_matrix[
                radical_pair2[0] : radical_pair2[1], radical_pair2[0] :
radical_pair2[1]
            ] = L
            kinetic0 = kinetics + kinetic_matrix0
            kinetic = kinetics + kinetic_matrix

            propagator0 = sp.sparse.linalg.expm(kinetic0 * dt)
            propagator = sp.sparse.linalg.expm(kinetic * dt)

            for j in range(0, dtpts):
                rho0 = init_state
                for k in range(0, len(ts)):
                    loop_rho[k, :] = rho0
                    if k>=j*pfpdecimate:
                        rho0 = propagator0 @ rho0
                    else:
                        rho0 = propagator @ rho0
                loop_yield[:,:,j] = loop_yield[:,:,j] + loop_rho

        total_yield[:, :, :, i] = loop_yield / num_samples
```

```
    return {"ts": ts, "Bs": Bs, "yield": total_yield}
```

Bhalf_withLFE_fit_fixLhalf function

```
def Bhalf_withLFE_fit_fixLhalf(
    B: np.ndarray, MARY: np.ndarray
) -> Tuple[float, np.ndarray, float, float]:
    """B_1/2 fit for MARY spectra.

    popt_MARY, pcov_MARY = curve_fit(
        DblLorentzianfixLhalf,
        B,
        MARY,
        p0=[MARY[-1], int(B[-1]/ 5),MARY[-1]/5],
        maxfev=10000000,
    )
# p0 initial values based on typical simulation
    fit_error = np.sqrt(np.diag(pcov_MARY))

    A_opt_MARY, Bhalf_opt_MARY,ALFE_opt_MARY = popt_MARY
    fit_result = DblLorentzianfixLhalf(B, *popt_MARY)
    Bhalf = np.abs(Bhalf_opt_MARY)

    y_pred_MARY = DblLorentzianfixLhalf(B, *popt_MARY)
    R2 = r2_score(MARY, y_pred_MARY)

    return Bhalf, fit_result, fit_error, R2
```

DblLorentzianfixLhalf function

```
def DblLorentzianfixLhalf(B: np.ndarray, MFEamplitude: float, Bhalf: float,
LFEamplitude: float) -> np.ndarray:
    """Double Lorentzian function for MARY spectra with substantial LFE.
Reference:

    Args:
            B (np.ndarray): The x-axis magnetic field values.
            MFEamplitude (float): The amplitude of the saturation field
value.
            Bhalf (float): The magnetic field strength at half the
                saturation field value.
            LFEamplitude (float): The amplitude of the LFE component
            BhalfLFE (float): The magnetic field strength at half the
            LFE saturation field value - this values is fixed for these
simulation

    Returns:
            np.ndarray: Double Lorentzian function for MARY spectrum.

    """
    Lhalf = 2
    return -
MFEamplitude*(B**2/(B**2+Bhalf**2))+LFEamplitude*(B**2/(B**2+Lhalf**2))
```

Main simulation code

```
import matplotlib.pyplot as plt
import numpy as np
import multiprocessing as mp
import os
from pebble import concurrent

from radicalpy.classical import Rate, RateEquations
from radicalpy.experiments import kine_quantum_mary2rp_onoff
from radicalpy.relaxation import SingletTripletDephasing
from radicalpy.simulation import Molecule, SemiclassicalSimulationJ
from radicalpy.utils import Bhalf_withLFE_fit_fixLhalf, is_fast_run

@concurrent.process
def JRWsemiclassicalPFP(Bmax=30, bpts=31, tmax=3e-6, tpts=1001,pfpdeci=10,
qconc=0):

    # Parameters
    time=np.linspace(0,tmax,num=tpts)
    dtime=np.linspace(0,tmax,num=int(1+(tpts-1)/pfpdeci))
    Bs = np.linspace(0, Bmax, num=bpts)
    num_samples = 20000
    kstd =1.6e7  # ST dephasing relaxation rate
    relaxation = SingletTripletDephasing(kstd)

    # Kinetic simulation of FAD at pH 2.8.
    # FAD kinetic parameters
    kex = Rate(0, "k_{ex}")  # groundstate excitation rate
    kfl = Rate(3.55e8, "k_{fl}")  # fluorescence rate
    kic = Rate(1.28e9, "k_{IC}")  # internal conversion rate
    kisc = Rate(3.64e8, "k_{ISC}")  # intersystem crossing rate
    kd = Rate(1e7, "k_d")  # protonated triplet to ground state
    k1 = Rate(7e6, "k_1")  # protonated triplet to RP
    km1 = Rate(2.7e9, "k_{-1}")  # RP to protonated triplet
    krt = Rate(0, "k^R_T")  # triplet state relaxation rate
    kbet = Rate(6.5e6, "k_{BET}")  # singlet recombination rate
    pH = 2.8  # pH of the solution
    Hp = Rate(10**-pH, "H^+")  # concentration of hydrogen ions
```

```
kesc = Rate(1.5e8, "k_{ESC}") #Rate of escape from intermolecular RP

# Quenching kinetic parameters
kqt = Rate(0 , "k_qt")   # triplet quenching rate
kqr = Rate(34e8, "k_qr") # RP quenching rate
kp = Rate(0, "k_p")  # 3.3e3  # free radical recombination(8e4*qconc/5e-
3
Q = Rate(qconc , "Q")  # 1e-3  # quencher concentration

# Rate equations
S0, S1, T1p, T10, T1m = "S0", "S1", "T1+", "T10", "T1-"
SS, STp, ST0, STm = "SS", "ST+", "ST0", "ST-"
TpS, TpTp, TpT0, TpTm = "T+S", "T+T+", "T+T0", "T+T-"
T0S, T0Tp, T0T0, T0Tm = "T0S", "T0T+", "T0T0", "T0T-"
TmS, TmTp, TmT0, TmTm = "T-S", "T-T+", "T-T0", "T-T-"
FRSS, FRSTp, FRST0, FRSTm = "FRSS", "FRST+", "FRST0", "FRST-"
FRTpS, FRTpTp, FRTpT0, FRTpTm = "FRT+S", "FRT+T+", "FRT+T0", "FRT+T-"
FRT0S, FRT0Tp, FRT0T0, FRT0Tm = "FRT0S", "FRT0T+", "FRT0T0", "FRT0T-"
FRTmS, FRTmTp, FRTmT0, FRTmTm = "FRT-S", "FRT-T+", "FRT-T0", "FRT-T-"
FR = "FR"

base = {}
base[S0] = {
    S0: -kex,
    S1: kfl + kic,
    T1p: kd,
    T10: kd,
    T1m: kd,
    SS: kbet,
    FRSS: kbet,
    FR: kp,
}
base[S1] = {
    S0: kex,
    S1: -(kfl + kic + 3 * kisc),
}
base[T1p] = {
    S1: kisc,
```

```
    T1p: -(kd + k1 + krt + kqt * Q),
    T10: krt,
    TpTp: km1 * Hp,
}
base[T10] = {
    S1: kisc,
    T1p: krt,
    T10: -(kd + k1 + 2 * krt + kqt * Q),
    T1m: krt,
    T0T0: km1 * Hp,
}
base[T1m] = {
    S1: kisc,
    T10: krt,
    T1m: -(kd + k1 + krt + kqt * Q),
    TmTm: km1 * Hp,
}
base[SS] = {
    SS: -(kbet + kqr * Q),
}
base[STp] = {
    STp: -(kbet + km1 * Hp + 2* kqr * Q) / 2,
}
base[ST0] = {
    ST0: -(kbet + km1 * Hp + 2* kqr * Q) / 2,
}
base[STm] = {
    STm: -(kbet + km1 * Hp + 2* kqr * Q) / 2,
}
base[TpS] = {
    TpS: -(kbet + km1 * Hp + 2* kqr * Q) / 2,
}
base[TpTp] = {
    T1p: k1,
    TpTp: -(km1 * Hp + kqr * Q),
}
base[TpT0] = {
    TpT0: -(km1 * Hp  + kqr * Q),
```

```
}
base[TpTm] = {
    TpTm: -(km1 * Hp + kqr * Q),
}
base[T0S] = {
    T0S: -(kbet + km1 * Hp  + 2* kqr * Q) / 2,
}
base[T0Tp] = {
    T0Tp: -(km1 * Hp  + kqr * Q),
}
base[T0T0] = {
    T10: k1,
    T0T0: -(km1 * Hp +kqr * Q),
}
base[T0Tm] = {
    T0Tm: -(km1 * Hp + kqr * Q),
}
base[TmS] = {
    TmS: -(kbet + km1 * Hp  + 2* kqr * Q) / 2,
}
base[TmTp] = {
    TmTp: -(km1 * Hp + kqr * Q),
}
base[TmT0] = {
    TmT0: -(km1 * Hp + kqr * Q),
}
base[TmTm] = {
    T1m: k1,
    TmTm: -(km1 * Hp + kqr * Q),
}

base[FRSS] = {
    FRSS: -(kbet+kesc),
    SS: kqr * Q,
}
base[FRSTp] = {
    FRSTp: -(kbet + 2*kesc)/2,
}
```

```
base[FRST0] = {
    FRST0: -(kbet + 2*kesc)/2,
}
base[FRSTm] = {
    FRSTm: -(kbet + 2*kesc)/2,
}
base[FRTpS] = {
    FRTpS: -(kbet + 2*kesc)/2,
}
base[FRTpTp] = {
    T1p: kqt * Q,
    TpTp: kqr * Q,
    FRTpTp: -(kesc),
}
base[FRTpT0] = {
    FRTpT0: -(kesc),
}
base[FRTpTm] = {
    FRTpTm: -(kesc),
}
base[FRT0S] = {
    FRT0S: -(kbet + 2*kesc)/2,
}
base[FRT0Tp] = {
    FRT0Tp: -(kesc),
}
base[FRT0T0] = {
    T10: kqt * Q,
    T0T0: kqr * Q,
    FRT0T0: -(kesc),
}
base[FRT0Tm] = {
    FRT0Tm: -(kesc),
}
base[FRTmS] = {
    FRTmS: -(kbet + 2*kesc)/2,
}
base[FRTmTp] = {
```

```
        FRTmTp: -(kesc),
    }
    base[FRTmT0] = {
        FRTmT0: -(kesc),
    }
    base[FRTmTm] = {
        T1m: kqt * Q,
        TmTm: kqr * Q,
        FRTmTm: -(kesc),
    }
    base[FR] = {
        FRSS: kesc,
        FRTpTp: kesc,
        FRT0T0: kesc,
        FRTmTm: kesc,
        FR: -kp,
    }

    rate_eq = RateEquations(base)
    mat = rate_eq.matrix.todense()
    rho0 = np.array([0, 0, 1/3, 1/3, 1/3, 0, 0, 0, 0, 0, 0, 0, 0, 0, 0, 0,
0, 0, 0, 0, 0, 0, 0, 0, 0, 0, 0, 0, 0, 0, 0, 0, 0, 0, 0, 0, 0, 0])

    #Simulation parameters:
    flavin=Molecule.all_nuclei("flavin_anion")#Molecule.fromdb("fad",
["N5", "N14", "N16", "H20","H21", "H22", "H23", "H24", "H25", "H26", "H27",
"H28", "H29", "H30", "H31", "N10"])
    adenine =Molecule.all_nuclei("adenine_cation") #Molecule.fromdb("fad",
["N6-H1", "N6-H2", "C8-H"])
    sim = SemiclassicalSimulationJ([flavin, adenine])

    #perform simulation
    resultssw=kine_quantum_mary2rp_onoff(
        sim,
        num_samples,
        rho0,
        radical_pair=[5, 21],
        radical_pair2=[21, 37],
```

```
        ts=time,
        switchpoint=0,
        pfpdecimate=pfpdeci,
        Bs=Bs,
        D=0,
        J=0,
        kinetics=mat,
        relaxations=[relaxation],)

    #Calculate matrix of deltadeltaF vs time vs magnetic field
    fluorescence_txb_off = np.zeros((len(time), len(Bs)), dtype=complex)
    fluorescence_txb = resultssw["yield"][:, 0, -1,:]
    for i in range(0, len(Bs)):
        fluorescence_txb_off[:, i] = fluorescence_txb[:, 0]
    ddf_txb = np.real((fluorescence_txb -
fluorescence_txb_off)/(1+fluorescence_txb_off))
    #Extract pp data for maximum field
    pp=ddf_txb[:,-1]

    #Calculate matrix of deltadeltaF vs on->off time vs magnetic field
(prob at last time point)
    onoff_delaytxb=resultssw["yield"][-1, 0, :, :]
    ddf_delaytxb=np.real((onoff_delaytxb-fluorescence_txb_off[-1,-
1])/(1+fluorescence_txb_off[-1,-1]))
    #Extract PFP data for maximum field
    PFP=ddf_delaytxb[:,-1]

    #Setup data for B_1/2 fitting
    bhalf_time_pp = np.zeros((len(ddf_txb)))
    bhalf_time_pfp = np.zeros(len(dtime))
    fit_time = np.zeros((len(Bs), len(ddf_txb)))
    fit_error_time = np.zeros((3, len(ddf_txb)))
    R2_time = np.zeros((len(ddf_txb)))

    #Perform B_1/2 fit for PP data
    for i in range(1, len(ddf_txb)):
        (
            bhalf_time_pp[i],
```

```
            fit_time[:, i],
            fit_error_time[:, i],
            R2_time[i],
        ) = Bhalf_withLFE_fit_fixLhalf(Bs, ddf_txb[i, :])

    #Perform B_1/2 fit for PFP data
    for i in range(1, len(dtime)):
        (
            bhalf_time_pfp[i],
            fit_time[:, i],
            fit_error_time[:, i],
            R2_time[i],
        ) = Bhalf_withLFE_fit_fixLhalf(Bs, ddf_delaytxb[i, :])

    return time,pp,bhalf_time_pp,dtime,PFP,bhalf_time_pfp

if __name__ == '__main__':
    qval=[0.4e-4,0.3e-3, 0.5e-3, 1e-3, 5e-3]
    Bmax=30
    bpts=61
    tmax=3e-6
    tpts=1001
    pfpdeci=10
    sim1=JRWsemiclassicalPFP(Bmax=Bmax, bpts=bpts, tmax=tmax, tpts=tpts,
pfpdeci=pfpdeci, qconc=qval[0])
    sim2=JRWsemiclassicalPFP(Bmax=Bmax, bpts=bpts, tmax=tmax, tpts=tpts,
pfpdeci=pfpdeci, qconc=qval[1])
    sim3=JRWsemiclassicalPFP(Bmax=Bmax, bpts=bpts, tmax=tmax, tpts=tpts,
pfpdeci=pfpdeci, qconc=qval[2])
    sim4=JRWsemiclassicalPFP(Bmax=Bmax, bpts=bpts, tmax=tmax, tpts=tpts,
pfpdeci=pfpdeci, qconc=qval[3])
    sim5=JRWsemiclassicalPFP(Bmax=Bmax, bpts=bpts, tmax=tmax, tpts=tpts,
pfpdeci=pfpdeci, qconc=qval[4])
    ppdata=np.zeros((tpts,11))
    pfpdata=np.zeros((int(1+(tpts-1)/pfpdeci),11))

    ppdata[:,0]=sim1.result()[0]
    ppdata[:,1]=sim1.result()[1]
```

```
ppdata[:,2]=sim2.result()[1]
ppdata[:,3]=sim3.result()[1]
ppdata[:,4]=sim4.result()[1]
ppdata[:,5]=sim5.result()[1]

ppdata[:,6]=sim1.result()[2]
ppdata[:,7]=sim2.result()[2]
ppdata[:,8]=sim3.result()[2]
ppdata[:,9]=sim4.result()[2]
ppdata[:,10]=sim5.result()[2]

pfpdata[:,0]=sim1.result()[3]
pfpdata[:,1]=sim1.result()[4]
pfpdata[:,2]=sim2.result()[4]
pfpdata[:,3]=sim3.result()[4]
pfpdata[:,4]=sim4.result()[4]
pfpdata[:,5]=sim5.result()[4]

pfpdata[:,6]=sim1.result()[5]
pfpdata[:,7]=sim2.result()[5]
pfpdata[:,8]=sim3.result()[5]
pfpdata[:,9]=sim4.result()[5]
pfpdata[:,10]=sim5.result()[5]

path = "FAD_TRP_kinetics_Bhalf"
np.savetxt(path+"_pp.csv",ppdata, delimiter=',')
np.savetxt(path+"_pfp.csv",pfpdata, delimiter=',')
```

### 9.2 PP-MARY measurement

**Figure S9-2, S9-3, S9-4, S9-5, and S9-6** show the PP-MARY spectra for FAD (10μM) alone and in the presence of tryptophan at concentrations of 0.3 mM, 0.5 mM, 1.0 mM and 5.0 mM, respectively. The $B_{1/2}$ values are estimated by performing curve fitting using a single Lorentzian function:

$$\mathcal{L}_1\left(B \middle| B_{1/2}, MFE_{sat}\right) = MFE_{sat}\frac{B^2}{B^2 + B_{1/2}^2} \tag{S9-1}$$

The corresponding fitting parameters are listed in **Table S9-1, S9-2, S9-3, S9-4, and S9-5**, respectively.

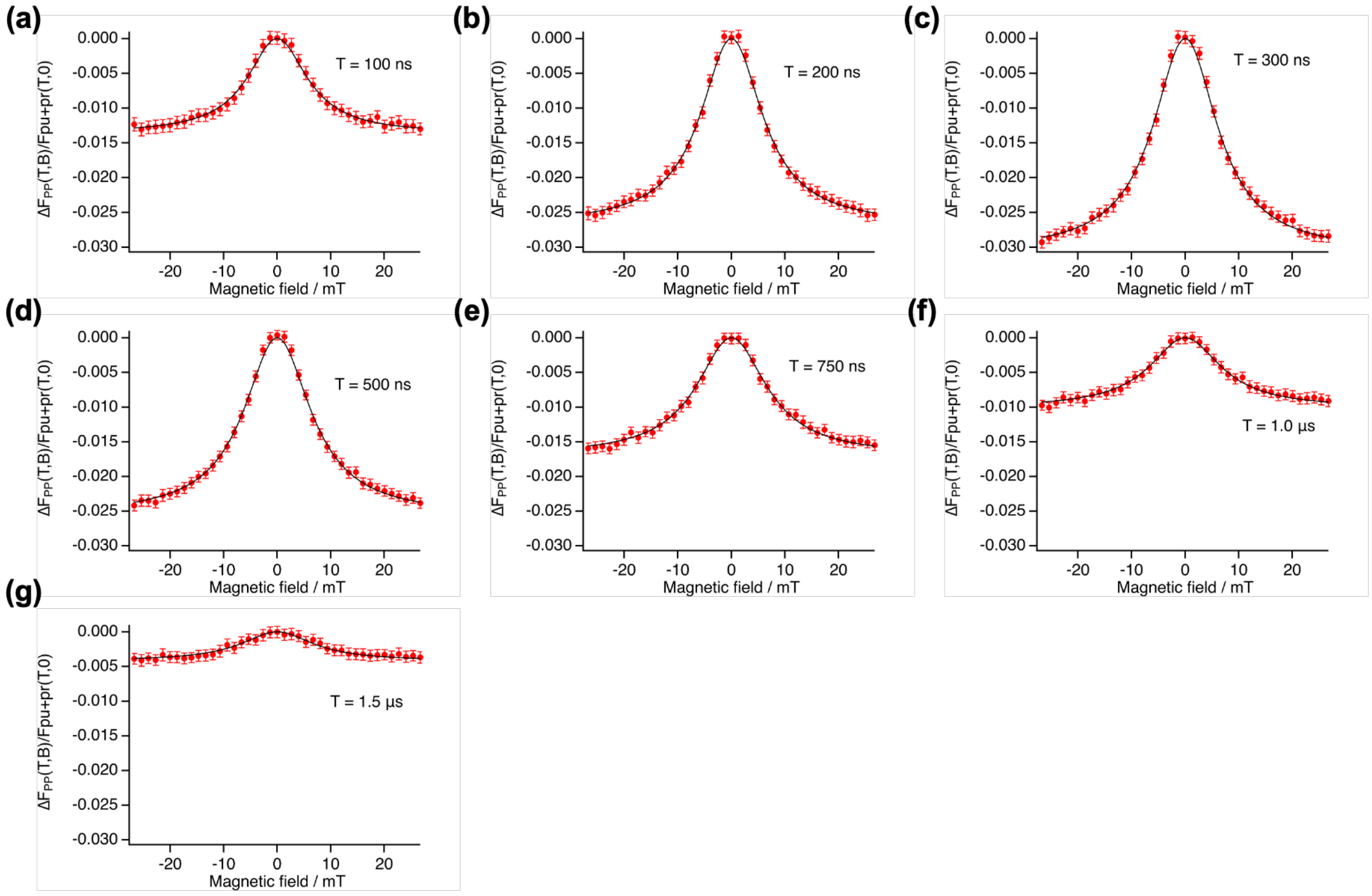

**Figure S9-2. PP-MARY spectra of FAD (10 μM) in pH 2.3 buffer (4.9 μm sample thickness), detected at different pump–probe laser delays of (a) 100 ns, (b) 200 ns, (c) 300 ns, (d) 500 ns, (e) 750 ns, (f) 1.0 μs, and (g) 1.5 μs.** The black lines represent fitting curves using a single Lorentzian function (eq. S9-2).

**Table S9-1** Fitting parameters of the PP-MARY spectra shown in **Figure S9-2.** Errors represent one standard deviation obtained from the fitting process.

| T / ns | $B_{1/2}$ / mT | MFEsat |
|---|---|---|
| 100 | 6.76 ± 0.15 | 0.01369 ± 0.00013 |
| 200 | 7.03 ± 0.10 | 0.02680 ± 0.00017 |
| 300 | 7.29 ± 0.12 | 0.03073 ± 0.00022 |
| 500 | 7.59 ± 0.11 | 0.02562 ± 0.00016 |
| 750 | 7.92 ± 0.17 | 0.01698 ± 0.00017 |
| 1000 | 8.08 ± 0.23 | 0.01011 ± 0.00014 |
| 1500 | 8.24 ± 0.45 | 0.00420 ± 0.00011 |

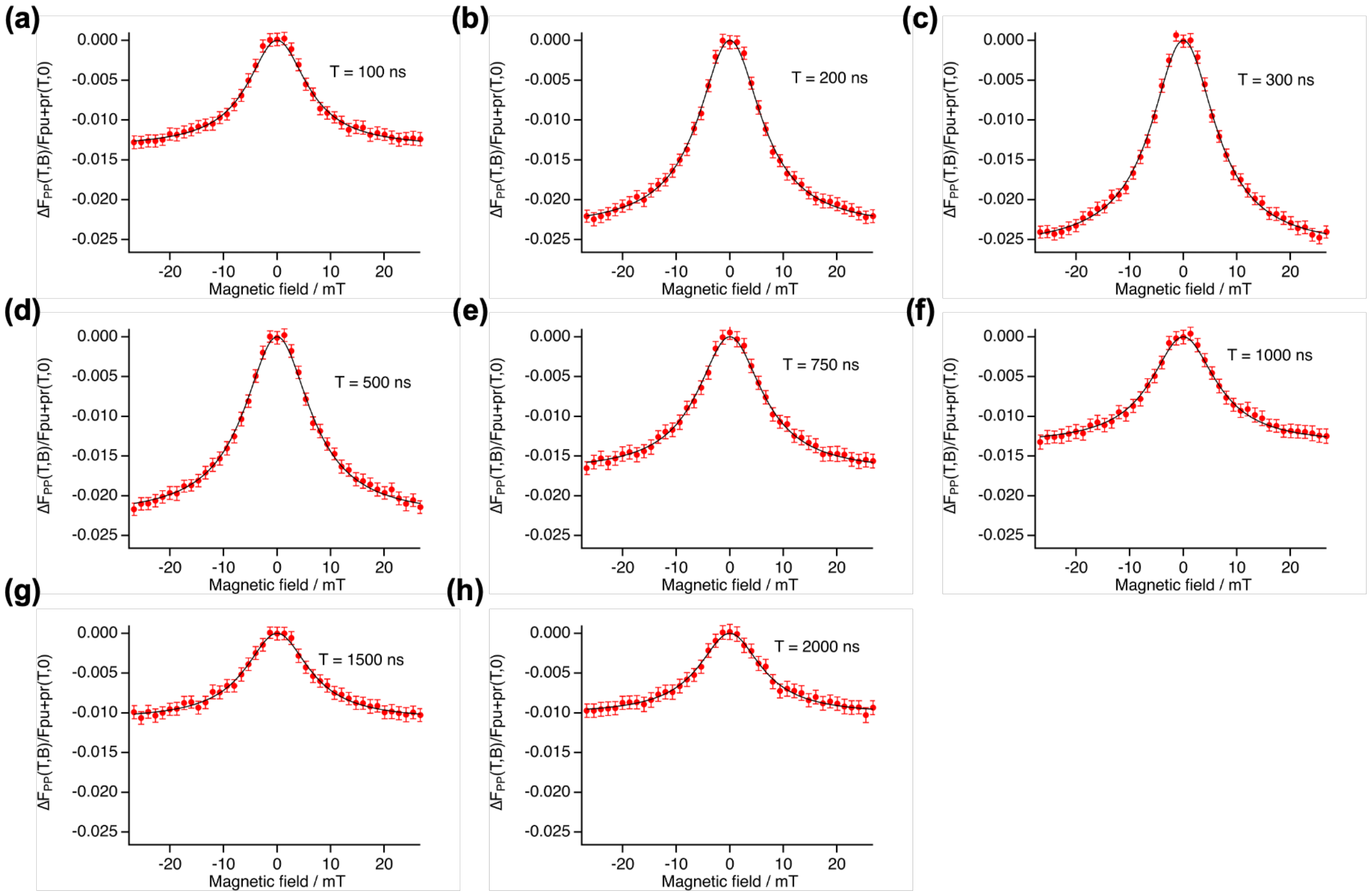

**Figure S9-3. PP-MARY spectra of FAD (10 µM) and tryptophan (0.3 mM) in pH 2.3 buffer (4.9 µm sample thickness), detected at different pump–probe laser delays of (a) 100 ns, (b) 200 ns, (c) 300 ns, (d) 500 ns, (e) 750 ns, (f) 1.0 µs, (g) 1.5 µs, and (h) 2.0 µs.** The black lines represent fitting curves using a single Lorentzian function (eq. S9-2).

**Table S9-2** Fitting parameters of the PP-MARY spectra shown in **Figure S9-3.** Errors represent one standard deviation obtained from the fitting process.

| T / ns | $B_{1/2}$ / mT | MFEsat |
|---|---|---|
| 100 | 6.73 ± 0.16 | 0.01343 ± 0.00013 |
| 200 | 7.11 ± 0.11 | 0.02359 ± 0.00015 |
| 300 | 7.27 ± 0.12 | 0.02607 ± 0.00018 |
| 500 | 7.52 ± 0.12 | 0.02263 ± 0.00017 |
| 750 | 7.36 ± 0.14 | 0.01700 ± 0.00014 |
| 1000 | 7.36 ± 0.17 | 0.01353 ± 0.00014 |
| 1500 | 7.23 ± 0.18 | 0.01090 ± 0.00012 |
| 2000 | 7.08 ± 0.20 | 0.01022 ± 0.00012 |

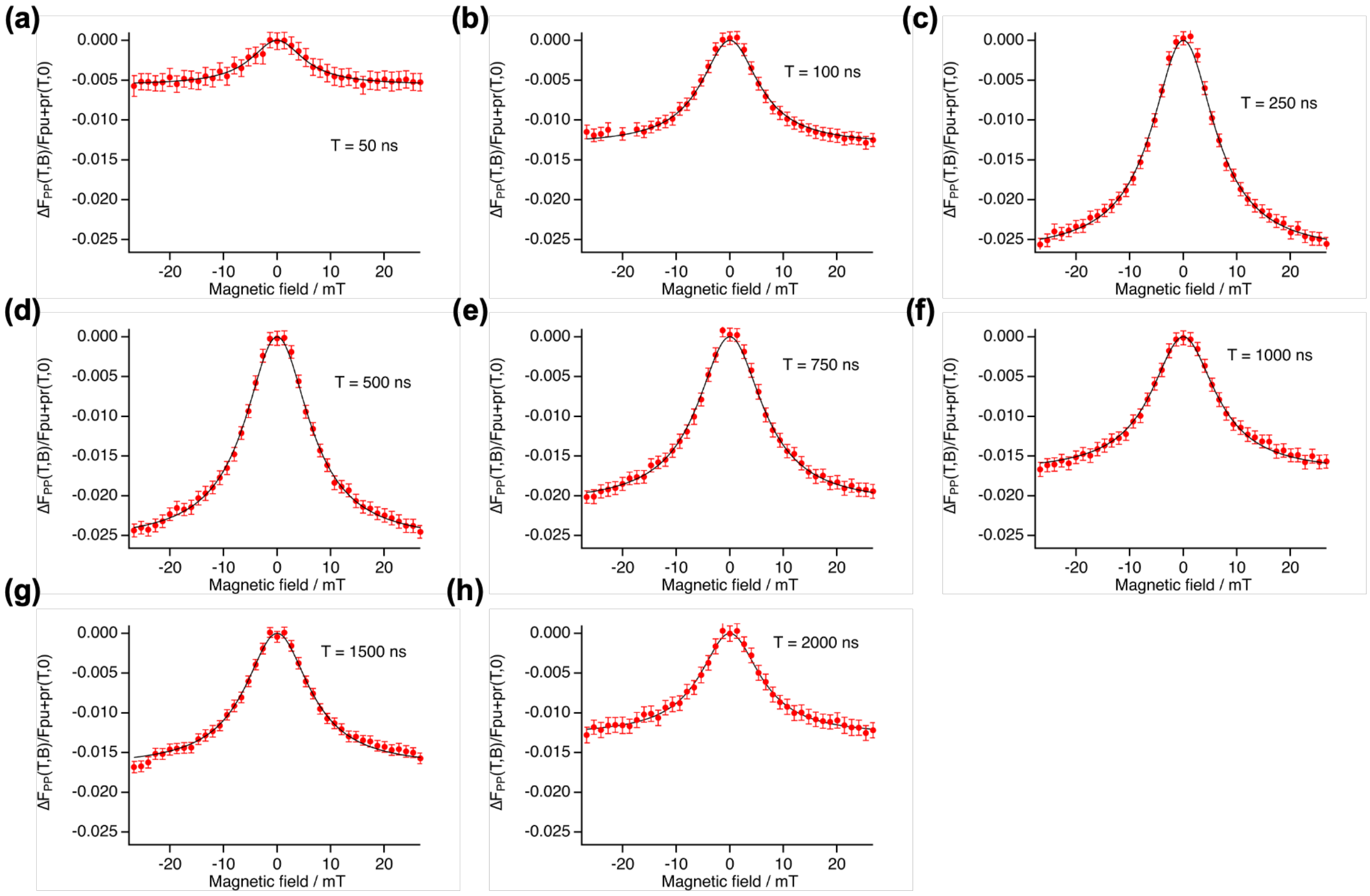

**Figure S9-4. PP-MARY spectra of FAD (10 µM) and tryptophan (0.5 mM) in pH 2.3 buffer (4.9 µm sample thickness), detected at different pump–probe laser delays of (a) 100 ns, (b) 200 ns, (c) 300 ns, (d) 500 ns, (e) 750 ns, (f) 1.0 µs, (g) 1.5 µs, and (h) 2.0 µs.** The black lines represent fitting curves using a single Lorentzian function (eq. S9-2).

**Table S9-3** Fitting parameters of the PP-MARY spectra shown in **Figure S9-4.** Errors represent one standard deviation obtained from the fitting process.

| T / ns | $B_{1/2}$ / mT | MFEsat |
|---|---|---|
| 50 | 5.96 ± 0.28 | 0.00562 ± 0.00010 |
| 100 | 6.50 ± 0.17 | 0.01311 ± 0.00014 |
| 250 | 7.12 ± 0.11 | 0.02673 ± 0.00018 |
| 500 | 7.28 ± 0.10 | 0.02578 ± 0.00016 |
| 750 | 7.38 ± 0.13 | 0.02108 ± 0.00017 |
| 1000 | 7.27 ± 0.14 | 0.01706 ± 0.00014 |
| 1500 | 7.32 ± 0.17 | 0.01678 ± 0.00017 |
| 2000 | 6.79 ± 0.16 | 0.01284 ± 0.00013 |

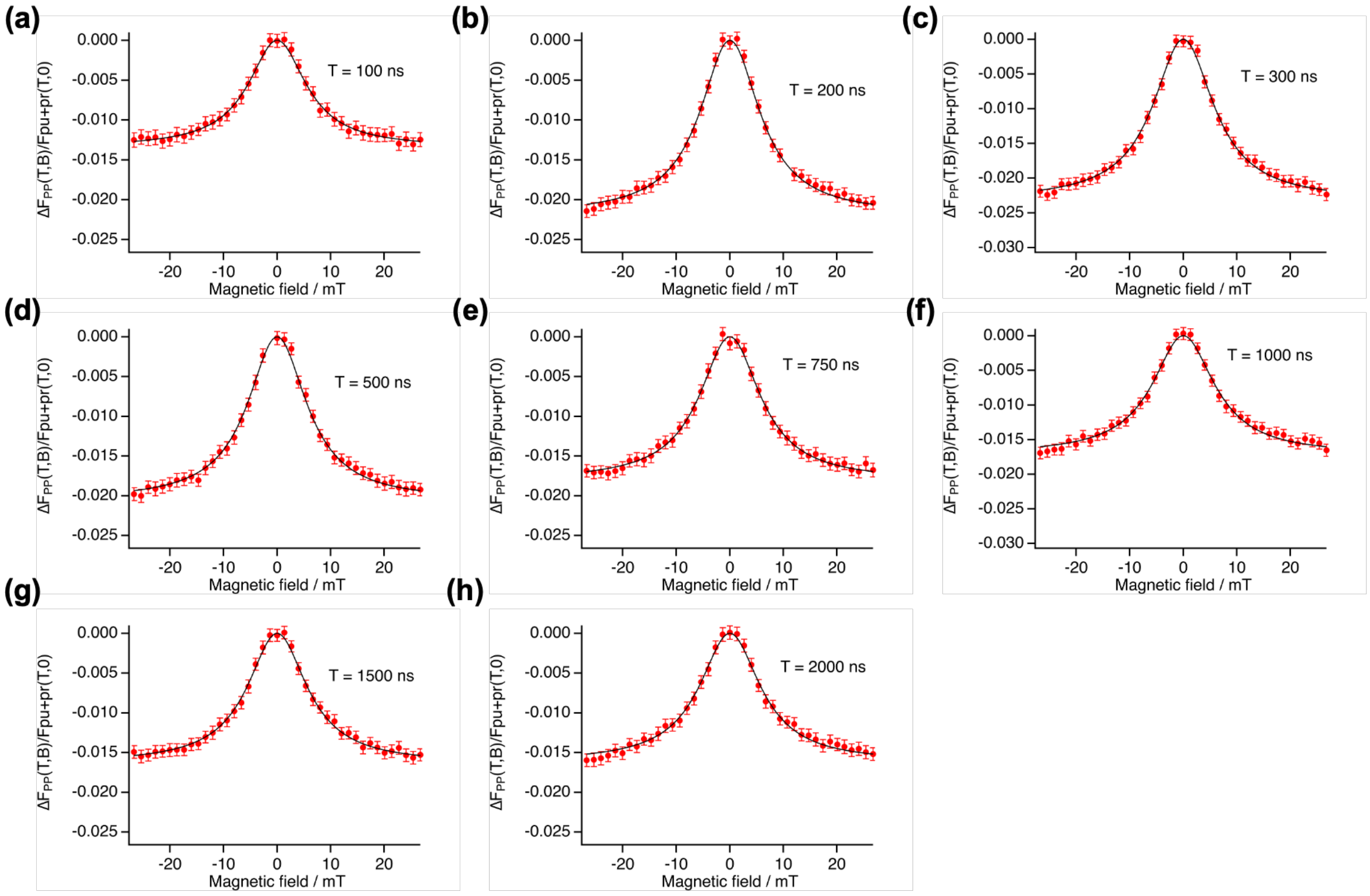

**Figure S9-5. PP-MARY spectra of FAD (10 µM) and tryptophan (1 mM) in pH 2.3 buffer (5 µm sample thickness), detected at different pump–probe laser delays of (a) 100 ns, (b) 200 ns, (c) 300 ns, (d) 500 ns, (e) 750 ns, (f) 1.0 µs, (g) 1.5 µs, and (h) 2.0 µs.** The black lines represent fitting curves using a single Lorentzian function (eq. S9-2).

**Table S9-4** Fitting parameters of the PP-MARY spectra shown in **Figure S9-5.** Errors represent one standard deviation obtained from the fitting process.

| T / ns | $B_{1/2}$ / mT | MFEsat |
|---|---|---|
| 100 | 6.54 ± 0.14 | 0.01343 ± 0.00012 |
| 200 | 6.76 ± 0.12 | 0.02189 ± 0.00016 |
| 300 | 6.89 ± 0.11 | 0.02318 ± 0.00016 |
| 500 | 6.71 ± 0.12 | 0.02058 ± 0.00015 |
| 750 | 6.89 ± 0.13 | 0.01807 ± 0.00014 |
| 1000 | 7.04 ± 0.17 | 0.01714 ± 0.00018 |
| 1500 | 6.74 ± 0.12 | 0.01635 ± 0.00012 |
| 2000 | 6.88 ± 0.16 | 0.01620 ± 0.00015 |

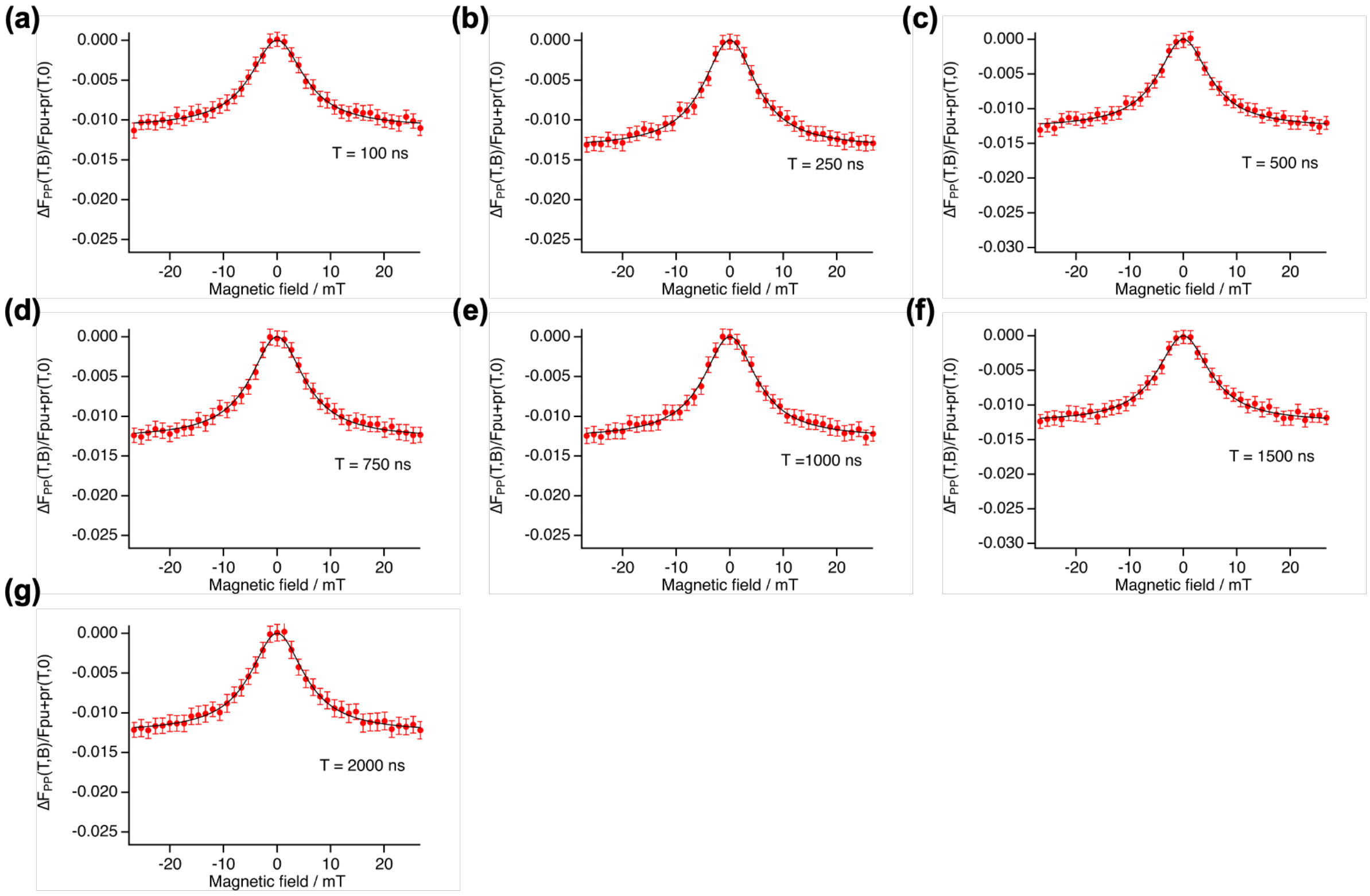

**Figure S9-6. PP-MARY spectra of FAD (10 µM) and tryptophan (5.0 mM) in pH 2.3 buffer (4.9 µm sample thickness), detected at different pump–probe laser delays of (a) 100 ns, (b) 250 ns, (c) 500 ns, (d) 750 ns, (e) 1.0 µs, (f) 1.5 µs, and (g) 2.0 µs.** The black lines represent fitting curves using a single Lorentzian function (eq. S9-2).

**Table S9-5** Fitting parameters of the PP-MARY spectra shown in **Figure S9-6.** Errors represent one standard deviation obtained from the fitting process.

| T / ns | $B_{1/2}$ / mT | MFEsat |
|---|---|---|
| 100 | 6.09 ± 0.16 | 0.01091 ± 0.00011 |
| 250 | 6.01 ± 0.14 | 0.01346 ± 0.00012 |
| 500 | 5.84 ± 0.15 | 0.01275 ± 0.00012 |
| 750 | 6.09 ± 0.14 | 0.01276 ± 0.00012 |
| 1000 | 6.03 ± 0.14 | 0.01270 ± 0.00011 |
| 1500 | 5.93 ± 0.14 | 0.01250 ± 0.00011 |
| 2000 | 6.11 ± 0.13 | 0.01247 ± 0.00011 |

### 9.3 PFP-MARY measurement

**Figure S9-7** shows the PFP-MARY spectra of the FAD (10μM) and Trp (0.3 mM) sample. Non-negligible LFEs were observed; therefore, the spectra were fitted using a double Lorentzian function to estimate the corresponding $B_{1/2}$ values:

$$\mathcal{L}_2(B|L_{1/2}, LFE_{sat}, B_{1/2}, MFE_{sat}) = LFE_{sat}\frac{B^2}{B^2+L_{1/2}^2} + MFE_{sat}\frac{B^2}{B^2+B_{1/2}^2} + offset \quad \text{(S9-2)}$$

The corresponding fitting parameters are listed **in Table S9-6**.

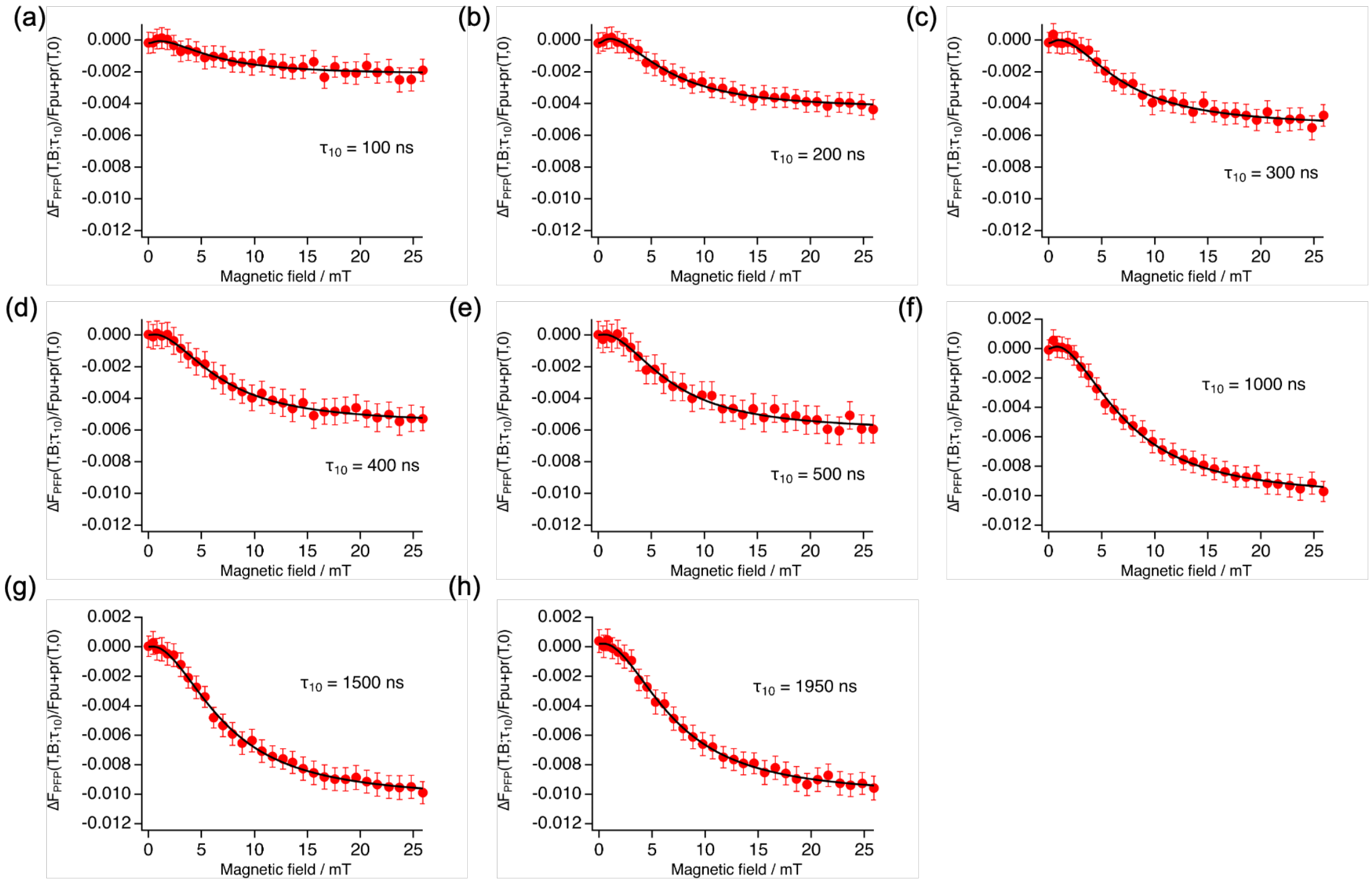

**Figure S9-7. PFP-MARY spectra of FAD (10 μM) and tryptophan (300 μM) in pH 2.3 buffer (5 μm sample thickness), detected at different RSMF On-Off shift delays of (a) 100 ns, (b) 200 ns, (c) 300 ns, (d) 400 ns, (e) 500 ns, (f) 1.0 μs, (g) 1.5 μs, and (h) 1.95 μs.** The black lines represent fitting curves using a double Lorentzian function (eq.S9-2).

**Table S9-6.** Fitting parameters of the PFP-MARY spectra shown in **Fig. S9-7**

| $\tau_{10}$ / ns | $L_{1/2}$ (fixed) / mT | LFEsat | $B_{1/2}$ / mT | MFEsat |
|---|---|---|---|---|
| 100 | 0.77 | 0.000298 ± 0.000044 | 6.14 ± 0.61 | 0.002262 ± 0.000095 |
| 200 | 0.77 | 0.000596 ± 0.000024 | 6.55 ± 0.17 | 0.004740 ± 0.000055 |
| 300 | 0.77 | 0.000507 ± 0.000045 | 6.87 ± 0.28 | 0.00576 ± 0.00011 |
| 400 | 0.77 | 0.000206 ± 0.000034 | 6.70 ± 0.20 | 0.005797 ± 0.000078 |
| 500 | 0.77 | 0.000198 ± 0.000052 | 6.74 ± 0.29 | 0.00627 ± 0.00012 |
| 1000 | 0.77 | 0.000502 ± 0.000036 | 7.18 ± 0.13 | 0.010666 ± 0.000088 |
| 1500 | 0.77 | 0.000211 ± 0.000039 | 7.07 ± 0.14 | 0.010537 ± 0.000094 |
| 1950 | 0.77 | 0.000201 ± 0.000039 | 7.01 ± 0.14 | 0.010523 ± 0.000094 |